\documentclass[12pt]{iopart}
\makeatletter
\def\set@curr@file#1{%
  \begingroup
    \escapechar\m@ne
    \xdef\@curr@file{\expandafter\string\csname #1\endcsname}%
  \endgroup
}
\def\quote@name#1{"\quote@@name#1\@gobble""}
\def\quote@@name#1"{#1\quote@@name}
\def\unquote@name#1{\quote@@name#1\@gobble"}
\makeatother
\usepackage{graphicx}
\usepackage{float}
\usepackage{wrapfig}
\expandafter\let\csname equation*\endcsname\relax
\expandafter\let\csname endequation*\endcsname\relax
\usepackage{color,soul}
\usepackage{amsmath}
\usepackage{listings}
\usepackage{amssymb}
\usepackage{relsize}
\usepackage{tikz}
\usepackage{siunitx}
\usepackage[linktocpage=true]{hyperref}
\hypersetup{colorlinks=true,citecolor=red,urlcolor=blue}
\usepackage[retainorgcmds]{IEEEtrantools}
\usepackage{pgfplots}
\usepackage{tabulary}
\usepackage{xfrac}
\usepackage{ctable}
\usepackage{xr-hyper}
\usepackage{gensymb}
\usepackage{comment}
\usetikzlibrary{patterns}
\pgfplotsset{width=10cm, compat=1.10}
\usepgfplotslibrary{fillbetween}
\pgfmathdeclarefunction{surf1}{0}{\pgfmathparse{0.2*sin(x)}}
\pgfmathdeclarefunction{surf2}{0}{\pgfmathparse{-4+0.2*sin(x)}}
\renewcommand*{\contentsname}{ELGAR}

\begin{document}

\title[ELGAR]{Technologies for the ELGAR large scale atom interferometer array}

\author{B.~Canuel$^1$, S. Abend$^2$, P. Amaro-Seoane$^{3,4,5,6,7}$, F.~Badaracco$^{8,9}$, Q.~Beaufils$^{10}$, A.~Bertoldi$^1$, K.~Bongs$^{11}$, P.~Bouyer$^1$, C.~Braxmaier$^{12,13}$, W.~Chaibi$^{14}$, N.~Christensen$^{14}$, F.~Fitzek$^{2,15}$, G.~Flouris$^{16}$, N.~Gaaloul$^2$, S.~Gaffet$^{17}$, C.~L.~Garrido Alzar$^{10}$,  R.~Geiger$^{10}$, S.~Guellati-Khelifa$^{18}$, K.~Hammerer$^{15}$, J.~Harms$^{8,9}$, J.~Hinderer$^{19}$, M.~Holynski$^{11}$, J.~Junca$^1$, S.~Katsanevas$^{20}$, C.~Klempt$^2$, C.~Kozanitis$^{16}$, M.~Krutzik$^{21}$, A.~Landragin$^{10}$, I.~L{\`a}zaro~Roche$^{17}$, B.~Leykauf$^{21}$, Y.-H.~Lien$^{11}$, S.~Loriani$^2$, S.~Merlet$^{10}$, M.~Merzougui$^{14}$, M.~Nofrarias$^{3,4}$, P.~Papadakos$^{16,22}$, F.~Pereira~dos~Santos$^{10}$, A.~Peters$^{21}$, D.~Plexousakis$^{16,22}$, M.~Prevedelli$^{23}$, E.~M.~Rasel$^2$, Y.~Rogister$^{19}$, S.~Rosat$^{19}$, A.~Roura$^{24}$, D.~O.~Sabulsky$^1$, V.~Schkolnik$^{21}$, D.~Schlippert$^2$, C.~Schubert$^2$, L.~Sidorenkov$^{10}$, J.-N.~Siem{\ss}$^{2,15}$,  C.~F.~Sopuerta$^{3,4}$, F.~Sorrentino$^{25}$, C.~Struckmann$^2$, G.~M.~Tino$^{26}$, G.~Tsagkatakis$^{16,22}$, A.~Vicer{\'e}$^{27}$, W.~von~Klitzing$^{28}$, L.~Woerner$^{12,13}$, X.~Zou$^1$}

\address{$^1$ LP2N, Laboratoire Photonique, Num{\'e}rique et Nanosciences, Universit{\'e} Bordeaux--IOGS--CNRS:UMR 5298, rue F. Mitterrand, F--33400 Talence, France}
\address{$^2$ Leibniz Universit{\"a}t Hannover, Institut f{\"u}r Quantenoptik, Welfengarten 1, D-30167 Hannover, Germany}
\address{$^3$ Institute of Space Sciences (ICE, CSIC), Campus UAB, Carrer de Can Magrans s/n, 08193 Cerdanyola del Vall\`es (Barcelona), Spain}
\address{$^4$ Institute of Space Studies of Catalonia (IEEC), Carrer del Gran Capit\`a, 2-4, Edifici Nexus, despatx 201, 08034 Barcelona, Spain}
\address{$^5$ Kavli Institute for Astronomy and Astrophysics, Beijing 100871, China}
\address{$^6$ Institute of Applied Mathematics, Academy of Mathematics and Systems Science, CAS, Beijing 100190, China}
\address{$^7$ Zentrum f{\"u}r Astronomie und Astrophysik, TU Berlin, Hardenbergstra{\ss}e 36, 10623 Berlin, Germany}
\address{$^8$ Gran Sasso Science Institute (GSSI), I-67100 L'Aquila, Italy}
\address{$^9$ INFN, Laboratori Nazionali del Gran Sasso, I-67100 Assergi, Italy}
\address{$^{10}$ LNE--SYRTE, Observatoire de Paris, Universit{\'e} PSL, CNRS, Sorbonne Universit{\'e}, 61, avenue de l'Observatoire, F--75014 PARIS, France}
\address{$^{11}$ Midlands Ultracold Atom Research Centre, School of Physics and Astronomy, University of Birmingham, Birmingham, B15 2TT, United Kingdom}
\address{$^{12}$ ZARM, Unversity of Bremen, Am Fallturm 2, 28359 Bremen, Germany}
\address{$^{13}$ DLR, German Aerospace Center, Linzer Strasse 1, 28359 Bremen, Germany}
\address{$^{14}$ ARTEMIS,  Universit{\'e} C{\^o}te d'Azur, Observatoire de la C{\^o}te d'Azur, CNRS, F--06304 Nice, France}
\address{$^{15}$ Institute for Theoretical Physics and Institute for Gravitational Physics (Albert-Einstein-Institute), Leibniz University Hannover, Appelstrasse 2, 30167 Hannover, Germany}
\address{$^{16}$ Institute of Computer Science, Foundation for Research and Technology - Hellas, 70013, Heraklion, Greece}
\address{$^{17}$ LSBB, Laboratoire Souterrain Bas Bruit,  CNRS, Avignon University - La grande combe, 84400 Rustrel, France}
\address{$^{18}$ Laboratoire Kastler Brossel, Sorbonne Universit\'e, CNRS, ENS-PSL Research University, Coll\`ege de France, 4 place Jussieu, 75005 Paris, France}
\address{$^{19}$ Institut de Physique du Globe de Strasbourg, UMR 7516, Universit{\'e} de Strasbourg/EOST, CNRS, 5 rue Descartes, 67084 Strasbourg, France}
\address{$^{20}$ European Gravitational Observatory (EGO), I-56021 Cascina (Pi), Italy}
\address{$^{21}$ Humboldt-Universit{\"a}t zu Berlin, Institute of Physics, Newtonstrasse 15, 12489 Berlin, Germany}
\address{$^{22}$ Computer Science Department, University of Crete, 70013, Heraklion, Greece}
\address{$^{23}$ Dept. of Physics and Astronomy, Univ. of Bologna, Via Berti-Pichat 6/2, I-40126 Bologna, Italy}
\address{$^{24}$ Institute of Quantum Technologies, German Aerospace Center (DLR), S\"oflinger Str.~100, 89077 Ulm, Germany}
\address{$^{25}$ Istituto Nazionale di Fisica Nucleare (INFN) Sezione di Genova, via  Dodecaneso 33,  Genova, Italy} 
\address{$^{26}$ Dipartimento di Fisica e Astronomia and LENS Laboratory, Universit\`a di Firenze and INFN-Sezione di Firenze, via  Sansone 1,  Sesto Fiorentino, Italy} 
\address{$^{27}$ Universit\`a degli Studi di Urbino ``Carlo Bo'', I-61029 Urbino, Italy and INFN, Sezione di Firenze, I-50019 Sesto Fiorentino, Firenze, Italy}
\address{$^{28}$ Institute of Electronic Structure and Laser, Foundation for Research and Technology - Hellas, 70013, Heraklion, Greece}

\ead{benjamin.canuel@institutoptique.fr}
\begin{abstract}
We proposed the European Laboratory for Gravitation and Atom-interferometric Research (ELGAR), an array of atom gradiometers aimed at studying space-time and gravitation with the primary goal of observing gravitational waves (GWs) in the infrasound band with a peak strain sensitivity of $3.3 \times 10^{-22}/\sqrt{\text{Hz}}$ at 1.7 Hz.
In this paper we detail the main technological bricks of this large scale detector and emphasis the research pathways to be conducted for its realization. We discuss the site options, atom optics, and source requirements needed to reach the target sensitivity. We then discuss required seismic isolation techniques, Gravity Gradient Noise reduction strategies, and the metrology of various noise couplings to the detector.

\end{abstract}
%Uncomment for PACS numbers title message
%\pacs{00.00, 20.00, 42.10}
% Keywords required only for MST, PB, PMB, PM, JOA, JOB? 
%\vspace{2pc}
%\noindent{\it Keywords}: Article preparation, IOP journals
% Uncomment for Submitted to journal title message
%\submitto{\JPA}
% Comment out if separate title page not required
\maketitle

\tableofcontents

\section*{Introduction}
%
%This is introduction
A new infrastructure based on a large scale Atom Interferometer (AI) network was proposed, the European Laboratory for Gravitation and Atom-interferometric Research (ELGAR) \cite{canuel2019elgar}.
With projections based on the latest trends in atom interferometry and atom optics, ELGAR stands as a terrestrial candidate for detecting GW in the mid-frequency band, but could also enable new tests of fundamental physics and have applications in the geosciences.
In this paper, we give details of the main technological bricks referenced in the ELGAR proposal \cite{canuel2019elgar}.
In the laboratory environment, AIs are extremely sensitive inertial sensing experiments performing accurate and precise measurements of accelerations \cite{Peters1999,Freier2016}, rotations \cite{Gustavson1997,Stockton2011,Savoie2018}, and fundamental physics \cite{Rosi2014,Bouchendira2011,Dimopoulos2007,Aguilera2014,Schlippert2014,Zhou2015,Rosi2017,Overstreet2018,Burrage2015,Jaffe2017,Sabulsky2019}.
Such techniques have been refined and adapted to fulfill instrumentation roles in the applied sciences outside of the laboratory \cite{Trimeche2019,deAngelis2009,Mnoret2018,Bidel2018,KasevichPatent2006,Barrett2019,Bongs2019}, moving this technology toward mobile, compact, and rugged experiments able to operate in varying and extreme environmental conditions. 
Based on such advances, large-scale AI experiments are already under development to pursue multiple scientific and technological developments; such as MIGA \cite{Canuel2018}, VLBAI \cite{Schlippert_2020}, ZAIGA \cite{Zhan_2019}, MAGIS \cite{magiswebsite,Coleman2018arXiv}, AEDGE \cite{AEDGE2020} or AION \cite{Badurina_2020}.
Motivated by the latest progress in sensitivity, reliability, and key advances in detection strategies with AI arrays \cite{Chaibi2016}, ELGAR envisions a transition from laboratory and field experiments to an decikilometric infrastructure to probe the mid-band (0.1 Hz to 10 Hz) GW spectrum. 
In this paper, we detail the technologies of the main systems of this instrument and identify the key developments needed. The realization a such large scale AI instrument will require new metrological studies: we also derive here the impact of various noise couplings to the detector.

\par We organize the manuscript as follows: 
Sec.~\ref{sec_Detector_configuration} briefly summarizes the measurement concept of large-scale atom interferometry and the geometry of ELGAR.
Sec.~\ref{sec_Detector_site} discusses three installation sites under consideration; these sites, located in Italy and France, are evaluated in terms of ambient noise. 
Sec.~\ref{sec_Atom_optics} and Sec.~\ref{sec_Atom_source} details the atom interrogation process and the properties of the individual atom sources. 
Sec.~\ref{sec_seis} gives insights to the suspension system required for the interrogation optics of the interferometer. 
Sec.~\ref{sec_NN_reduction} presents the sensitivity of ELGAR to different sources of Newtonian Noise and presents a mitigation strategy. 
Sec.~\ref{sec_other_noise_couplings} gives a complete view of the required metrology of the instrument, identifying and projecting the different noise sources in terms of equivalent strain. 

\section{The ELGAR detector}\label{sec_Detector_configuration}
\begin{figure}[ht!]
\centering
\includegraphics[width=.5\linewidth]{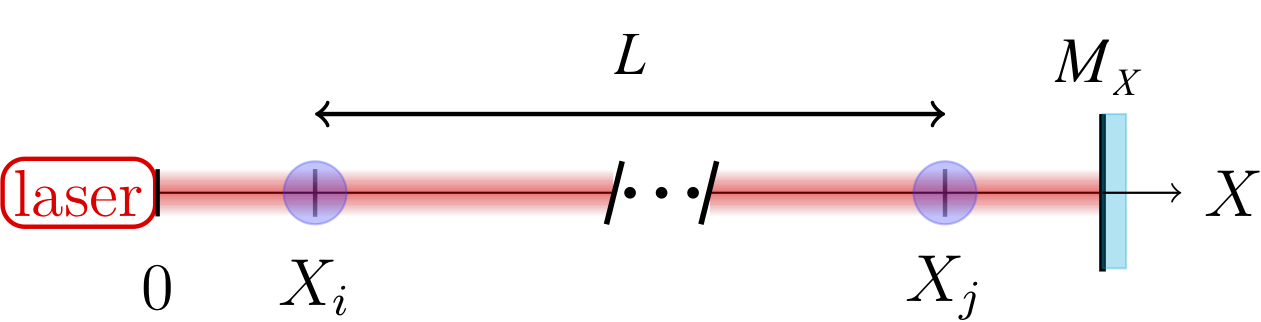}
\caption{Scheme of the atom gradiometer configuration, from \cite{canuel2019elgar}.
\label{fig:AIgradio}}
\end{figure}

Here we give a brief summary and schematic description of how the ELGAR detector is sensitive to gravitational waves. A more complete description of ELGAR geometry and signal extraction can be found in \cite{canuel2019elgar}.
The basis of the detector is the atom gradiometer configuration shown in Fig.~\ref{fig:AIgradio}: two free-falling atom interferometers placed at positions $X_{i,j}$ along the $x$-axis are interrogated by a common laser beam which is retro-reflected by a mirror placed at position $M_X$. The gradiometer signal is obtained by a differential measurement between the two AIs. Sensitivity to GWs along the gradiometer baseline $L$ arise from the accumulated phase difference between the two counter-propagating beams.
Referring to \cite{Canuel2018,canuel2019elgar} the strain sensitivity $S_{h}$ of the gradiometer is obtained by the sum :
\begin{align}\label{eq:Sh}
S_{h}&=\frac{4S_{\delta \nu}(\omega)}{\nu^{2}}+\frac{4S_{NN_1}(\omega)}{L^2}+\frac{4\omega^{2}S_{\delta x_{M_X}}(\omega)}{c^{2}}+\frac{8S_{\epsilon}(\omega)}{(2nk_\mathrm{l})^2L^{2}|\omega G(\omega)|^{2}}\, ,
\end{align}
where ${k_\mathrm{l}=\frac{2\pi\nu}{c}}$ is the wave number of the interrogation laser, $S_{NN_1}(\omega)$ is the is the Gravity Gradient Noise: PSD of the relative displacement of the atom test masses with respect to the interrogation laser, $S_{\delta x_{M_X}}$ is the PSD of the seismic position noise of the retro-reflecting mirror, $S_{\delta \nu}(\omega)$ is the PSD of the interrogation laser frequency noise; $S_{\epsilon}$ is the PSD of the detection noise, and $G(\omega)$ is the Fourier transform of 
AI sensitivity function \cite{Cheinet08}.

\begin{figure}[htp!]
\centering
\includegraphics[width=1\linewidth]{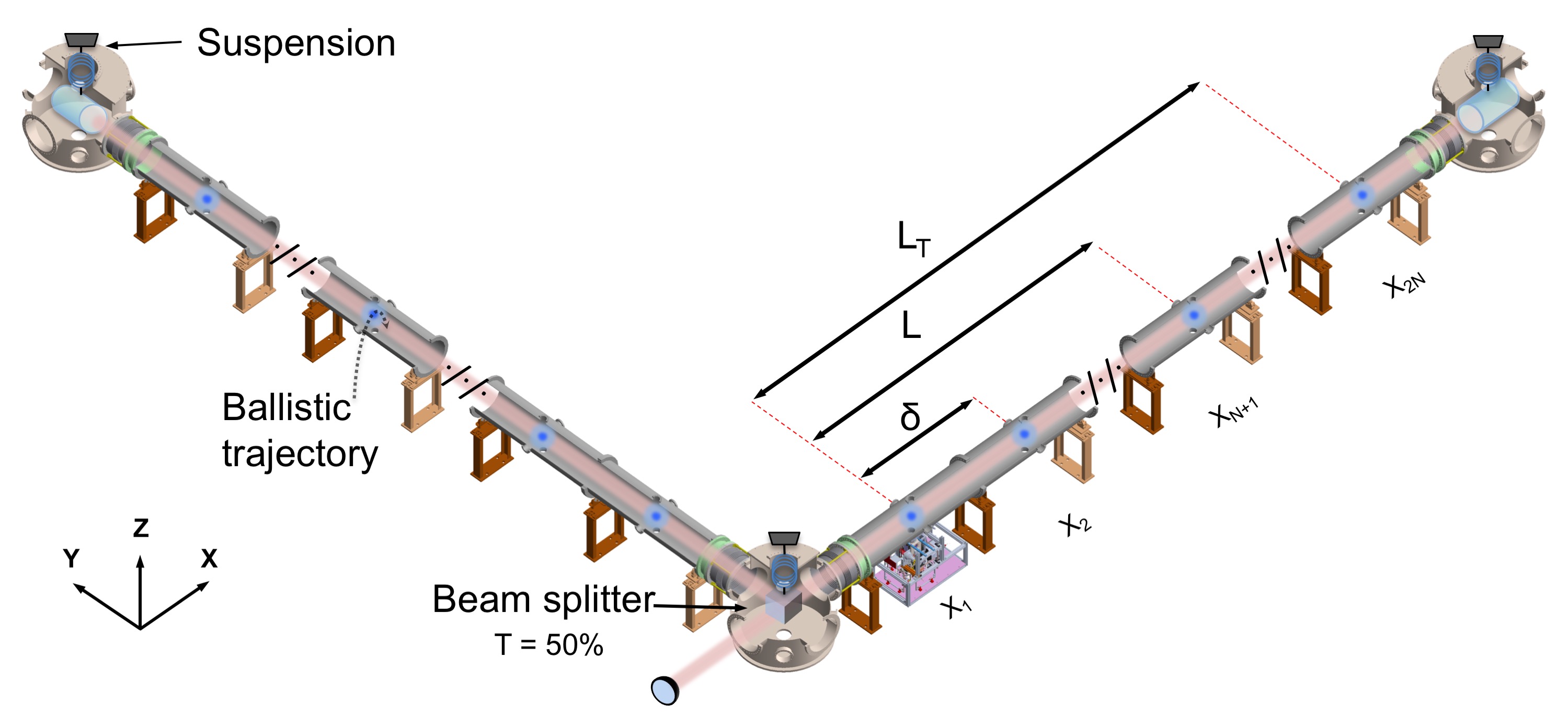}
\caption{Schematic diagram of the proposed ELGAR infrastructure, from \cite{canuel2019elgar}.
Each arm uses an array of $N$ gradiometers of baseline $L$ regularly spaced by a length  $\delta$ along a total baseline of $L_{T}$.
The retro-reflection mirrors and the beam splitter are placed on suspension systems.
\label{fig:ElgarGeo}}
\end{figure}

Based on this gradiometric configuration, the full ELGAR instrument geometry is shown in Fig.~\ref{fig:ElgarGeo}. Each arm of the detector is 
composed of an array of $N$ gradiometers regularly spaced by a length  $\delta$ along a total baseline of $L_{T}$. In this configuration, the GW signal is extracted from the difference between the averaged signals of the gradiometers of each arms, which enables to reduce the influence of  Gravity Gradient Noise (see Sec.\ref{sec_NN_reduction}).
Using this averaged signal the strain sensitivity of the detector is~\cite{canuel2019elgar}:
\begin{equation}\label{eq:SHnetwork}
S_h(\omega)=\frac{4\omega^{2}}{c^{2}}S_{\delta x_{M_X}}(\omega)+\frac{S_{NN}(\omega)}{L^2}+\frac{4S_{\epsilon}(\omega)}{N(2nk_\mathrm{l})^2L^2|\omega G(\omega)|^2}\, , 
\end{equation}
where $S_{NN}(\omega)$ is the is the averaged Gravity Gradient Noise in each single arm.

\section{Detector site}\label{sec_Detector_site}
Following this, we discuss the properties of candidate sites in France and Italy which could host ELGAR: In France, the Laboratoire Souterrain \`{a} Bas Bruit (LSBB), an underground low-noise laboratory located in Rustrel, east of Avignon, which hosts the MIGA prototype antenna \cite{Canuel2018}; in Italy, two candidate sites on Sardinia contained within former mining concessions. 
\subsection{Site requirements}\label{sec_site_requirements}
Previous metrological studies carried out for both AI and GW detection have shown that a candidate site for ELGAR should account for stringent environmental requirements which otherwise could impact its functionality.
This includes controlling or monitoring signals that spoil GW detection and those that affect the individual atom interferometers.
Practical aspects are also important and must be examined, as well as feasibility of installation. 
The detector requires an analysis of environmental impact and that it be far from present and future anthropogenic disturbances, environmental pollution originating from human activity.
The construction must consider preexisting infrastructure, additions to infrastructure, impact on the environment and the local community, as well as identify suitable infrastructure for users. 
Finally, a cost model for each candidate site is required to compare the differing cost between suitable sites.
%%%
\par Regarding metrological aspects, seismic noise has proven to be a major concern for both AI and GW detection.
On an AI, vibrations during the launch and preparation of an atomic sample translate into fluctuation of the atomic readout but the most important impact is to imprint spurious atomic phase into the AI from movement of the retroreflection mirror.
A gradiometric configuration reduces the sensitivity of the detector to vibrations, see section~\ref{sec_Detector_configuration}, but given ELGAR's projected strain sensitivity, vibrations and rotations must be carefully managed.
It is necessary to counter the noise in rotation and in displacement to measure GWs.
Another metrological aspect with impact on AI are time-varying stray magnetic fields and field gradients and the difference in these fields between each AI. 
All atomic species, regardless of their nuclear spin, have some dependency on magnetic fields in all AI techniques. 
Magnetic fields and magnetic field gradients changing in time present a technical noise that can only be met with magnetic shielding or active compensation.
A candidate site requires mapping of magnetic fields and field gradients to mitigate this technical noise concern.
Related to seismic noise, local gravity gradient noise is a important  source of technical noise for GW detection in the ELGAR observation band, see sections~\ref{sec_NN_reduction}.
This noise can be separated into a seismic component and an atmospheric component, which places importance on local geological features and the local climate of the candidate site. 
Finally, we must consider the feasibility of installing the detector at a candidate site; construction of such a detector requires significant anthropogenic disturbances, the need to ensure future ecological protection, the availability or construction of a host facility, and local infrastructure for staff and installation of the detector.
It is crucial to examine the impact of the construction and operation of the detector to the candidate site and surrounding community.
%%% 
\par A candidate detector site requires characterization and monitoring of the local seismic activity.
Correlation of this seismic noise across the detector site poses another technical noise problem - this movement of mass around the detector, amplified or attenuated by the local soil/rock composition, is seismic contribution to gravity gradient noise along one arm.
In these ways, a site that has low seismic noise properties year round is advantageous; continuous monitoring by a network of seismic detectors is required.
The data provided by a network is required to model the seismic noise and design seismic isolation units for the detector - the noise can be either up-converted or down-converted in frequency to remove it from within the observation band, see section~\ref{sec_seis}.
Further compounding the problem, broad contrasts in seismic impedance, e.g. ore deposits or geological levels, at a site complicate seismic analysis via reflection, conversion, or diffraction of surface waves - geological study of a site is required to mitigate such phenomena. 
In addition to concerns about seismic properties at the candidate site, anthropogenic noise is endemic throughout the ELGAR observation band - sites with nearby industrial and/or human activity must be ruled out due to vibrations. 
As a starting point, we consider sites appropriate for an underground detector as it provides a well controlled environment for large scale atom interferometry \cite{Canuel2018} in that the magnitude of seismic noise and GGN is reduced throughout the ELGAR observation band.
%%%
\par Seismic noise considerations are related to the geology and surrounding environment of the candidate site.
Soil and rock composition of the detector site is critical for both seismic GGN contributions, but also for magnetic noise; mechanical properties like the density, homogeneity, and other geological aspects have an impact on infrastructure works, changing the feasibility and cost. 
Geological features such as nearby flowing water are of interest to geophysicists and hydrologists, which enhances a candidate site's potential.
For example, while water flow near to the detector could constitute a large GGN background, it is a potential candidate for a major hydrogeological studies~\cite{Henry, Geiger2015, karsth20} like ground water transfer in the critical zone and the layout of underground water resources.
The technical noise added from such a background can be mitigated with a network of seismometers, tiltmeters, flowmeters, and gravimeters and the models created through study of these hydrogeological phenomena. 
%%%
\par The atmospheric contribution to gravity gradient noise places restrictions on the detector location.
Any atmospheric gravity gradient noise contribution is attenuated in an underground detector; monitoring, modeling, and characterization are still required.
Atmospheric temperature, propagating infrasound fields ~\cite{Fiorucci,Junca19}, barometric pressure readings, and wind direction/magnitude all must be monitored to account for their effect within the detector's observation band. 
A candidate site should not be at the confluence of multiple weather systems, to avoid large barometric pressure fluctuations.
In addition, we seek to mitigate the risk posed by storms, floods, and excessive humidity to an underground detector by analyzing climate data. 
%%%
\par Feasibility of construction and operation of a detector must be considered in choosing a candidate site.
Rock with strong impedance contrasts, like a high heterogeneity with composite rock of varying mechanical resistances are more difficult to bore through and could prove problematic for reasons of AI metrology and GW detection. 
Rural regions where candidate sites are located may lack the infrastructure required to host and support a detector; it could prove costly to transport the tunnel boring equipment along safe roads.
Removing the material from boring requires road access that local infrastructure may not be well equipped for; this waste material must be carefully managed and stored appropriately considering its possible environmental impact. 
This strain on local communities in remote locations must be studied for each candidate site.
A study on the impact of construction at candidate site must be performed to ensure minimum disruption to the ecosystem, safety of ground water, the power grid, and the local population. 
%%%
\par An output of candidate site surveys would be to approximate installation costs through a cost model at various locations for an ELGAR detector.
These costs include labor, power requirements, travel, geological studies, legal and environmental administration, as well as maintenance for existing facilities and the construction of new ones to help facilitate a large detector and it's community - interactions with existing initiatives could reduce costs significantly. 
Practical considerations necessitate that the detector not be too far from local facilities for staff, like schools and residences, as well as reasonable access to the facility via road, rail, and access to the detector via access shafts or tunnels. 
In addition, there are water, power, and sewage requirements for a site, as well as environmental restrictions and managing the risk of developing anthropogenic noise in the future.
The site would need to be located on protected land or land that could be declared as such to mitigate the risk of nearby human development that could impede or jeopardize the detector's operation. 
All these requirements, in addition to construction and maintenance, constitute a major disruption to any local community and considerations must be made about how the ELGAR detector affects the detector site. 
A full survey of candidate sites could be conducted, like what was accomplished for the Einstein Telescope~\cite{ET-design}.
\subsection{Candidate Sites in France and Italy}
In this paper, we include two candidate sites that are presently under study. The first site considered is the Laboratoire Souterrain \`{a} Bas Bruit (LSBB), located in the hamlet of Rustrel, in southern France; it is the location of the MIGA experiment. 
The second set of sites are located on the Italian island of Sardinia and are all former mining concessions considered for both GW detection and gravity gradiometry with AI.
\subsubsection{The LSBB facility.}
\begin{figure}[htp!]
\centering
\includegraphics[width=1\linewidth]{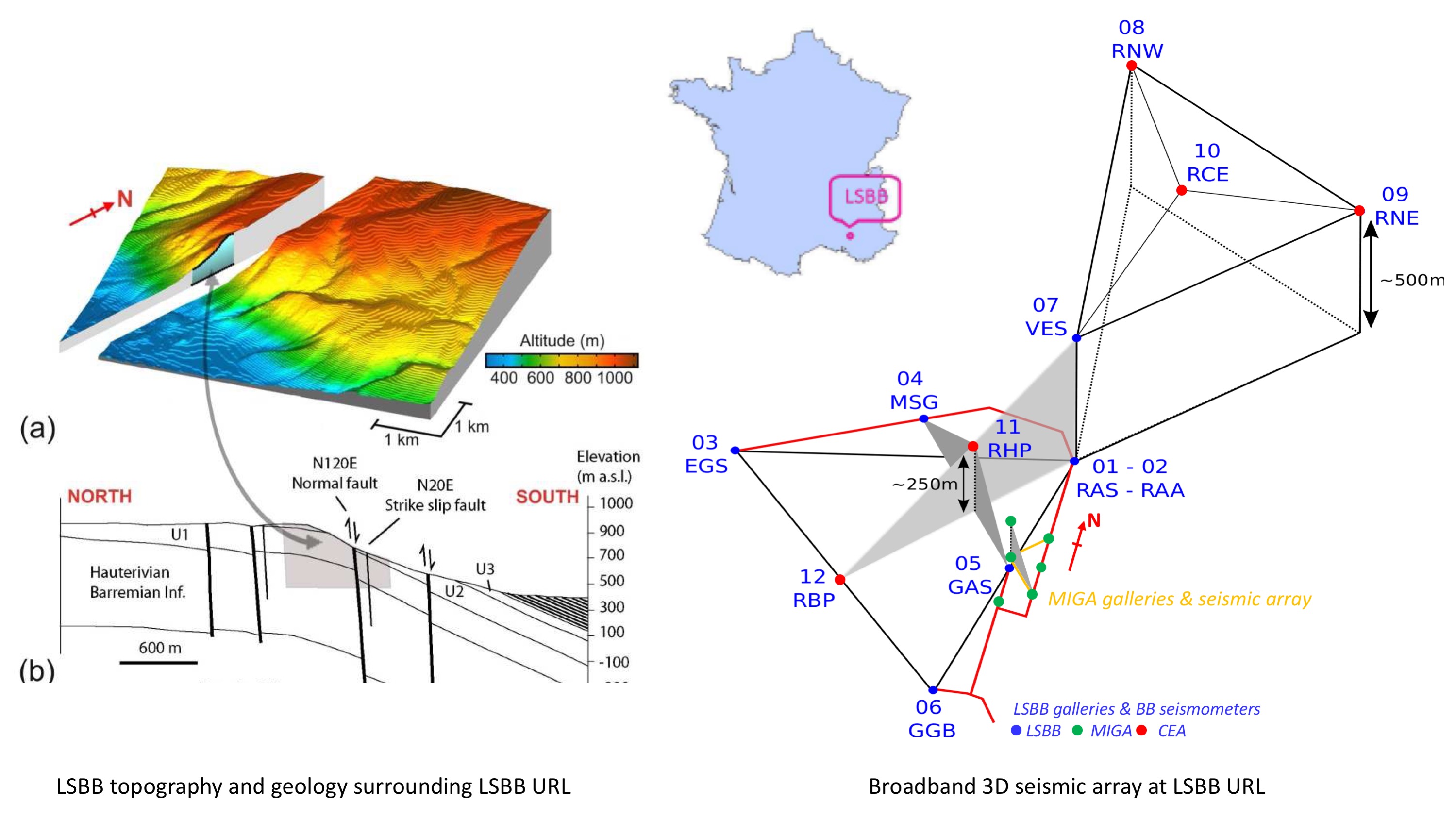}
\caption{The LSBB facility. (A) The laboratory is underneath a carbonate massif in southeastern France. (B) The underground facility is shown in thin red lines with a schematic representation of the permanent 3D broadband seismic array~\cite{labonne_seismic_2016, labonne_seismic_2016-1} - installed at the surface and in the galleries. Inset; the LSBB overlooks the hamlet of Rustrel, 60 km east from the high-speed train station in Avignon and 100 km north of Marseille, within the Fontaine-de-Vaucluse watershed, part of the Regional Natural Park of Luberon - this region is sparsely inhabited. Modified from~\cite{maufroy_travel_2014}, with permission. 
\label{fig:LSBB1}}
\end{figure}
The Laboratoire Souterrain \`{a} Bas Bruit (LSBB), a Low Background Noise Underground Research Laboratory located in Rustrel, near the city of Apt in Vaucluse, France, and about 100~km from the international airport in Marseille, is a European-scale interdisciplinary laboratory for science and technology created in 1997.
The laboratory was formed after the decommissioning of an underground launch control  facility for the French nuclear deterrence.
The LSBB is now a ground and underground based scientific infrastructure~\cite{gaffet2019} characterized by an ultra-low noise environment, both seismic and electromagnetic, as a result of the distance of the site from anthropogenic disturbances and its location under a large massif. 
The LSBB fosters multidisciplinary interactions and interdisciplinary approaches, pursuing both fundamental and applied research and has broad scientific and industrial expertise.
For these reasons, this site was chosen for the location of the MIGA project~\cite{Canuel2018}, which aims to study gravity gradient noise and test advanced detector geometries for its mitigation~\cite{Chaibi2016}.
New galleries were blasted specifically for the MIGA project, giving LSBB and the MIGA consortium expertise in the administration of large infrastructure works.
Further infrastructure work at this facility would benefit from this acquired experience.
%%%here
\par The LSBB facility is in the Fontaine-de-Vaucluse watershed, the fifth largest karst aquifers on Earth, covering 1115 km$^{2}$ in surface area. 
LSBB has 4.2~km long horizontal drifts at depths below ground ranging from 0~m to 518~m; its orientation lies primarily north to south and north-east to south-west. 
Carbonate rock surrounding the facility result in a thermal blanketing effect - a passive temperature stability better than 0.1~$\degree$C is observed. 
Internal air pressure and circulation are controlled via a series of air locks (SAS) that protect underground areas against anthropogenic disturbances.
The whole facility is connected to power lines, high-speed telecommunications (RENATER), and GPS time. 
The facility's location in the karst system has generated interest from hydrogeologists and geophysicists~\cite{carriere_role_2016, gaillardet_ozcar:_2018, jourde_sno_2018, cappa_stabilization_2019, barbel-perineau_karst_2019} studying such watershed systems; this has resulted in a series of sensors installed at the facility to help measure the tilting of rock mass \cite{lazaro_roche_design_2019, hivert_muography_2017, lesparre_new_2017} and the influence of gravity gradient variations - the local environment around LSBB is completely monitored, from wind speed, temperature, humidity, local gas composition (CO, O$_{2}$, CO$_{2}$, Rn) and microbarometric variations to local seismic and hydrogeological activities, as well as gravimetric perturbations measurements via superconducting magnetometers and gravimeters \cite{pozzo_di_borgo_minimal_2012, collot_operation_2013, henry_simultaneous_2016, Rosat2018}.
%%%
\par
\begin{figure}[htp!]
\centering
\includegraphics[width=.9\linewidth]{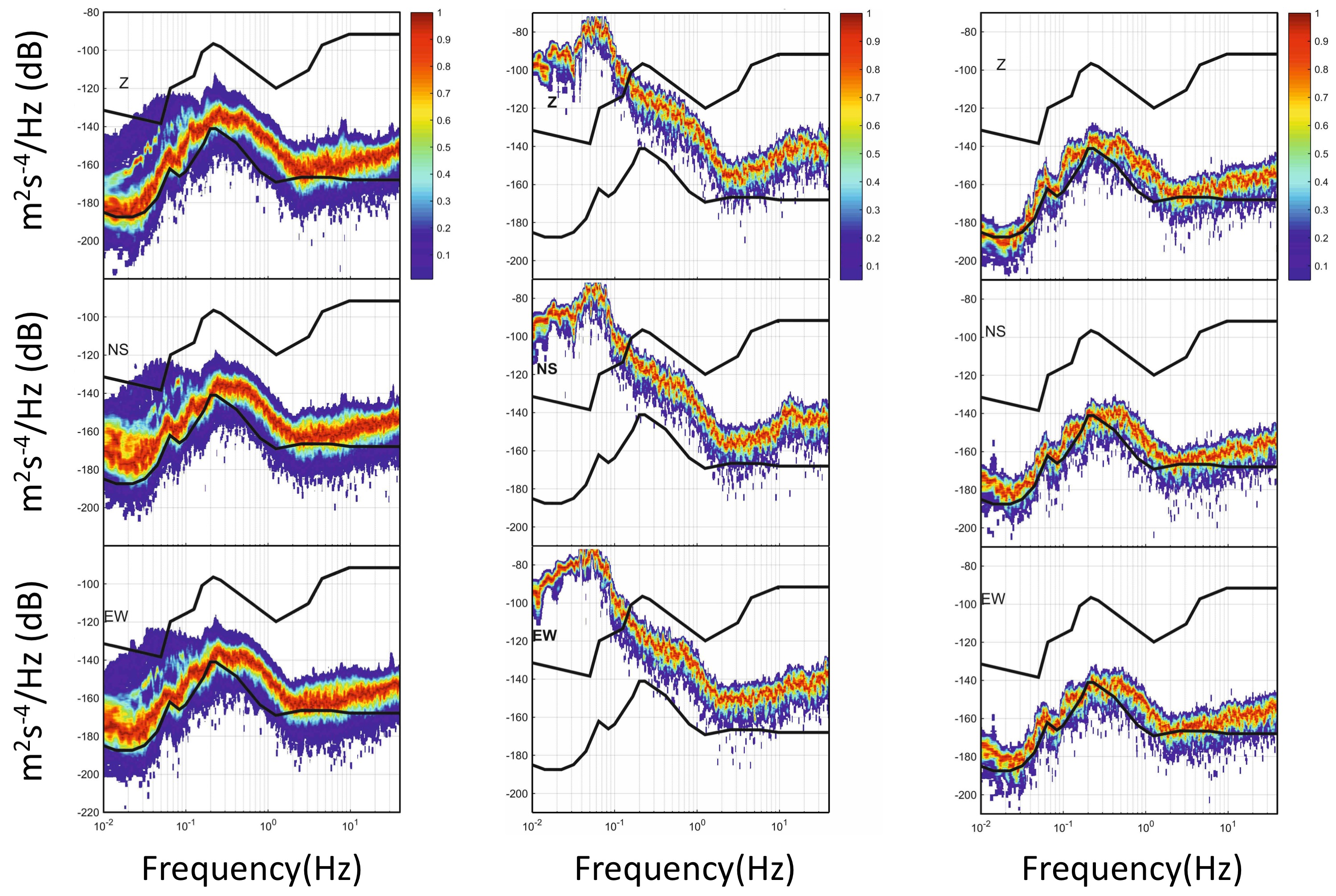}
\caption{Probability density functions samples of ambient seismic noise (color scale) of RUSF.01 broadband station for year 2011, including quiet days and days with earthquakes (left column); 6 hours of ground motion which include the Tohoku-Oki earthquake of March 11, 2011 (center column); a quiet day (June 26th, 2011) without seismic events (right column). The seismic noise PSD for three components (top row, Z; middle row, N-S; bottom row, E-W) is compared to the Peterson’s high and low noise models (black lines).
\label{fig:LSBB3}}
\end{figure}

 A network of 17 broadband seismometers monitors the seismic ground motion at the facility.
A sample of seismic noise data recorded at the station RUSF.01 (the underground seismometer named RAS in Fig.~\ref{fig:LSBB1}) is depicted in Fig.~\ref{fig:LSBB3}. 
This figure shows the probability density function (PDF) compared to Peterson's models for this site.
These data show three orthogonal components for three separate measurement intervals: the first column includes, roughly, the whole year of 2011, including the Mw 9.1 Tohoku-Oki mega-thrust earthquake in Japan from March 11$^{\rm{th}}$ of that year, the second column is a six hour interval during the earthquake, and the third column is a 24 hour period with no seismic activity. 
These nine plots show low seismic noise, baring large transient signals from earthquakes. 
The highest probability of noise occurrence, over a long time interval, is close to Peterson's low-noise model for all three components over the entire frequency band considered. 
Below 200~mHz, the PDF at one year spreads between the high and low noise models.
This low frequency band includes worldwide seismic activity and, more specifically, all surface waves that superficial earthquakes produce; considering a quiet day, the PDF can dip significantly below the low noise model with high probability.
The seismic noise spectra recorded are close to worldwide minima.
%%%
\par The low-noise seismic properties of LSBB have been confirmed using superconducting gravimeters (model iOSG from GWR Instruments Inc.~\cite{Rosat2018}).
This device was installed at the underground site in 2015 in order to define a gravimetric baseline dedicated and complementary to the MIGA experiment. 
These instruments installed at LSBB have a demonstrated stability of 1.8 $\text{nm} / s^{2} / \text{Hz}^{1/2}$ at 1~mHz, a noise performance among the best in a worldwide network of superconducting gravimeters~\cite{farah_underground_2014, Rosat2018}.
%
%%%
\par A measurement campaign near to the location of the MIGA project found that the observed magnetic field in the frequency band 1~mHz to 400~Hz is much lower than the expected background~\cite{Canuel2018, henry_simultaneous_2016}.
Magnetic field fluctuations around 2 pT/$\sqrt{\text{Hz}}$ at 1 Hz were observed in the tunnels near to the MIGA installation - throughout the measurement galleries in the facility, fluctuations ranged from 10 to 0.08 pT/$\sqrt{\text{Hz}}$, depending upon the shielding in each measurement hall. 
%%%
\par LSBB is protected against anthropogenic disturbances within the Regional Natural Park of Luberon, which is lightly industrialized.
The location of LSBB beneath a massif nets a unique sheltering effect with respect to electromagnetic noise~\cite{waysand_first_2000}.
There is a two-kilometer exclusion zone near the facility, reducing magnetic interference from high-voltage power lines and railways. 
The facility is 500~m below carbonate rocks loaded with water - this gives a high frequency cut-off around 200~Hz for incoming electromagnetic waves. 
\par The LSBB has established maintenance facilities, running water, power, sewage, nearby villages and towns with residences, schools, restaurants and shops, as well as the reputation of an established pan-European large scale research facility.
The facility's low-noise characteristics as well as the suitability of the surrounding area has already been extensively studied for the MIGA project, among other European projects.
\subsubsection{Sardinia facility.}
\par
\begin{figure}[htp!]
\centering
\includegraphics[width=.9\linewidth]{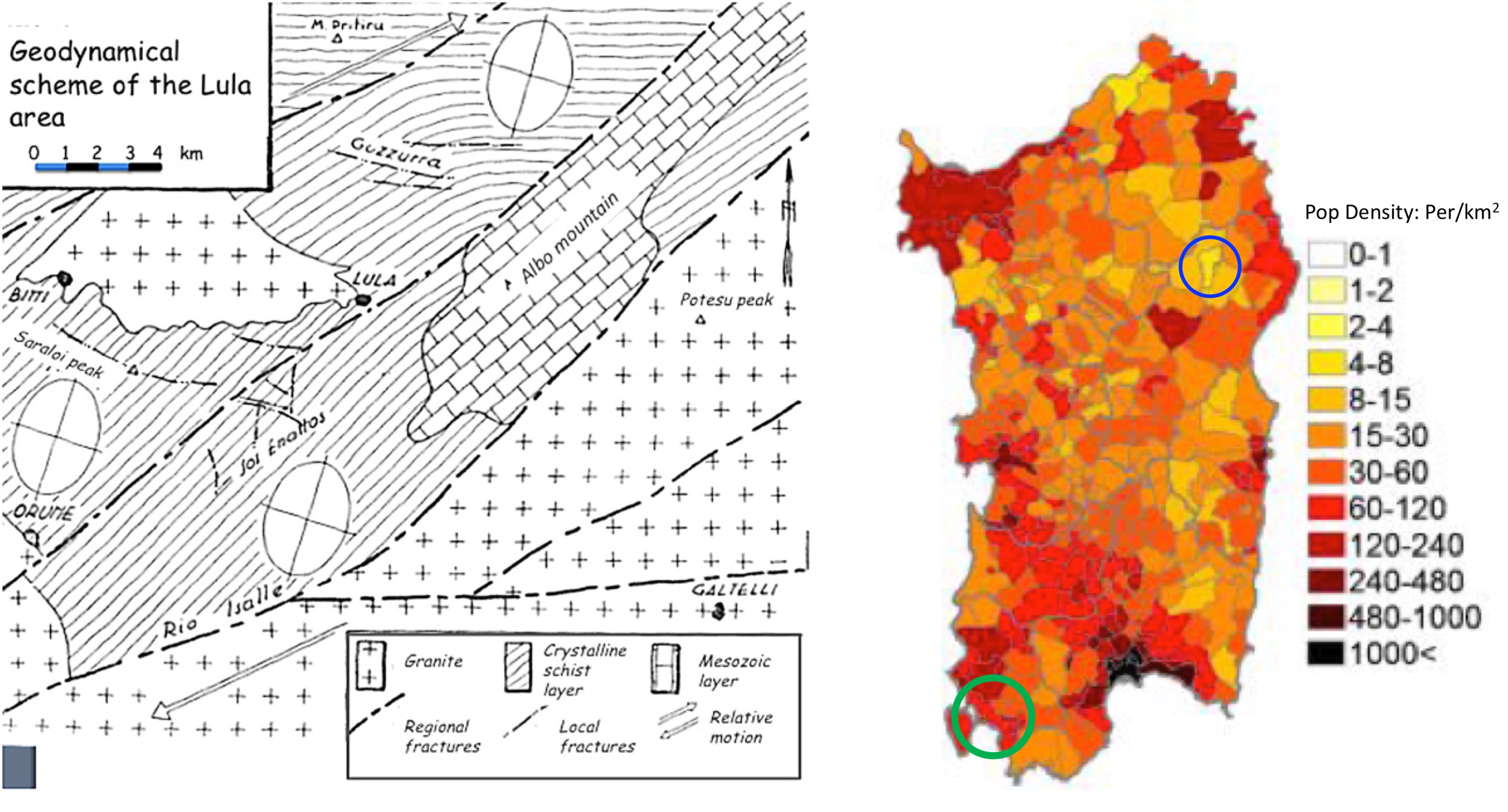}
\caption{The Sos Enattos site on Sardinia. Left: Geodynamical scheme of the Lula area, around Sos Enattos on Sardinia.
Right: distribution of population density in Sardinia. 
The blue circle indicates the location of the Sos Enattos site and the green circle indicates the location of the Seruci site.     
\label{fig:Sard1}}
\end{figure}

 The island of Sardinia in Italy offers several candidate sites which could host a detector of the scale of ELGAR. 
It contains areas with low anthropogenic disturbances, given a population density on the island that is among the lowest in Europe. 
The geological structure of the island is ancient; due to this, micro-seismic activity is among the lowest on Earth. 
The continential landmass, the Sardinia-Corsica block, is isolated and partially removed from the Alps block.
Addition, this block is located on the European tectonic plate and far from any fault lines.
From a feasibility standpoint, the island offers several interesting locations - former mine shafts have appealing characteristics.
Two potential candidate sites are under study; the Sos Enattos site near Lula, in the north-east, and the Seruci/Nuraxi Figus mining sites in the southern area of Sulcis, see Fig.~\ref{fig:Sard1}.
%%%
\par
\begin{figure}[htp!]
\centering
\includegraphics[width=1\linewidth]{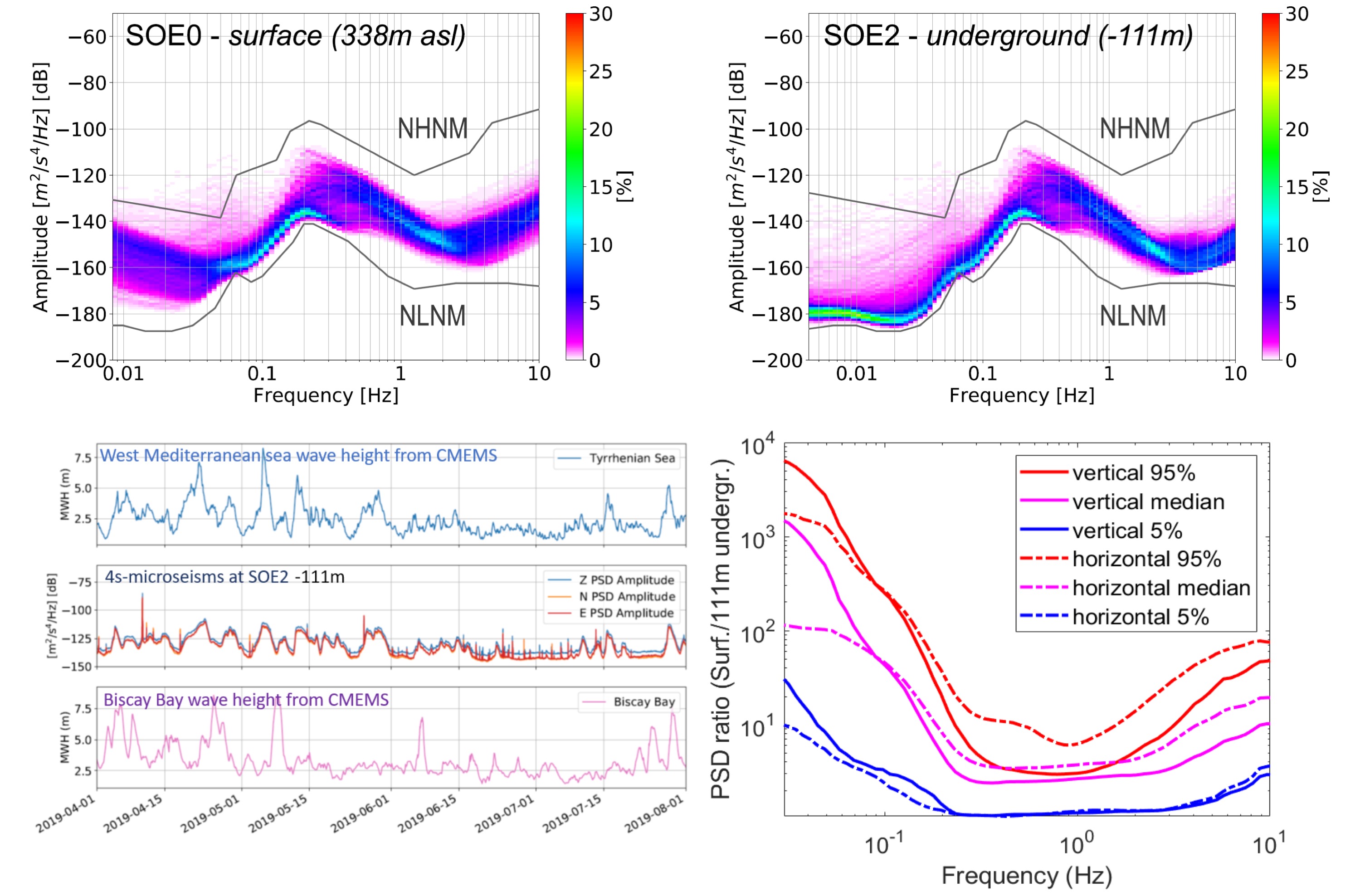}
\caption{Seismic noise at the Sos Enattos site. 
Figures courtesy of \cite{bignat19}.
Top: Power spectrum percentiles of seismic noise (vertical component), for the period April-August 2019. 
Surface measurements (338 m above sea level) on the left, underground measurement below 111 m of rock on the right. 
The continuous lines represents Peterson's models. 
Bottom left: Sea wave height from CMEMS and microseismic amplitude at Sos Enattos. 
Bottom Right: Ratio between surface and underground seismic noise. 
\label{fig:Sard3}}
\end{figure}

 The center and eastern sections of the island are dominated by carbonate rock.
The first site consider is located there, the Sos Enattos mine, and is located 40 km north by northwest from the city of Nuoro and 5 km from the village of Lula. 
The mine is composed of sphalerite ([Ze,Fe]S) and galena (PbS) rock.
The mine focused on the excavation of lead and zinc deposits for 130 years - well maintained underground caverns have left the site usable since 1864. 
This site is particularly favorable due to ancient geology; a carbonate platform covers a base of Hercynian granites. 
The layers have few fractures and are more likely to bend than create breaks. 
Despite these favorable mining conditions, the population density of the Albo mountain area is a factor seven lower than the mean for the island of Sardina. 
The site is managed by the I.G.E.A. S.p.A. company, which has entertained the Einstein Telescope collaboration's interest in this site for their detector.
A series of seismic, acoustic, and magnetic measurement campaigns are presently underway to fully characterize this facility \cite{bignat19}.
%%%
Fig.~\ref{fig:Sard3} shows the seismic power spectral density (PSD) at the Sos Enattos site. 
Seismic noise is close to the Peterson's New Low-Noise Model (NLNM). 
Underground measuring stations at depths of 84 and 111 m show a large attenuation of anthropic noise above few Hertz, and of slow thermal and pressure fluctuations below 80 mHz. 
Seismic noise below 1 Hz is dominated by microseismic peaks from sea waves. 
Correlation of the seismic PSD with wave height from Copernicus Marine Environment Monitoring Service (CMEMS) in the western Mediterranean sea and in Biscay bay shows that that the dominant contribution comes from the closer Mediterranean sea, in particular for a period of 4.5~s \cite{bignat19}. 
Population density in this region is among the lowest in Europe, so the anthropic noise background, which is usually dominant above 1~Hz, is low even at the Earth's surface, and is additionally attenuated underground as shown in Fig.~\ref{fig:Sard3}.
%%%
\par On the southwestern portion of Sardinia, the carbonate rock gives way to coal bearing substrate. 
Here, the Monte Sinni mining concession C233, extracts coal; it covers an area of 59.4 km$^{2}$, falling entirely within the municipalities of Gonnesa, Carbonia, and Portoscuso in the province of Carbonia-Iglesia, South Western Sardinia. 
The company Carbosulcis S.p.A., owner of the mining concession, resumed mining coal in 1976, which had been interrupted by ENEL a few years prior. 
Until recently, mining and industrial activities were concentrated at the two work sites of Seruci and Nuraxi Figus. 
After the centralization of the services in Nuraxi Figus, the mining activities of Seruci were halted.
\par The mining site includes an underground area for the primary phases of the coal extraction cycle, comprising excavation and preparation of tunnels, cultivation of coal, and transportation of crude coal to the surface.
The coal mine extends underground through a network of tunnels whose total length is about 30 km. 
At present, the structure of the mine is developed to a depth between 350 and 500 m below the surface (between 200 m and 400 m above sea level).
The connection between the surface and underground is maintained through four main shafts (two in Seruci and two in Nuraxi Figus) and a winze with inclined ramps. 
The ventilation system is operated by two aspiration fans. 
The connection between the shafts by structural galleries.
The dimensions of the main galleries allow for the transport and installation of mining and excavation equipment. 
The shafts and the main galleries have lighting, electricity, water, and compressed air.
%%%
\par
\begin{figure}[htp!]
\centering
\includegraphics[width=0.9 \linewidth]{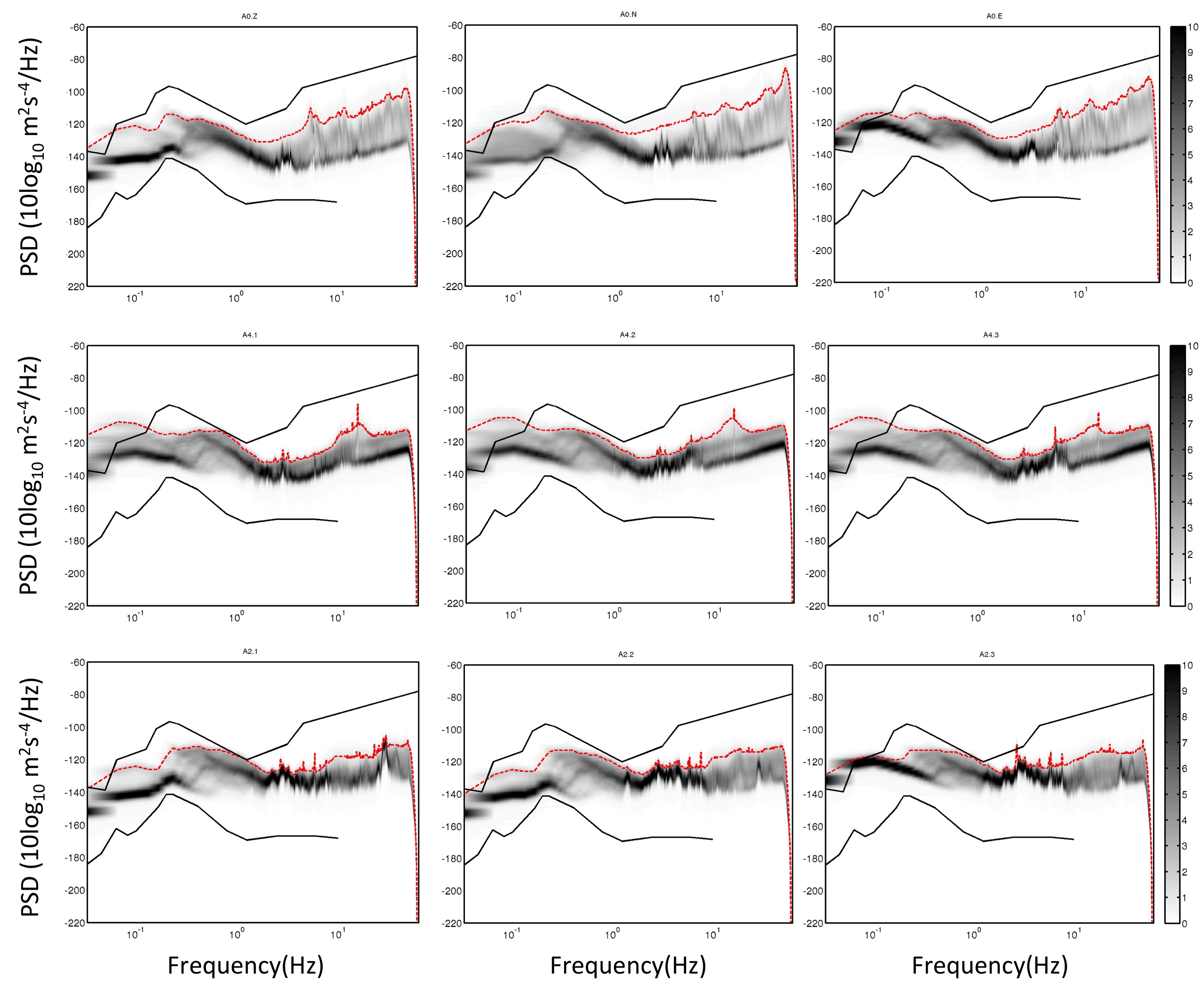}
\caption{Probability density function of the spectral powers for the three stations near the Seruci site; figure courtesy of INGV.
The distributions refer to 30 days of continuous ground motion recording. From left to right, the graphs refer to the Z, N-S and E-W components. 
From top to bottom, the plots refer to the A0, A4, and A2 stations respectively. 
The red line indicates the 95th-percentile of the spectral power. 
The probability density is indicated in gray scale. 
The black lines indicate Peterson's upper and lower models of Earth's seismic noise.
\label{fig:Sard2}}
\end{figure}

 Forced ventilation guaranteed everywhere in the mine for safety reasons; the air speed through the gallery section, depending if it is a primary, secondary or service gallery, is between 0,8 and 1,6 m/s. 
The typical section of a primary gallery is up to 24 m$^{2}$ (7.2 m wide and 4.9 m high), decreasing to 18 m$^{2}$ for a secondary gallery. 
In working areas, there can be special electrical devices or instruments.
It is possible to smooth and level the gallery pavement, or cover the ceiling by sputtering concrete. 
The roof supports are made with bolts and iron arches,
%.
Transport of people and materials through the mine is available dedicated underground vehicles. 
Special teams are trained and ready for emergency and support. 
%%%
There is a system of continuous underground environmental monitoring in place, which consists of analyzers in fixed locations near air inlets of the reflux wells, in the secondary reflux, and in the active cultivation yards - all places where harmful gases may develop. 
Presently, the control station is outfit to monitor the following gases: CH4, CO, O2, CO2, NOX, and the following parameters: ambient temperature, relative humidity, and, air velocity.
We show data from continuous seismic monitoring of the site to illustrate the location's suitability. 
The data in Fig.~\ref{fig:Sard2}, power spectral densities, were obtained from the average of the Fourier transforms calculated on 10 consecutive windows of length 30~s, with 50~$\%$ overlap. 
The estimates were then repeated on successive intervals along 30 days of continuous recording, leading to a total of over 16,000 independent spectral estimates (523 PSD / day x 30 days). 
The results, represented by the probability densities (PDF) of the spectral power versus frequency, are compared with Peterson's Earth seismic models NLNM and NHNM of Low- and High-Noise. 
Before PSDs computation, the time series were corrected for the instrumental transfer function, regularizing the deconvolution by band-pass filtering in the 0.1-50 Hz interval. 
Out of this interval the spectra powers must be taken with caution.
The data show that site displays quiet seismic properties. 
%%%
\par The mine maintains significant infrastructure and is now transitioning into scientific exploration.
The Seruci 1 shaft is presently being adapted for a 350 m vertical cryogenic column, required for the $^{40}$Ar distiller for the ARIA project \cite{Aria}. 
This requires an upgraded bearing framework that will be able to hold more than one column - for this reason, it is under consideration as a potential distributed vertical arm of the ELGAR detector at reduced cost. 
Other main shafts are also available and will be compared for available column depth, environmental noise, and crossing with horizontal galleries.
Such studies have been performed for the proposed Einstein Telescope \cite{ET-design}, another entity considering Sardinia and the forming mining concessions as scientific infrastructure.  
\section{Atom optics}\label{sec_Atom_optics}
Two atom interferometers in a gradiometric configuration separated by a baseline $L$, and coherently manipulated by the same light fields with a wavenumber $k_{\mathrm{eff}}$ can be utilized as a differential phasemeter~\cite{dimopoulos_atomic_2008,Canuel2018}.
An incident gravitational wave with amplitude $h$ and frequency $\omega$ modulates the baseline, leading to a differential phase shift between the two atom interferometers.
Single-loop (3 pulses)~\cite{dimopoulos_atomic_2008,Canuel2018,Hogan2016PRA}, double-loop (4 pulses)~\cite{Hogan2011}, and triple-loop (5 pulses)~\cite{Hogan2011} geometries in gradiometric configuration were proposed for vertically~\cite{dimopoulos_atomic_2008} or horizontally~\cite{Canuel2018} oriented gravitational wave detectors on ground and in space.
The beam splitters, mirrors, and recombiners therein forming the interferometer typically consist of composite pulse and / or higher-order processes based on Bragg- / Bloch transitions to enhance the wavenumber $k_{\mathrm{eff}}=2 n k_{\mathrm{l}}$ corresponding to the relative momentum between the two trajectories in the atom interferometer, $n$ two-photon transitions, and a (single-photon) wavenumber $k_{\mathrm{l}}$ of the driving light field.
Single photon processes were also proposed to relax requirements on laser frequency noise~\cite{Graham2013PRL}.
All these geometries share similarities as free falling atoms, linear scaling of the phase shift in the effective wavenumber, the distance between the two atom interferometers, and a frequency response dependent on the pulse separation time $T$, leading to a general form of the phase shift $\Delta\phi\sim h k_{\mathrm{eff}} L f(\omega T)$.
\begin{figure}
\centering
\includegraphics[width=0.48\linewidth]{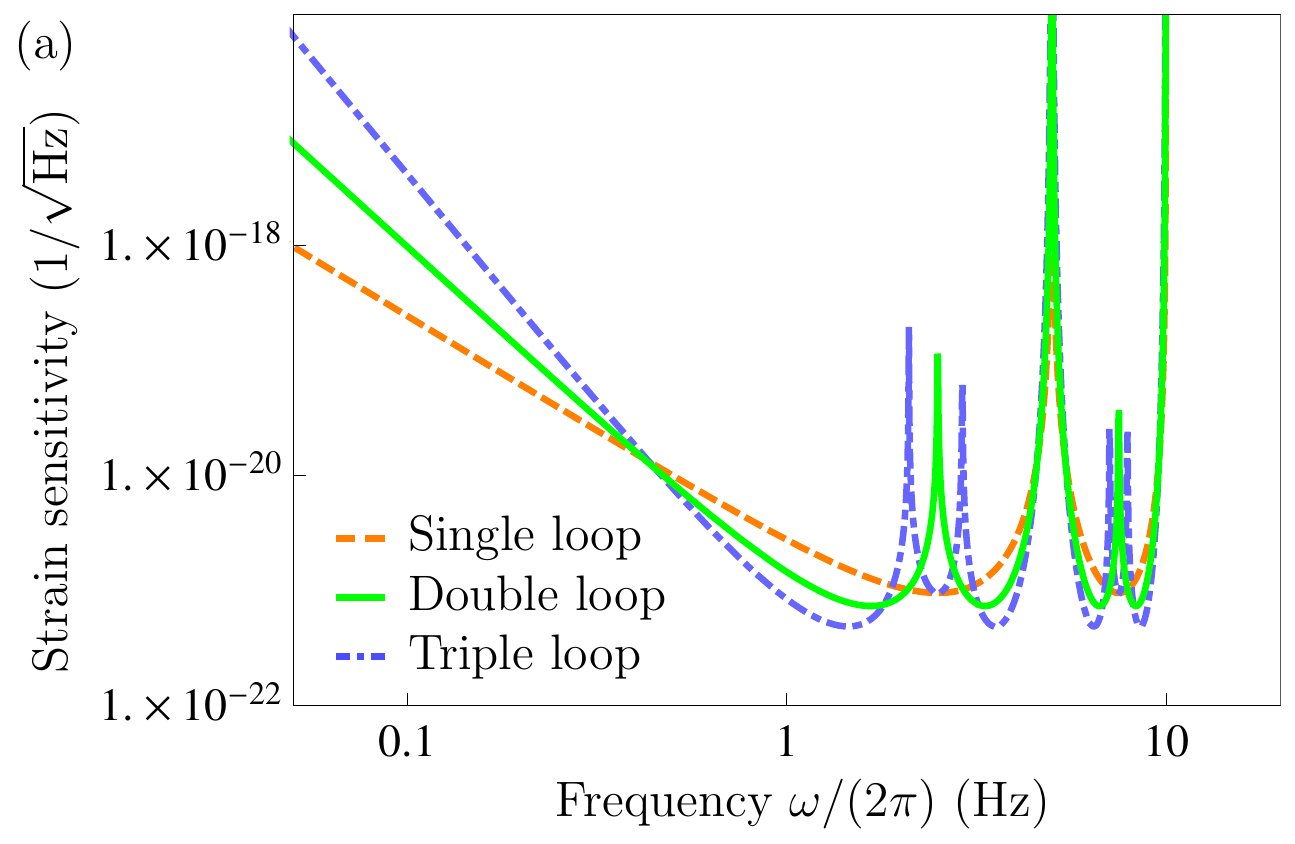}
\includegraphics[width=0.48\linewidth]{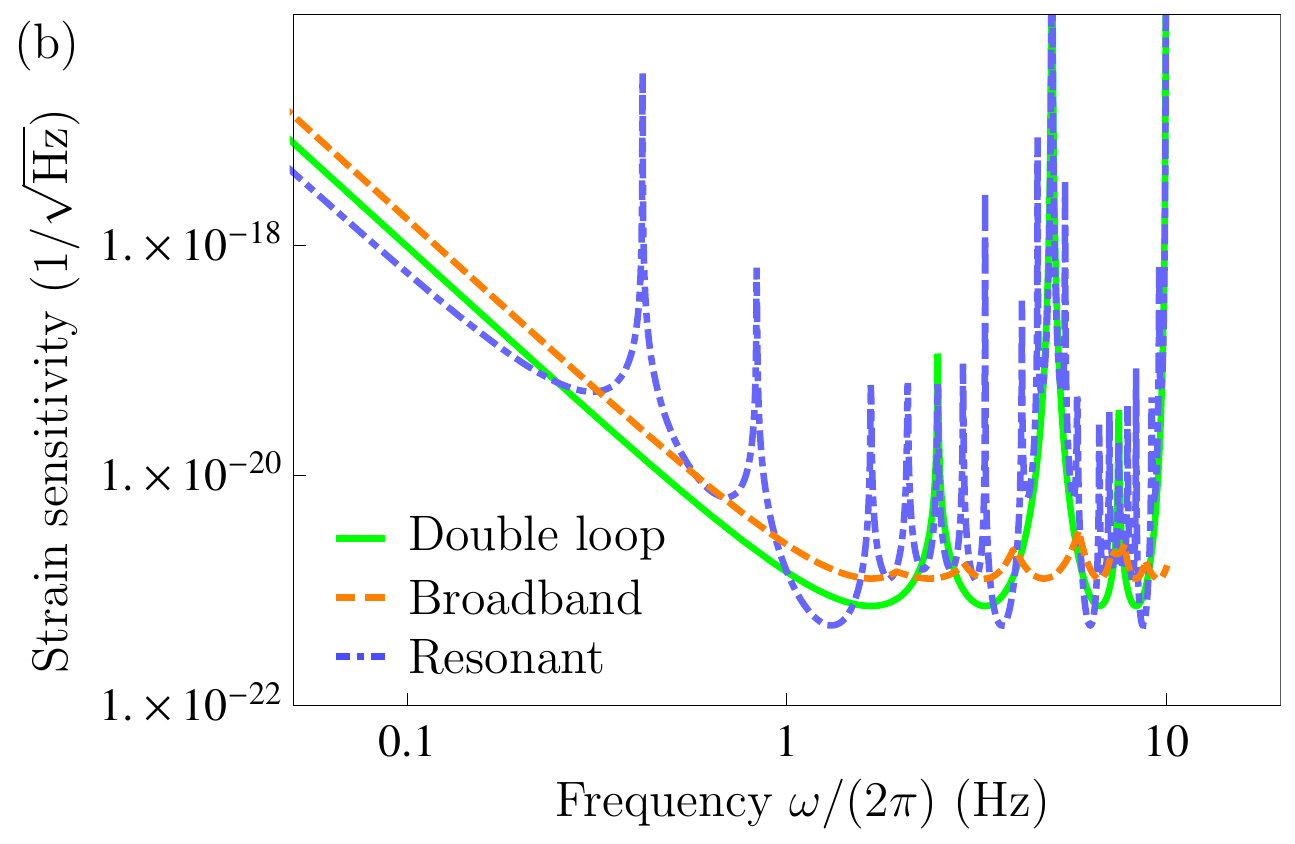}
\caption{Shot-noise limited strain sensitivities for different geometries. 
(a) Comparison of single-, double-, and triple-loop interferometers.
The graph assumes an effective wavenumber $k_{\mathrm{eff}}=1000\cdot2\pi/(780\,\mathrm{nm})$, a pulse separation time $T=200\,\mathrm{ms}$, a baseline $L=16.3\,\mathrm{km}$, a phase noise of $1\,\mu\mathrm{rad}/\sqrt{\mathrm{Hz}}$, and 80 gradiometers per arm, as well as sky and polarization averaging~\cite{Moore2015CQG}.
(b) Comparison of a double-loop interferometer as in (a) to a double-loop configuration in broadband mode and in resonant mode with a tripled number of loops.
The broadband mode uses interleaved interferometers with the three different pulse separation times $0.7\,T$, $0.9\,T$, and $T$. 
}
\label{fig:atom-optics-strain-sensitivities}
\end{figure} Differences are that multiple loops change the response $f(\omega T)$ at low frequencies (Fig.~\ref{fig:atom-optics-strain-sensitivities} (a)), lead to a resonant enhancement~\cite{Graham2016PRD} (Fig.~\ref{fig:atom-optics-strain-sensitivities} (b)), and can suppress spurious phase terms~\cite{Hogan2011}.  
Depending on the implementation, one laser link for a vertical~\cite{dimopoulos_atomic_2008} or more for a horizontal setup~\cite{Canuel2018} are required for the geometries.

\subsection{Overview of beam splitters}
Being the cornerstone of atom interferometry, the atom optics essentially does two tasks: imprinting the phase of light field into atoms and manipulating the quantum states of the atoms.
Currently there are three popular choices of atom optics based on stimulated Raman transition~\cite{Kasevich1991}, Bragg diffraction~\cite{Giltner1995}, and Bloch oscillations~\cite{Cadoret2008}.
Recently, there are new development using single-photon transitions in demonstration experiments~\cite{Hu2017} including large momentum transfer~\cite{Rudolph2020}.
The Raman atom optics employs the stimulated Raman transition between the long-lived ground hyperfine states. Two counter propagating light beams give the atoms a momentum transfer \(k_{\mathrm{eff}}\approx 2 k_{\mathrm{l}}\).
Both Bragg diffraction and Bloch oscillations utilize the ac Stark shift induced by the optical lattice to interact the atoms.
In the Bragg case, the atoms are coherently scattered by the spatially periodic potential created by the ac Stark shift and gain the momentum transfer from the lattice.
In the Bloch case, the atoms are put into a moving lattice by slightly detuning the frequencies of two light fields.
Under certain conditions, the atoms will be coherently accelerated by the moving lattice and at the end of each oscillation period the atoms will receive the momentum transfer \(2 \hbar k_{\mathrm{l}}\).
Comparing with Raman atom optics, which affect both the internal and external quantum states, the Bragg and Bloch atom optics will only interact with the external quantum states and therefore the systematic effects caused by the environment can be greatly suppressed.
However, the requirements of the optical power for consecutive Bragg pulse and Bloch lattices are increased to suppress spontaneous emission by having the individual laser frequencies far-off resonant.
The power requirement for higher order Bragg diffraction is even more stringent~\cite{Szigeti2012NJP,Muller2008}.   
All these atom optics can be combined with other techniques including a rapid adiabatic passage to improve the efficiency and reduce the contrast loss~\cite{Weitz1994,Peik1997,Kovachy2012}.
One major research direction for atom optics is the large-momentum-transfer optics~(LMT) because the sensitivity of an atom interferometer is typically proportional to the transferred momentum.
Several LMT schemes including consecutive pulses sequence~\cite{Jaffe2018,McGuirk2000}, higher-order atom optics~\cite{Muller2008}, Bloch oscillations~\cite{Cadoret2008}, the combination of schemes~\cite{Gebbe2019arxiv,Muller2009PRL}, and dual-lattice beam splitters~\cite{Pagel2017arXiv} have been demonstrated.
So far the highest LMT separating the trajectories in an atom interferometer utilises the combination of double Bragg diffraction~\cite{ahlers2016} and a twin lattice driving Bloch oscillations for symmetric beam splitting to reach an effective wave number of \(k_{\mathrm{eff}}=408\ k_{\mathrm{l}}\)~\cite{Gebbe2019arxiv}, where $k_{\mathrm{l}}=2\pi/(780\,\mathrm{nm})$ is the wave number for a single photon.
By mitigating the technical limitations due to spontaneous emission and intensity inhomogeneities across the atom-light interaction zone, effective wave numbers 
\begin{equation}
  k_{\mathrm{eff}}\equiv 2n k_{\mathrm{l}} =1000\,k_{\mathrm{l}} 
  \label{eq:definition_keff}
\end{equation}
appear to be feasible for ELGAR.

\subsection{Suppression of spurious phase terms}
In a single-loop geometry, rotations and gravity gradients give rise to phase terms $2\,\mathbf{k}_{\mathrm{eff}} \cdot (\mathbf{v}_0 \times \boldsymbol{\Omega}) T^2$ and $\mathbf{k}_{\mathrm{eff}}^\text{T}\,\Gamma{T^2}(\mathbf{x}_0 + \mathbf{v}_0 T)$ with couplings to the initial central position $\mathbf{x}_0$ and central velocity $\mathbf{v}_0$~\cite{Bongs2006,Hogan2009}. %($n=1$).
Here $\boldsymbol{\Omega}$ denotes the angular velocity and $\Gamma$ is the gravity gradient tensor, which corresponds to minus the Hessian of the gravitational potential $U(\mathbf{x})$ with components $\Gamma_{ij} = - \partial^2 U / \partial x^i \partial x^j$.
Moreover, we have employed a convenient vector-matrix notation where boldface characters correspond to vector quantities and the superscript T denotes the matrix transpose.
If the gravity gradient is known, the corresponding phase term can be suppressed in a gradiometric configuration by adjusting the wave number of the mirror pulse to $\mathbf{k}_{\mathrm{eff}}+\Delta{\mathbf{k}_{\mathrm{eff}}}$ with $\Delta{\mathbf{k}_{\mathrm{eff}}}=(\Gamma{T^2}/2)\,\mathbf{k}_{\mathrm{eff}}$~\cite{Roura2017PRL}.
The Sagnac term $2\,\mathbf{k}_{\mathrm{eff}} \cdot (\mathbf{v}_0 \times \boldsymbol{\Omega}) T^2$, however, remains and implies stringent requirements on the residual expansion rate which is connected to the mean velocity~\cite{Schubert2019arxiv}.
This challenge is mitigated by multi-loop geometries.
A double-loop geometry retains sensitivity to DC rotations, but decouples the leading order term from the initial velocity~\cite{Hogan2011,Canuel2006,Dubetsky2006PRA}.
Contributions from gravity gradients can again be suppressed by adjusting the wave number of the two mirror pulses~\cite{Roura2017PRL}.
Further details are provided in sec.~\ref{sec_other_noise_couplings}.
To first order, the triple-loop scheme suppresses both the coupling of rotations and gravity gradients to the initial conditions and shows no sensitivity to DC accelerations and rotations for adjusted wave numbers of the three mirror pulses~\cite{Hogan2011}, but requires an additional atom-light interaction zone as compared to the other two geometries if no techniques for suspension or relaunch are employed.

\subsection{Broadband and resonant detection modes}\label{sec_operating_mode}
Intrinsically, the transfer function of an atom interferometer features peaks at frequencies corresponding to multiples of the pulse separation time~\cite{Cheinet08}, as depicted in Fig.~\ref{fig:atom-optics-strain-sensitivities}~(a).
For the detection of gravitational waves, this can be an undesired feature if the frequency of the gravitational wave is not known, and the pulse separation time cannot be adjusted accordingly.
The solution is to drive $m$ interleaved interferometers~\cite{Savoie2018,Biedermann2013PRL} with different pulse separation times, so that the peaks are slightly shifted between subsequent cycles, effectively leading to a broadband detector with a flat response over a chosen frequency range~\cite{Hogan2016PRA} at the cost of a factor of $\sim\sqrt{m}$ in the strain sensitivity (Fig.~\ref{fig:atom-optics-strain-sensitivities}~(b)).
Provided, that the signal of a gravitational wave is identified, the pulse separation time can be adjusted to its frequency to maximise the response.
In addition, the geometry can be extended by adding additional pulses before the final recombination pulse, so that the total number of loops gets multiplied by an integer factor $b$ (Fig.~\ref{fig:atom-optics-strain-sensitivities}~(b)).
This leads to an amplification of the response by the multiplication factor $b$ and corresponds to a resonant detection mode~\cite{Graham2016PRD}.
Depending on the implementation, e.g. the size of the vacuum vessel or the beam diameter in a horizontal detector, the tuneability of the pulse separation time $T$ will be limited.
The resonant detection is also affected by the total free-fall time that the vacuum vessel can support, may need relaunching for a ground detector, requires highly efficient beam splitters due to the added pulses, and has implications for the detection due to the extended time of flight.
For pulse separation times $T=200\,\mathrm{ms}$ and $b=3$, the total time of flight would be $2.4\,\mathrm{s}$, constraining residual expansion rates to $\sim100\,\mu$m/s, which requires a well-collimated atomic ensemble~\cite{Rudolph2016Diss,Kovachy2015PRL,Muntinga2013PRL}.

\subsection{Vertical and horizontal arms}
In a vertical arm~\cite{Coleman2018arXiv,dimopoulos_atomic_2008}, the free fall of the atoms is aligned with with the axis of the laser link for coherent manipulation.
This implies a tuneability in the pulse separation time $T$ enabling the broadband detection mode, adjusting $T$ to the frequency of interest, and the possibility of resonant detection within the limit of the area in which the atoms can efficiently be manipulated.
An interleaved mode requires a labelling of the concurrent interferometer, e.g. by different Doppler shifts~\cite{Hogan2016PRA,Hogan2011}.
The accessibility of deep boreholes may limit the maximum baseline $L$, in the case of Ref.~\cite{Coleman2018arXiv} reported to 1\,km.
If beam splitting techniques other than single-photon transitions~\cite{Graham2013PRL} are implemented, this implies the requirement of an additional horizontal arm to suppress laser frequency noise.
For horizontal arms, baselines of several kilometres as in LIGO~\cite{Abbott2016} or VIRGO~\cite{Acernese2014} and possibly more~\cite{Chaibi2016} appear feasible.
This relaxes the requirements on the beam splitting order and the intrinsic phase noise of the interferometer.
In the horizontal configuration, the atoms travel orthogonal to the beam splitters, constraining the minimum beam size for efficient manipulation, and defining the pulse separation $T$ if more than one atom-light interaction zone is required.
Depending on the chosen geometry, this may limit the possibilities of a broadband or resonant detection mode.
An advantage of spatially separated atom-light interactions zones which can address atoms only at specific times of flight is the easier accommodation of interleaved operation~\cite{Savoie2018}.

\subsection{Double-loop geometry}\label{sec_double_loop}
The double-loop geometry~\cite{Marzlin1996PRA} consists of an initial beamsplitting pulse, two mirror pulses which invert the momenta, and finally a recombination pulse.
\begin{figure}
\centering
\includegraphics[width=0.48\linewidth]{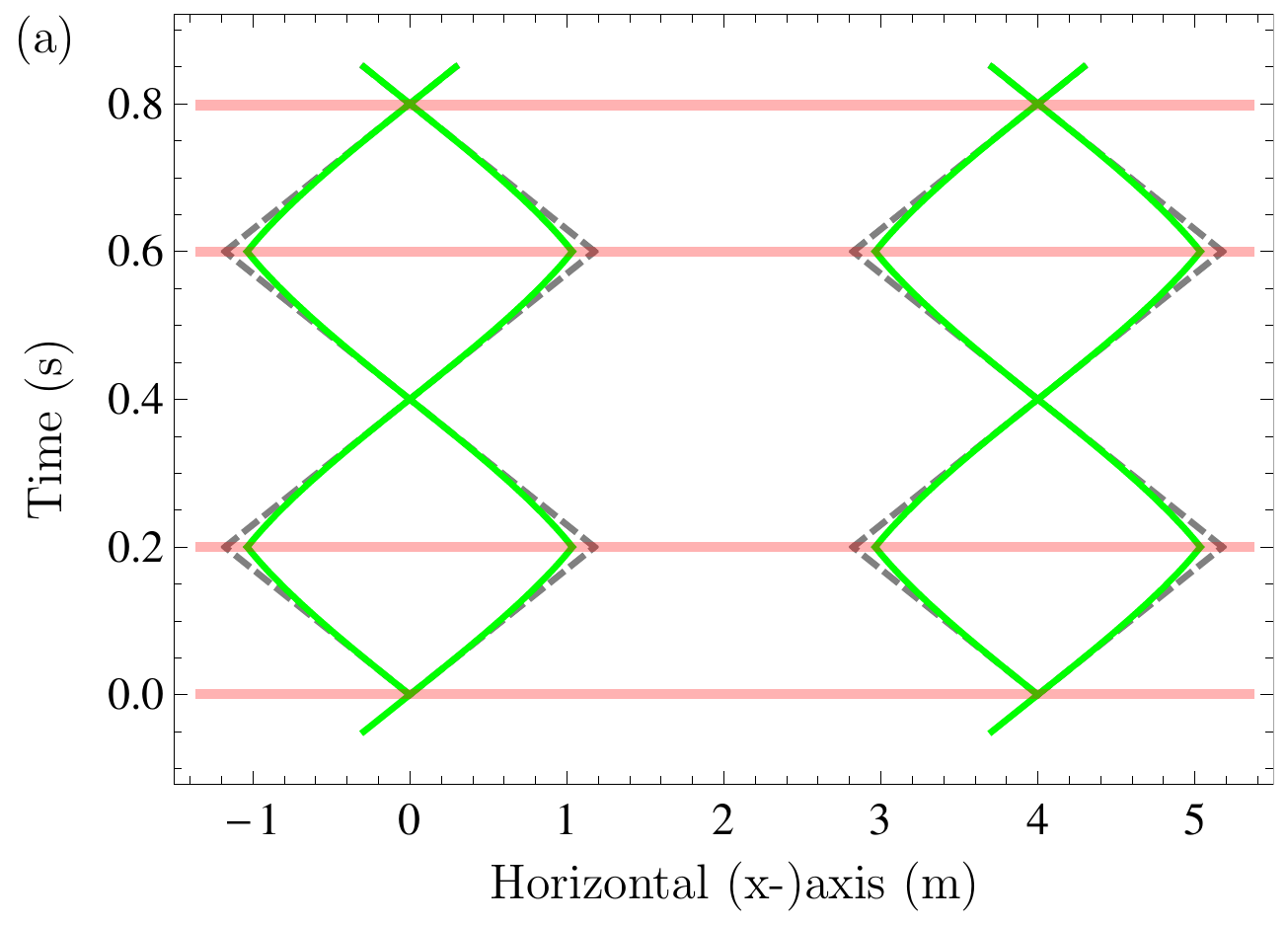}
\includegraphics[width=0.48\linewidth]{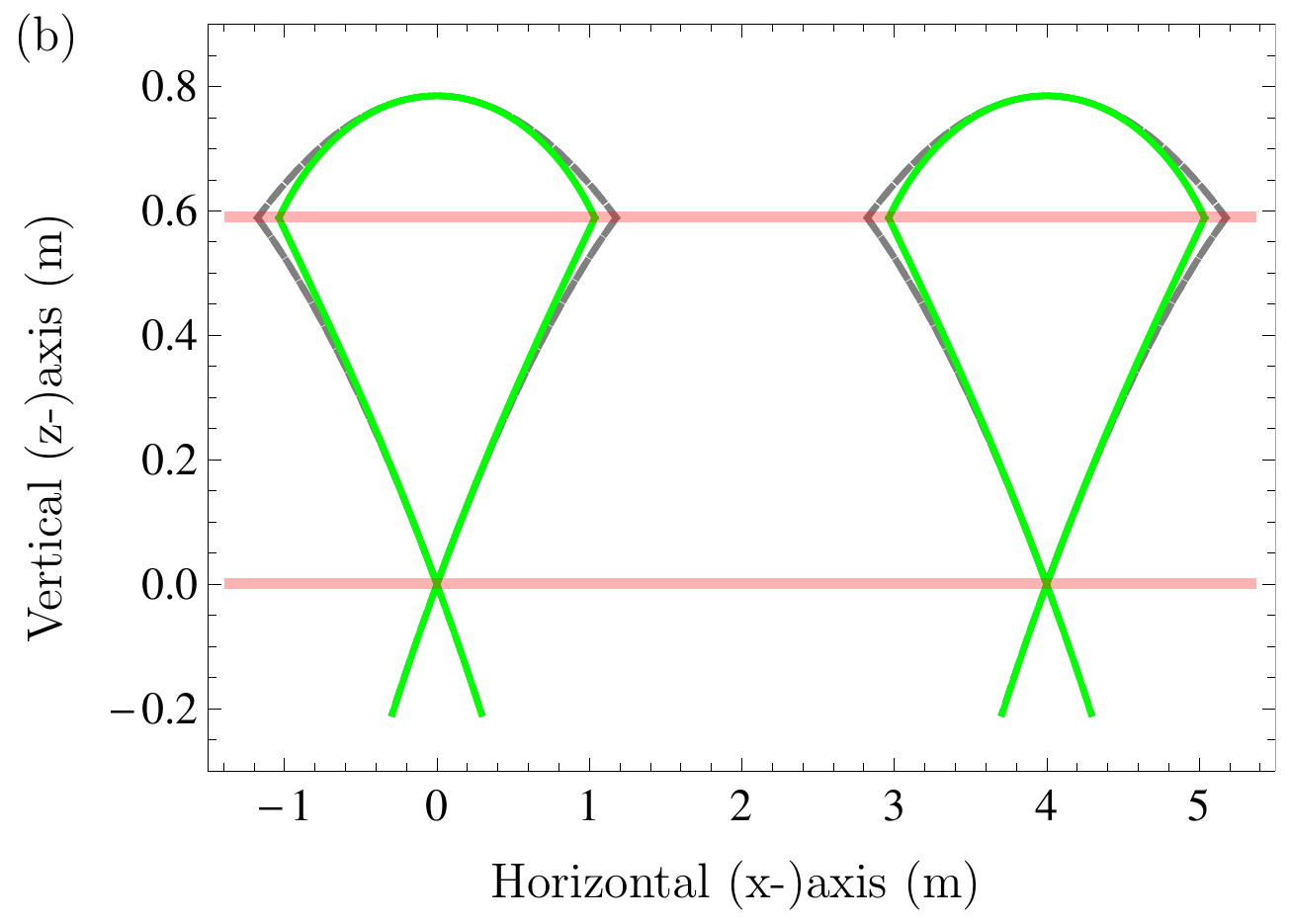}
\caption{Sketch of the double-loop geometry. Two interferometers (gray dashed / green lines) are manipulated by the same beam splitters (red lines).
Gray dashed lines correspond to the central trajectories in absence of gravity gradients, and green lines to the central trajectories with gravity gradients and the compensation technique.
(a) Time evolution of the trajectories in horizontal axis.
(b) Trajectories in vertical and horizontal axis. Herein, the input port and one output port overlap.
For clarity, the impact of the gravity gradient is exaggerated, and the baseline between the two interferometers is shortened.
}
\label{fig:atom-optics-double-loop-geometry}
\end{figure} Herein, the pulse spacing is $T$ -- $2T$ -- $T$, leading to a total interferometer time of $4T$.
A sketch of the geometry is shown in Fig.~\ref{fig:atom-optics-double-loop-geometry}.
Neglecting other terms, the differential phase between two double-loop interferometers induced by an impinging gravitational wave is $8k_{\mathrm{eff}}hL\sin^2(\omega{T}/2)\sin(\omega{T})$~\cite{Hogan2011}. 
As discussed previously, the double-loop geometry suppresses spurious phase shifts, especially when adjusting the wavenumber of the two central mirror pulses to cancel effects of gravity gradients~\cite{Roura2017PRL}.
To implement the double-loop geometry~\cite{Savoie2018,Hogan2011,Canuel2006,Dubetsky2006PRA}, the atoms are initially launched upwards and subsequently interact with the two spatially separated beamsplitting zones as depicted in Fig.~\ref{fig:atom-optics-double-loop-geometry}.
At the lower zone, the initial beam splitter and the final recombiner are applied, whereas the two mirror pulses are flashed onto the atoms in the upper zone.
The distance between these two zones defines the free fall time $T$.
Depending on the size of the interaction zones, $T$ may be tuneable within a limited range.
The double-loop geometry is also compatible with the implementation of a single-loop sequence.
Fig.~\ref{fig:atom-optics-strain-sensitivities} compares the strain sensitivities of these schemes and depicts the broadband and resonant detection modes for the case of a double-loop interferometer.

\subsection{Folded triple-loop geometry}
The folded triple-loop geometry is an alternative which features symmetric beam splitters~\cite{Gebbe2019arxiv,Abend2017Diss,ahlers2016} in the horizontal direction and relaunches~\cite{Abend2016PRL} at the intersections of the trajectories so that only a single laser link is required~\cite{Schubert2019arxiv}.
This enables a scalability in $T$, consequently a broadband detection mode~\cite{Hogan2016PRA}, and a resonant detection mode~\cite{Graham2016PRD} by adding additional relaunches and beam splitting pulses.
In addition, the triple-loop geometry is robust against fluctuations of the mean position and velocity of the wave packet entering the interferometer.
The scheme requires a pointing stability of the relaunch vectors at the level of $\sim$\,prad, comparable to the requirement on the initial launch vector in a single-loop geometry~\cite{Schubert2019arxiv}.
Omitting other terms, the differential phase shift between two triple-loop interferometers in a gradiometric configuration caused by a gravitational wave is $8k_{\mathrm{eff}}hL\sin^4(\omega{T}/2)\left[ (7+8\cos(\omega{T}))/2 \right]$~\cite{Hogan2011}.
A shot-noise limited strain sensitivity curve is shown in Fig.~\ref{fig:atom-optics-strain-sensitivities}~(a) (dash-dotted blue line).
As an option, this geometry can be implemented at a later stage for broadband and resonant detection modes.

\subsection{Beam-splitter performance estimation}
Common to all configurations described here is the need for large momentum transfer of around $k_{\mathrm{eff}}=1000\,k_{\mathrm{l}}$ while the atoms are moving perpendicular to the light beams at velocities in the order of a few m/s. Supposing atom-light interaction in the order of several $\mu$s per transferred photon recoil implies centimeter beam waists to cover the motion of the atoms. In the following, we estimate the power and waist of the beam required to realize 1000\,$k_{\mathrm{l}}$ momentum transfer for transverse atomic motion of $v_a = 4$~m/s, with numerical models both for sequential first-order Bragg diffraction and for accelerated optical lattices driving Bloch oscillations. 

\begin{figure}
\centering
\includegraphics[width=1\linewidth]{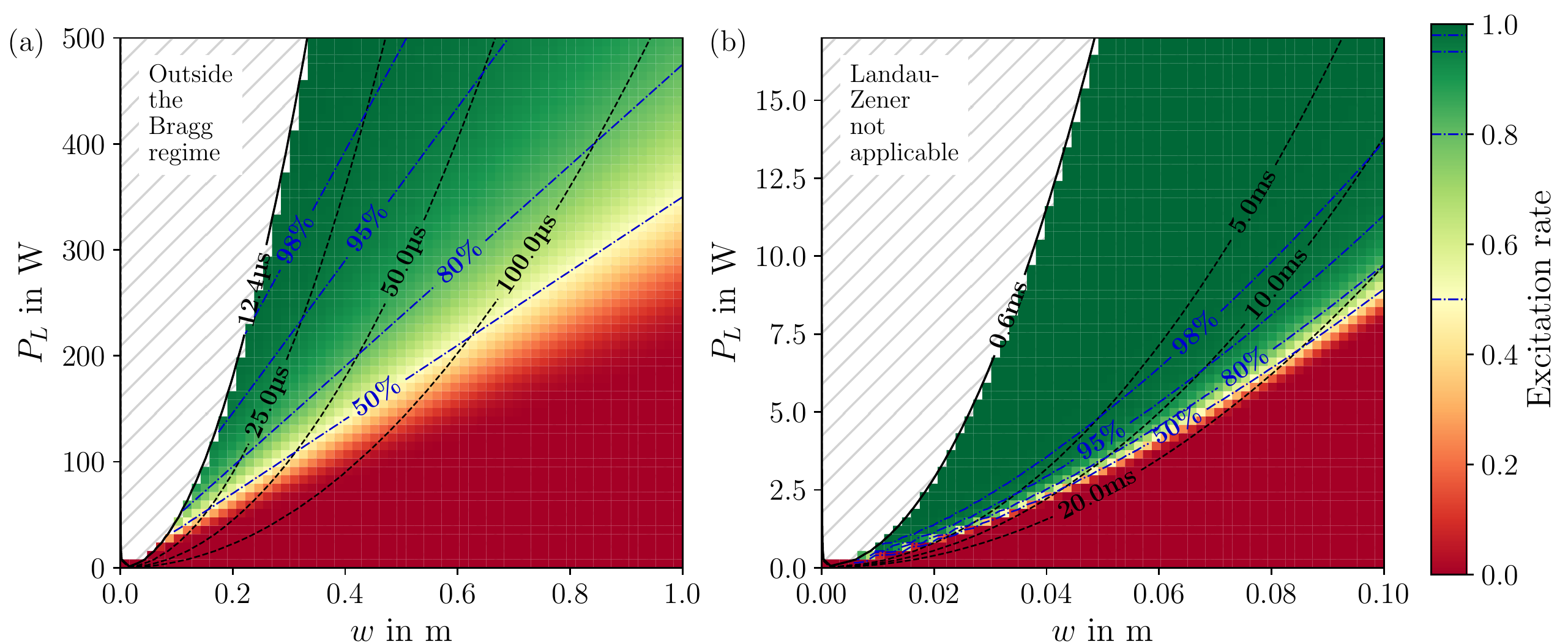}
\caption{Estimated efficiency of a $1000\,k_{\mathrm{l}}$ momentum transfer, assuming a transverse atom velocity of 4~m/s, as a function of laser waist $w$ and power $P_L$. The dashed blue lines denote parameters resulting in equal probabilities, which value is written on the line. \textbf{(a)} 500 subsequent first-order Bragg pulses. The dashed lines denote the pulse duration of a $\pi$-pulse, which is determined by the peak Rabi frequency $\Omega_0=\Omega_0(P_L,w)$. The gray area excludes the parameters outside the deep Bragg regime. \textbf{(b)} Accelerated Bloch lattice. The dashed lines denote the full duration of the acceleration sequence. The gray area excludes parameters for which the Landau-Zener model is not valid.}
\label{fig:bs_eff}
\end{figure}

To this end, we first consider an effective two-level system in the deep Bragg regime, which is the case when Rabi frequencies are small compared to the recoil frequency~\cite{Mueller2008}. The efficiency  $P_i$ of a single Bragg pulse is determined by the atoms' position at time $t_i$ within the transverse intensity profile of the beam. Therefore, we evaluate the integral~\cite{CheinetPhD}
\begin{equation}\label{eq:p_i}
P_{i} = \int dr \frac{1}{\sqrt{2\pi}\sigma_r}e^{-\frac{(r-r_0(t_i))^2}{2\sigma_r^2}} p(r)
\end{equation}
to weigh the spatially dependent excitation rate
\begin{equation}
p(r) = \sin^2\left(\frac 1 2 \int_0^\infty \Omega(t,r)~dt\right)
\end{equation}
with the density distribution of an atomic cloud of width $\sigma_r$, which is centered around $r_0(t_i)=r_0(0)+v_a t_i$. $\Omega(t,r)$ is the Rabi frequency given by the transverse intensity distribution of the light beam and the temporal shape of the light pulse~\cite{szigetiPhD}. Here, we consider Gaussian distributions in time as well as in position. The total efficiency of a $1000\,k_{\mathrm{l}}$ beam splitter is calculated by multiplying the individual single-pulse excitation rates. 
Evidently,  the pulses, for which the atomic density distribution is centered around the wing of the beam, are less efficient than the ones where the cloud passes its center. As a consequence, the beam waist needs to be sufficiently large to accommodate the atoms at all times, with appropriately enhanced power. Fig. \ref{fig:bs_eff} (a) illustrates the required parameters range. Typically, efficient Bragg pulses are realized with pulses of about $100~\mu$s duration, such that the atoms are expected to travel about 20~cm perpendicular to the beam. Indeed, in order to provide a sufficient light intensity over this distance, the required beam waist is in the order of several 10~cm, which in turn implies a high laser power. 

Moreover, we similarly study the performance of a beam splitter based on Bloch oscillations in a moving optical lattice with an adapted Landau-Zener model. To account for the inhomogeneities given by the beam divergence and atomic motion, we decompose the coherent lattice acceleration into single Bloch oscillations, of which the efficiency is quantified by 
\begin{equation}
\eta_{LZ}(r) = 1-\exp\left(-\frac{\pi \Omega_{bg}^2(r)}{2\alpha}\right). 
\end{equation}
The band gap $\Omega_{bg}(r)$ is numerically determined through the energy difference of the lowest Bloch bands at the boundary of the Brillouin-zone~\cite{Abend2017Diss}, and is a function of beam intensity. The lattice acceleration time $\tau_{acc}$, which enters the lattice chirp rate $\alpha = 1000\,k_{\mathrm{l}}/\tau_{acc}$, is chosen such that $1000\,k_{\mathrm{l}}$  are transferred with 99\% probability at the center of the beam. 
By replacing $p(r)$ with $\eta_{LZ}(r)$ in Eq.~\ref{eq:p_i}, we can convolute the time-varying spatial distribution of the atoms with the inhomogeneous excitation rate. Fig.~\ref{fig:bs_eff} (b) displays the results of this study. Notably, accelerated Bloch lattices transfer momentum in a shorter time than Bragg pulses, namely in the order of a few $10~\mu$s per photon recoil. As a consequence, the total momentum transfer can be achieved faster, such that the transverse distance travelled by the atoms is around one order of magnitude smaller as compared to Bragg. For similar intensities $I\sim P/w^2$, the required power is hence relaxed by two orders of magnitude. Comparing the requirements in power and waist of the two operation modes (a) and (b) of Fig.~\ref{fig:bs_eff}, one concludes that achieving the needed beam splitting is more realistic when Bloch lattices are implemented.  

\subsection{Conclusion on geometry and beam-splitting type}
Motivated by the available sites for ELGAR, the possibility to exploit baselines of several kilometres, and the differential suppression of laser frequency noise between two arms, the horizontal configuration is favoured.
To date, the highest momentum transfer between the two trajectories of an interferometer was realised by the combination of Bragg transitions and Bloch oscillations, and an effective wave number of $k_{\mathrm{eff}}=1000\,k_{\mathrm{l}}$ can be expected.
A double-loop interferometer with a gravity gradient compensation scheme offers robustness against spurious phase terms.
Therefore, the baseline design foresees a double-loop interferometer with a gravity gradient compensation scheme, an effective wavenumber of $k_{\mathrm{eff}}=1000\,k_{\mathrm{l}}$, a baseline $L=16.3\,\mathrm{km}$, and a pulse separation time $T=200\,\mathrm{ms}$ for an intrinsic phase noise of $10^{-6}$\,rad/$\sqrt{\mathrm{Hz}}$.
This design keeps the possibility open to implement other geometries such as the single-loop or folded triple-loop interferometer.
A vertical arm can be added in addition, with a trade-off in sensitivity due to a shorter baseline.
The latter is expected to prevent the full exploitation of the rejection scheme~\cite{Chaibi2016} for Newtonian Noise.

\section{Atom source}\label{sec_Atom_source}
The design of the atomic source is essential for the successful operation of the  interferometer as it determines the sensor sensitivity as well as the susceptibility to environmental effects. 
The intrinsic noise of a two-mode sensor with uncorrelated input states is characterized by the standard quantum limit (SQL) $\delta \phi_{\rm{SQL}} \sim 1/\sqrt{n_{\rm{meas}}~N}$, scaling with the number of atoms $N$ and interferometric measurements $n_{\rm{meas}}$. Therefore, the generation of large ensembles of atoms at a fast rate is desirable to attain sufficient stability level in a shortest possible integration time. Matching the target sensitivity of $\SI{1}{\mu rad/\sqrt{Hz}}$ for a single atom interferometer operating at SQL, requires the flux of $10^{12}$ atoms per second.

In order to mitigate various systematic and statistical contributions to the measurement uncertainty, it is necessary to engineer the external (spatial and momentum distribution) as well as the internal (e.g. magnetic) states of the atoms. The interferometry scheme under consideration requires the uncertainties in the center-of-mass position and velocity of the ensemble to be kept below $\SI{400}{nm}$ and $\SI{3}{nm/s}$, respectively, given by coupling of wave front imperfections, beam misalignment and pointing jitter etc. to the phase-space properties of the atoms (see section \ref{sec_other_noise_couplings}). Therefore, the careful design of the atomic source lies at heart of both experimental and theoretical efforts as will be outlined in the following.

\subsection{Species choice}
Cold-atom experiments typically employ alkaline species (Li, Na, K, Rb, Cs) as they offer multiple pathways to ultra-cold temperatures and quantum degeneracy.
Being the workhorse solution for atom interferometry for a wide range of applications, Rb is the prime choice for the gravitational wave antenna under consideration and is the work hypothesis for the entire paper. State-of-the-art atom optics techniques, as outlined in the previous section, have been primarily explored with Rb, which provides comfortable wavelengths in the optical range. Moreover, the lowest effective expansion energies in the order of a few tens of pK, expressed in temperature unit, have been demonstrated with Rb~\cite{Kovachy2015PRL,Rudolph2016Diss}. 
Another promising candidate is Cs, as employed for example in experiments for measurements of the hyperfine structure constant~\cite{Parker2018}, where efficient large momentum transfer was demonstrated.

Alkaline-earth and alkaline-earth-like atoms such as Yb, Sr and Ca are routinely utilized in optical atomic clocks due to the very narrow linewidth of their intercombination transition. 
Additionally, their immunity to quadratic Zeeman shifts is a distinct asset and renders them auspicious for high-precision experiments.
Indeed, these species are well-suited~\cite{Loriani2019} for space-borne gradiometers with large baselines based on single-photon transitions, which mitigate laser phase noise to serve as gravitational wave antennas~\cite{Graham2013PRL, Hogan2016PRA}. In these configurations, the baselines in the order of $10^9$~m alleviate the need for ambitious atom optics, requiring a few photon recoils per beam splitter only.
The distinct advantage of these schemes over two-photon excitation mechanisms is the inherent mitigation of laser phase noise, which is common-mode in a gradiometric setup. Schemes based on Raman- or Bragg-diffraction are, even in a gradiometric setup, prone to laser phase noise due to the finite speed of light, which prohibits simultaneous interaction of the two interferometers with the light over large baselines~\cite{Graham2013PRL}.
Recently, large momentum transfer through consecutive application of single-photon $\pi$-pulses was realized~\cite{Rudolph2020} in $^{88}$Sr on the weakly-allowed $^1$S$_0$--$^3$P$_1$ transition. However, the extension to higher momentum transfer is limited by the finite life time of the excited state, such that working on the narrower $^1$S$_0$--$^3$P$_0$ line is required. In bosons, this forbidden transition is enabled by application of a static magnetic field~\cite{Taichenachev2006}. However, the need to reach reasonable Rabi frequencies and to mitigate the second-order Zeeman shift put strict requirements on the magnitude and stability of the magnetic field~\cite{Hu2019}. In the fermionic isotopes, the intercombination line is allowed due to the nuclear spin and they seem to be a more promising solution in the long term despite their low isotopic abundance. Overall, in terms of heritage, the prime candidates for for ambitious experiments based on alkaline-earth species are Sr and Yb, both of which are being explored in large fountain interferometers~\cite{Hartwig2015,Rudolph2020}.

\subsection{Atomic source preparation}
Atomic sources for alkaline atoms are routinely operated in a double MOT configuration.
While a 2D+MOT can provide a high flux of atoms loaded from background vapor, either fed by dispensers or ovens, the 3D MOT is separated via a differential pumping stage. 
This has the advantage of maintaining good ultra-high vacuum in the main experimental chamber. 
Current 2D+MOT setups achieve loading rates in the order of $10^{11}$ atoms per second~\cite{Rudolph_2015}.
In the case of alkaline-earth-like elements, the 2D+MOT may be exchanged with a Zeeman slower, while the source has the same functionality and can achieve a comparable flux.
In the 3D MOT, atoms are first cooled down to the Doppler temperature, followed by sub-Doppler cooling in a molasses configuration almost reaching the recoil temperature. Cooling protocols based on gray molasses are inherently faster, and maintain a higher atomic density~\cite{Rosi2018,Naik2020}. Alkaline-earth-like atoms can be directly cooled on a narrow line achieving much lower Doppler temperatures. 
The laser-cooled sample is then loaded either in a magnetic or optical trapping potential, where the sample is evaporated to quantum degeneracy. 
The fastest evaporation rates are so far achieved either using atom chips~\cite{Rudolph_2015} or painted optical potentials \cite{Roy_2006,Condon2019} reaching condensation in $\simeq$1~s or less.
There are also recent examples of experiments which reach quantum degeneracy with direct laser-cooling, for example, by Raman sideband cooling in an optical lattice~\cite{Urvoy_2019}, these do not, however, reach performances comparable to conventional methods so far.
After reaching quantum degeneracy the samples expansion rate can be further reduced by delta-kick collimation.
Residual expansion rates in the pK regime have been previously achieved~\cite{Rudolph2016Diss,Kovachy2015PRL,Muntinga2013PRL}.
If the chosen isotope provides a magnetic sub-structure, the atoms may be transferred into the non-magnetic sub-state by pulsed or chirped RF fields prior to further manipulation.

From the region where the atomic sample has been prepared, it has to be transported to the interferometry region.
Atomic transport via shifting or kicking in trapping potentials suffers of several limitations due to required adiabaticity and delicate experimental implementation.
Therefore, the atomic transport to reach large enough distances away from the atom chip via an optical lattice is the preferred option, although requiring additional optical access. 
Bloch oscillations in an optical lattice~\cite{Peik1997} can be driven with very high efficiency and thanks to the discrete momentum transfer, the transport can be controlled very accurately.

\subsection{Atomic Flux}
The high degree of control over the atomic ensemble required to mitigate stochastic and systematic effects ultimately suggests the use of quantum degenerate sources. 
However, the flux of condensed atoms is arguably one of the most challenging aspects of the present proposal. In order to reach target performance of $\SI{1}{\mu rad/\sqrt{Hz}}$ phase sensitivity, a flux of $10^{12}$ atoms per second is required in shot-noise limited operation. This exceeds state-of-the-art flux \cite{Rudolph2016Diss} by six orders of magnitude; however, there are several promising pathways for improvement, and their combination is expedient to reach that goal. 
The natural effort is a higher atom number per experimental shot. Multiple techniques aim at increasing significantly the atom number before evaporation, such as improved mode-matching through time-averaged potentials. At the same time, in the present scheme, the preparation and interrogation zones in the interferometer are spatially separated, allowing for the concurrent operation of multiple interferometers~\cite{Savoie2018}. This might be accompanied by multiple source chambers and a system, which transfers the atomic sources to the interferometric region, for example based on moving optical lattices~\cite{Dahan1996}. Instead of a mere increase of the atom number, the phase sensitivity can  be also increased by employing entangled atomic ensembles (section \ref{subsec:squeezing}). Eventually, the realisation of an atom source needs a careful trade-off between sensor performance requirements and technical possibilities in terms of flux. Ultra-cold sources close to degeneracy might suffice to comply with the requirements and naturally offer a higher atom number and shorter preparation time as investigated in~\cite{Loriani2019}.

\subsection{Atomic sources for entanglement-enhanced interferometry}
\label{subsec:squeezing}
The requirements on the atomic flux can be reduced by employing atomic sources that can surpass the SQL. 
Sensitivities beyond the SQL can only be reached with nonclassical input states.
Such entangled atomic source can, to date, be generated by two main methods: Atom-light interaction (quantum non-demolition measurements or cavity feedback) or atomic collisions~\cite{Pezze2018}.
The best results so far were obtained with cavity feedback~\cite{Hosten2016, Cox2016} and allowed for a demonstrated enhancement of 18.5 dB, which is equivalent to a 70-fold increase of the atom number.
These techniques were so far only demonstrated with thermal ensembles, the best results with Bose-Einstein condensates were obtained with atomic collisions~\cite{Hamley2012} with a metrological gain of 8.3 dB.
A technologically interesting concept, that is also followed in laser interferometry, is the application of squeezed vacuum~\cite{Kruse2016}, where only the formerly empty input port is squeezed.
The development of a reliable, high-flux, quantum degenerate atomic source with a metrological gain of 20~dB for a 100-fold reduction of the required atom number presents an active research goal.

\subsection{Source engineering and transport} 

The operation of the atom interferometer requires the transport of the atoms from the cooling and evaporation chamber to a science or interferometry chamber where several interleaved interferometers are occurring. In order not to affect the cycling time, and hence the performance of the interferometer, this transport has to be time efficient. To this end shortcut-to-adiabaticity protocols, as proposed in~\cite{Corgier2018} and implemented in~\cite{Rudolph2016Diss}, will be engineered. These fast transports induce very low residual dipole oscillations in the final traps, with a typical amplitude of the order of 1\,$\mu$m. If a quantum control is needed i.e. to ensure the gases are in the ground state of the target traps, optimal control solutions are available and proved to be equally fast~\cite{amri2019} for relevant degenerate atomic systems. 

Moreover, the atomic ensembles need to be slowly expanding with expansion energies down to 10--100\,pK. This is realized by implementing the so-called delta-kick collimation technique. The atomic gases are freely expanding for a few tens of ms before being subject to an optical or magnetic trap flashed for a few ms that collimates the ensemble~\cite{Chu1986,Amman1997}. This drastically reduced the expansion rate of the cloud to the desired level in recent experiments~\cite{Muntinga2013PRL,Kovachy2015PRL,Rudolph2016Diss}.

Moving the atomic ensembles into the interferometry region might require the displacement over rather long distances (few tens of cm) in a short time. This is possible using an accelerated optical lattice driving Bloch oscillations. In a few ms, velocities corresponding to 200 recoils~\cite{Cadoret2008,Gebbe2019arxiv} could be imparted. Thanks to the atomic cloud collimation step, it would be possible to restrict the Bloch beam waists to 1--2 mm keeping the power usage at a reasonable level. In combination with the Bloch beams, symmetric schemes, that reduce the impact of spatially dependent systematics, will be sought for. They would rely on double-Raman-~\cite{Leveque2009} or double-Bragg-~\cite{ahlers2016} diffraction beams. 

\section{Seismic Isolation}
\label{sec_seis}
\bigskip

GW detectors, either using laser interferometry or matter-wave interferometry, share the same basic principle: the distance between two “free-fall” inertial test masses is precisely measured with a highly stable laser in order to detect tiny modulations that can be attributed to a GW. Noise due to seismic vibrations or to fluctuating gravitational forces can induce motions of the test masses, which limit the sensitivity of ground based detectors. For laser interferometry, the “free-fall” inertial test masses are mirrors suspended from high performance vibration isolation systems presently allowing for the detection at frequencies 10 Hz and above. 
%At lower frequencies, gravitational force fluctuations are important and will add with seismic noise, in which the ground's ambient motions, filtered through the detector’s vibration isolation system, produce motions of the test masses~\cite{PhysRevD.86.102001}. This terrestrial gravitational noise, namely Newtonian Noise (NN), will pose one of the main sensitivity limitations. It will need to be suppressed by orders of magnitude in the frequency band 10 mHz to 1 Hz in order to allow for GW observations~\cite{PhysRevD.92.022001}. 

In the ELGAR concept a matter-wave interferometer relies on atoms that are naturally in free fall and represent ideal test masses that are naturally isolated from seismic vibrations. Using as test masses free falling atom sources instead of suspended mirrors enables overcoming most of the technical limitations presented by optical GW detectors at low frequency. 
See Eq.~\ref{eq:SHnetwork} for the strain sensitivity of the ELGAR detector.
An examination of Eq.~\ref{eq:SHnetwork} displays an important difference between the AI GW detector sensitivity and that of laser interferometric GW detectors. The GW strain noise, $S^{1/2}_h(\omega)$ is related to the mirror and beam-splitter displacement noise, $S^{1/2}_{\delta x_{M_X}}$ by a factor of $4 \pi f_{GW}/c$. For laser interferometric GW detectors the displacement noise is divided by the arm length (3 km for Virgo and 4 km for LIGO) to produce the noise on the strain signal, so a factor of about $3 \times 10^{-4}$ m$^{-1}$. For the AI GW detectors the factor of $4 \pi f_{GW}/c$, and a frequency of $f_{GW} = 1$ Hz, provides a numerical factor of $\sim 2 \times 10^{-8}$ m$^{-1}$. This is a definite advantage in terms of seismic isolation. % for AI GW detectors.  
Still, it will be important to have adequate vibration isolation for the components of ELGAR. %This will be summarized below.
%The isolation of the optical components will be critically important for a GW detector based on atom interferometry. 
Indeed, the retro-reflection mirror
%or optical cavity for the beamsplitter laser link 
acts as an inertial reference for the atom interferometers in the detector. Although the differential signal between two atom interferometers, which contains the signal of the GW, already implies a rejection of mirror movement, a spurious coupling due to imperfections will remain.
%The targeted quality of the cavity adds additional requirements. 
Therefore, a sophisticated suspension system will need to be realized.

%\subsection{Seismic Noise}
%A section of the level of seismic noise itself. Be exhaustive. Get a geophysicist to do this. 

\subsection{Status of Seismic Isolation Presently for Gravitational-Wave Detectors and Test Systems}
%The Advanced LIGO~\cite{TheLIGOScientific:2014jea} and Advanced Virgo~\cite{TheVirgo:2014hva} GW detectors are currently operating. LIGO and Virgo have published 10 detections of GWs produced by binary black hole mergers, and a GW event from a binary neutron star merger~\cite{LIGOScientific:2018mvr}. These 11 detections came from the first two observational runs for Advanced LIGO and Advanced Virgo. In April 2019 LIGO and Virgo started their third obsevational run. As these detectors approach their design sensitivity there will be further observational runs~\cite{Aasi:2013wya}.

%The operating frequency band for Advanced LIGO and Advanced Virgo are not the same as that for the atom interferometer GW detector discussed in this paper. Advanced LIGO and Advanced Virgo were designed to observed in the 10 Hz to a few kHz band~\cite{TheLIGOScientific:2014jea,TheVirgo:2014hva}. Still, the seismic isolation systems for Advanced LIGO and Advanced Virgo are important to understand within the context of lower frequency applications.

The requirements for the isolation of the Advanced LIGO~\cite{TheLIGOScientific:2014jea} test masses (consisting of the input and end mirrors of the Fabry-Perot cavities in each arm of the interferometer) are such that the longitudinal noise must not exceed $1 \times 10^{-19}$ m Hz$^{-1/2}$ at 10 Hz. The Advanced LIGO test masses are suspended from a quadruple pendulum suspension with three stages of cantilevered blade springs~\cite{Aston_2012}. This configuration isolates the test mass in six degrees-of-freedom for frequencies of 1 Hz and above. In addition, there is an active damping system for suspension rigid body modes below 9 Hz, as acted up from the top stage of the suspension system. This is accomplished with a magnetic actuation system for the upper three levels of the suspension system. At the lowest level, an electrostatic actuation system is used. Much noise can be generated from the friction at the clamping points of the suspension~\cite{doi:10.1063/1.1148692,CAGNOLI1999230}. The use of silica wires has been determined to be the optimal way to minimize these sources of noise; these wires are attached to the mirror by welding or the use of silica bonding. This will replicate the contact between materials at the molecular level~\cite{doi:10.1063/1.1150040,PhysRevLett.85.2442,Grote_2008}. This is called a monolithic suspension~\cite{TheVirgo:2014hva}. The Advanced LIGO test masses are suspended via such a monolithic fused silica assembly. The suspension system is mounted upon seismically isolated optics tables. For the Advanced LIGO beamsplitter, it is a triple pendulum isolation. The requirement for the isolation of the beamsplitter is such that the longitudinal noise must not exceed $6 \times 10^{-18}$ m Hz$^{-1/2}$ at 10 Hz. A comprehensive description of the Advanced LIGO seismic isolation system can be found in~\cite{TheLIGOScientific:2014jea}.

Advanced Virgo has similar seismic isolation goals to that of Advanced LIGO~\cite{TheVirgo:2014hva}. The test masses (mirrors) are suspended in a monolithic suspension configuration. The mirrors are suspended via four wires from the {\it marionette}. The marionette is suspended by a single steel the so called super attenuator.
The super attenuator is a combination (both active and passive) seismic isolation system. The super attenuator for Advanced Virgo reduces the seismic noise by more than 10 orders of magnitude in six degrees of freedom above a few Hz. 
The super attenuator for Virgo consists of three principal systems: an inverted pendulum, a chain of seismic filters, and the mirror suspension.
The inverted pendulum~\cite{doi:10.1063/1.1149783} is made up of three 6-m long aluminum monolithic legs, each of which is connected to ground via flexible joints and supports an inter-connecting structure on the very top~\cite{TheVirgo:2014hva}. 
The chain of seismic filters consists of an 8 m long set of five cylindrical passive filters (each of which will reduce the seismic noise by 40 dB in both the horizontal and vertical degrees of freedom from a few Hz). Magnetic actuators are used to adjust the final alignment of the test masses~\cite{TheVirgo:2014hva}.
A comprehensive summary of the Virgo super attenuator and its performance characteristics are presented in~\cite{ACERNESE2010182}.

Research is already underway to extend the GW detection technology past that of Advanced LIGO and Advanced Virgo. For example, the Einstein Telescope (ET) is proposed to be a third generation ground-based gravitational wave detector with a sensitivity 10 times better than that of Advanced LIGO and Advanced Virgo~\cite{Punturo2010,Hild_2011,ET-design}. The current plan for ET is to have an equilateral triangle configuration, with arms lengths of 10 km. There will be 3 low-frequency interferometers, and three high frequency interferometers~\cite{Punturo2010,Hild_2011,ET-design}. The observational band for ET will start at a few Hz. The design of ET is such that seismic, gravity gradient, and suspension thermal noise will be the limitation at the very low frequencies of 1 - 10 Hz~\cite{Punturo2010}.
%, while suspension thermal noise and quantum noise (photon radiation pressure on the suspended mirrors in the Fabry–Perot cavities of the arms will be the limitation in the  10 - 40 Hz band~\cite{Punturo2010}. 
The upper part of the ET suspension for the mirrors will the super attenuator, similar to that of Advanced Virgo. This will provide the needed seismic and acoustic isolation. In order to obtain a sufficiently low cut-off frequency the height of the individual pendulum stages of the super attenuator will be 2 m per stage. The proposed modified super attenuator will consist of of six pendulum stages, each stage providing horizontal and vertical isolation). The total height of the super attenuator will be 17 m~\cite{Hild_2011}. The last suspension stage, or payload, will be crucial for setting the thermal noise performance and the mirror control. The payload is the system that couples the test mass to the super attenuator; it compensates the residual seismic noise and steers the mirror through internal forces exerted from the last super attenuator element. The payload is similar to that of Advanced Virgo, namely a marionette, a recoil mass, and the test mass (mirror). The marionette is the first stage used to control the mirror position by means of coil-magnet actuators acting between the last stage upper part of the suspension and the marionette arms, while the recoil mass is used to protect and to steer the mirror. On the mechanical structure electro-magnetic actuators (coils) act on the mirror~\cite{ET-design}. The goal of the ET seismic isolation system will be to have a strain noise sensitivity of $5 \times 10^{-21}$ Hz$^{-1/2}$ at 1 Hz, and $1 \times 10^{-24}$ Hz$^{-1/2}$ at 6 Hz~\cite{Hild_2011}.  

Another important method for observing low-frequency GWs is the Torsion-Bar Antenna (TOBA), which will attempt to observe GWs in the 1 mHz to 10 Hz band~\cite{PhysRevLett.105.161101,Shimoda2019}. The target GW strain sensitivity for TOBA is $10^{-19}$ Hz$^{-1/2}$ at 0.1 Hz. Because a mechanical oscillator behaves as a freely falling mass above its resonant frequency, the low resonant frequency of the torsion pendulum could provide for the observation of GWs, and in the case of TOBA, this would correspond to observational frequencies of a few tens of mHz. The TOBA design is to have two orthogonal bars suspended horizontally by wires. The tidal force from a GW would produce differential rotations on the bars, thereby providing the means to observe GWs through the measurement of the rotation of the bars and observed via interferometric sensors~\cite{Shimoda2019}. Because of the low-frequency attempt to measure GWs, a sophisticated seismic isolation system would be needed. The current suspension and isolation system for TOBA consists of the torsion pendulum suspended from an intermediate mass, and then the whole system is connected to an active vibration isolation table. The goal is to produce a near-term demonstration of a sensitivity of $10^{-15}$ Hz$^{-1/2}$ at 0.1 Hz. See~\cite{Shimoda2019} for more explicit details on the seismic isolation goals for TOBA.

\subsection{Seismic Isolation for Atomic Interferometer Systems}
Vibration isolation systems are extremely important for experiments involving atom gravimeters~\cite{doi:10.1063/1.4919292} and atom interferometers~\cite{doi:10.1063/1.4895911}. A recent study displayed the results of a three-dimensional active vibration isolator (essentially an isolated table) applicable for atom gravimeters. The results were the suppression of vertical vibration noise by a factor of 50, and a factor of 5 for horizontal noise in the 0.1 Hz and 2 Hz band~\cite{doi:10.1063/1.4919292}. 
%Atom interferometers using ultracold atoms have been demonstrated to make excellent gravimeters, allowing for gravity measurements statistical uncertainties of 10 nm s$^{-2}$~\cite{Karcher_2018}. 
Experiments like these use a both passive isolation and vibration correction methods~\cite{LeGouet2008}. Another study demonstrated a programmable broadband low frequency active vibration isolation system for atom interferometry~\cite{doi:10.1063/1.4895911}. They started with a passive isolation platform where the vertical resonance frequency was to 0.8 Hz. A digital control system was then activated to provide feedback, and in doing so, the intrinsic resonance frequency of the system was reduced to 0.015 Hz. It was then demonstrated that the use of this active isolation improved the vertical vibration in the 0.01 to 10 Hz band by up to a factor of 500 over the passive isolation. In the end this experiment had a vibrational noise spectrum density as low as $6 \time 10^{-10} g$ Hz$^{-1/2}$in the 0.01-10 Hz band.
% thereby allowing their atom interferometer to achieve an uncertainty of $6.6 \times 10^{-9} g$~\cite{doi:10.1063/1.4895911}.

\begin{figure}[ht]
\centering
\includegraphics[width=.8\linewidth]{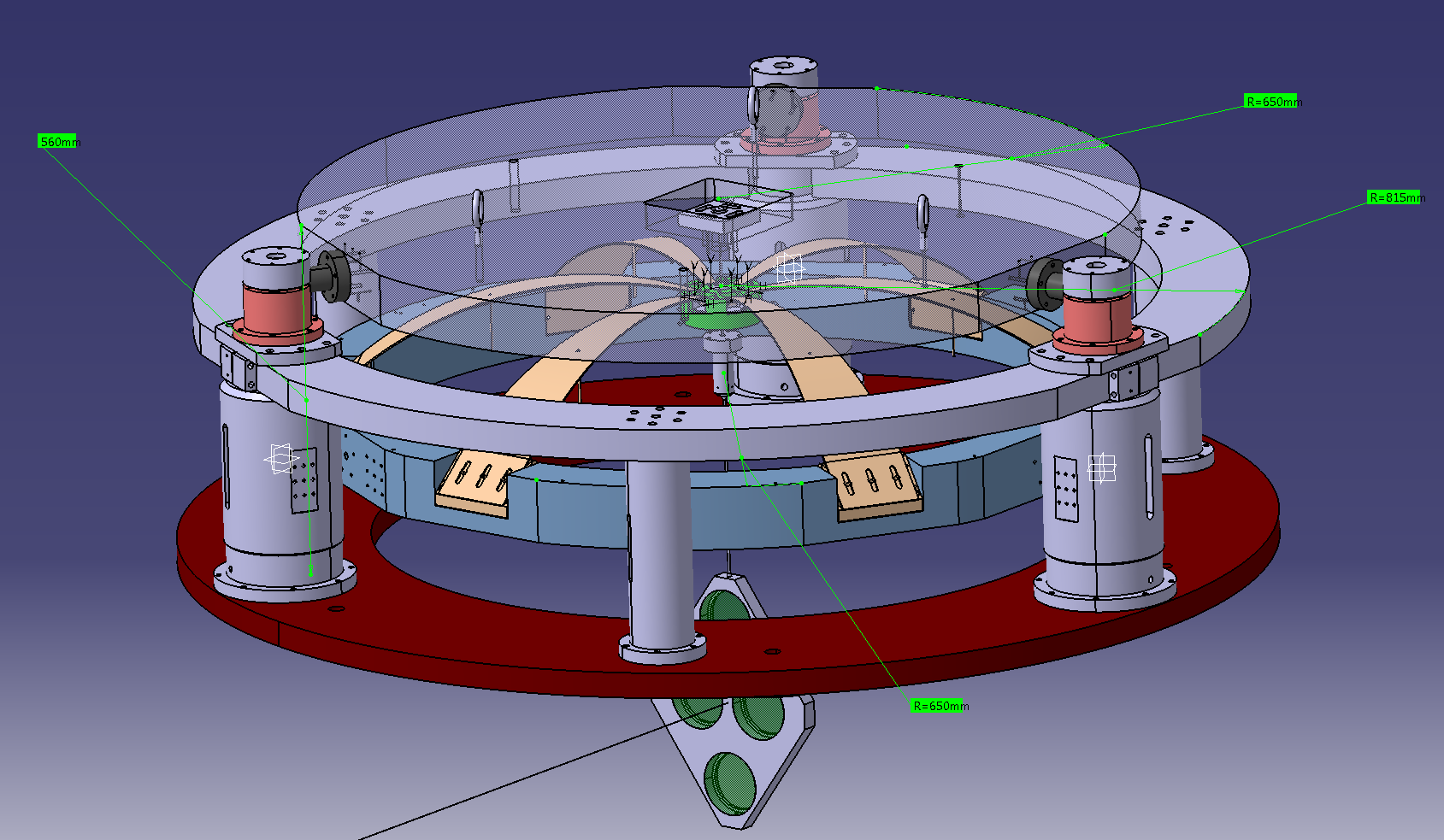}
\caption{A drawing of the MIGA seismic isolation system for its beamsplitter and mirrors. This system has been constructed and its performance is presently being measured. The system consists of two 20 cm pendula that provide isolation for horizontal noise, and pre-constrained blades for vertical isolation.}
\label{fig:MIGA_seismic}
\end{figure}

There is already a very active research program for seismic isolation for atom interferometric GW detectors. As part of the research for the matter-wave laser interferometer gravitation antenna (MIGA) there has been the design and construction of an isolation system for the MIGA mirrors and beamsplitter~\cite{MIGO-Canuel-2014}. The prototype is a similar to Virgo's super attenuator isolation system~\cite{ACERNESE2010182}. The MIGA prototype isolation system has two 20 cm pendula that provide isolation for horizontal noise, and pre-constrained blades for vertical isolation. A drawing of this system can be seen in Fig.~\ref{fig:MIGA_seismic}. The characteristics of this system are presently being measured and its actual performance will be published in the near future.

\subsection{Requirements for the suspension of ELGAR's optics}
%The ELGAR detector will need to take advantage of innovative and effective seismic isolation techniques if it is to succeed at low frequencies. Possible scenarios are discussed below.
%The isolation of the optical components will be critically important for a GW detector based on atom interferometry. The retro-reflection mirror or optical cavity for the beamsplitter laser link acts as an inertial reference for the atom interferometers in the detector. Although the differential signal between two atom interferometers, which contains the signal of the GW, already implies a rejection of mirror/cavity movement, a spurious coupling due to imperfections will remain. The targeted quality of the cavity adds additional requirements. Therefore, a sophisticated suspension system will need to be realized. 
%\subsubsection{Isolation of the Cavity Mirrors.}
The motion of the retroreflecting mirrors and the beamsplitter will contribute to the noise of the detector, and could be a limitation on its sensitivity.
As can be seen from Eq.~\ref{eq:SHnetwork}, the spectral density of the optics motion, $S_{\delta x_{M_X}}$, will be the limity on the sensitivity if
\begin{equation}\label{eq:Sh_mirror}
S_{h}=\frac{4\omega^{2}}{c^{2}}S_{\delta x_{M_X}}(\omega).
\end{equation}
Consider a target sensitivity of $3.3 \times 10^{-22}$ Hz$^{-1/2}$ at 1 Hz.
The requirement for on the noise for the motion of the retroreflecting mirror is be to less than $8 \times 10^{-15}$m Hz$^{-1/2}$ at 1 Hz.
Similarly, for a target of a strain sensitivity of $2 \times 10^{-19}$ Hz$^{-1/2}$ at 0.1 Hz, the limit on the retroreflecting mirror noise would be $5 \times 10^{-11}$m Hz$^{-1/2}$.
To reach such performances, the ELGAR detector will need to take advantage of innovative and effective seismic isolation techniques. Possible scenarios are discussed below.

%\subsubsection{Beamsplitter Isolation}
%When there is position noise (created, for example, by seismic noise) there can be the creation of frequency noise in the light entering one cavity of the GW detector but not the other cavity. Consider a detector where its arms define the $x$ and $y$ axes, and consider light traveling in the $x$ arm. The beamsplitter will be susceptible to seismic noise, which will induce motion in the $x$ direction of $\delta x$. This will then cause the light that interacts with the beamsplitter to acquire a change in its phase $\phi_{L}$, which can also be interpreted as light frequency noise, $\Delta \nu = \frac{1}{2 \pi} \frac{d \phi_{L}}{dt}$. As derived in~\cite{Chaibi2016}, a minimum sensitivity (SNR = 1) gives the condition for the minimum displacement noise of $\delta x_{min} = h c /2 \pi f$. Assuming a target strain sensitivity for the GW detector of $4.6 \times 10^{-22}$ Hz$^{-1/2}$ at 1.7 Hz, the requirements for the seismic noise isolation is $\delta x_{min} \sim 1.3 \times 10^{-14}$ m Hz$^{-1/2}$. For a frequency of 0.18 Hz with a strain sensitivity of $1 \times 10^{-19}$ Hz$^{-1/2}$ it will be necessary to reduce the displacement noise of the beamsplitter to $\delta x_{min} \sim 2.65 \times 10^{-11}$ m Hz$^{-1/2}$~\cite{Chaibi2016}.

\subsection{Possible Suspension Designs}
The requirements for the development of adequate vibration isolation system will be challenging, but not impossible. This will be an active and important research area for the types of AI GW detectors discussed in this paper. One could naively imagine, for example, a double suspension~\cite{doi:10.1063/1.1142030} with a resonance frequency at 10 mHz. Such a system would have a transfer function for seismic noise of $10^{-4}$ at 0.1 Hz,  $10^{-8}$ at 1 Hz, and  $2 \times 10^{-12}$ at 10 Hz. However a simple pendulum with a resonant frequency of 10 mHz would require a length of 25 m, which becomes both impractical and expensive. Other solutions must clearly be found.

\begin{figure}[ht]
\centering
\includegraphics[width=.8\linewidth]{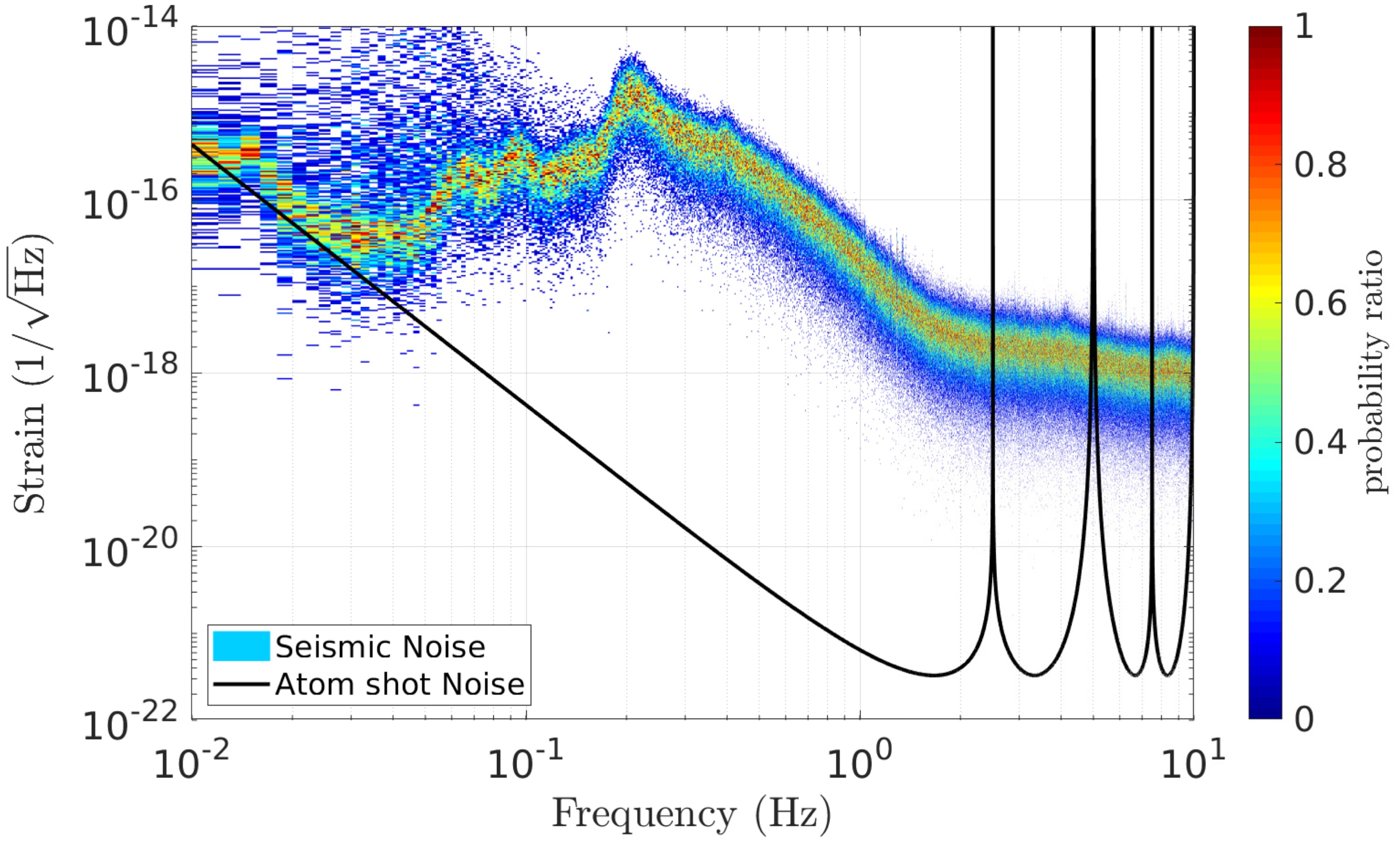}
\caption{An example of input seismic noise (as measure in the galleries of the Laboratoire Souterrain \`a Bas Bruit, Rustrel, France) is displayed in colors. The black curve shows the estimated atom shot noise.}
\label{fig:seismic}
\end{figure}

As an example of the isolation requirements derived by considering the desired sensitivity and the seismic noise of a quiet location, see Fig.~\ref{fig:seismic}. The seismic noise was measured in the galleries of the Laboratoire Souterrain \`a Bas Bruit, Rustrel, France. Also presented is the target sensitivity of ELGAR. The isolation transfer function required is approximately $1.4 \times 10^{-3}$ at 0.1 Hz, $4 \times 10^{-5}$ at 1 Hz, and $3 \times 10^{-4}$ at 10 Hz. 
%One can imagine, for example, a double suspension~\cite{doi:10.1063/1.1142030} with a resonance frequency at 10 mHz. Such a system would have a transfer function for seismic noise of $10^{-4}$ at 0.1 Hz,  $10^{-8}$ at 1 Hz, and  $2 \times 10^{-12}$ at 10 Hz.

%The retro-reflection mirror or optical cavity for the beam splitter laser link acts as an inertial reference for the atom interferometers in the detector. Although the differential signal between two atom interferometers, which contains the signal of the GW, already implies a rejection of mirror/cavity movement, a spurious coupling due to imperfections will remain. The targeted quality of the cavity adds additional requirements. Therefore, a sophisticated suspension system will need to be realized. The initial design assumption is a target of $x(f) = 10^{-14} \textrm{m}/\sqrt{\textrm{Hz}}$ at 1 Hz, but an ultimate goal of of $x(f) = 10^{-15} \textrm{m}/\sqrt{\textrm{Hz}}$ at 1 Hz is a target. To achieve these goals will require a dedicated research program. 

One promising strategy is to adapt the design developed for the Australian International Gravitational Observatory (AIGO)~\cite{AIGO_2009,doi:10.1063/1.3250841}. 
The AIGO Compact vibration isolation and suspension system demonstrated two stages of horizontal pre-isolation and a single stage of vertical pre-isolation with resonant frequencies around 100~mHz. 
The demonstrated system contained multiple pre-isolation stages, self-damping pendulums, Euler springs~\cite{Winterflood_2002}, and niobium ribbons suspension. There were three pre-isolation stages consisting of an inverse pendulum for horizontal pre-isolation, a LaCoste linkage for vertical pre-isolation~\cite{274fe9face5f4a5fbcd42fb43eb7db16}, a second stage of horizontal pre-isolation based on a Roberts linkage~\cite{doi:10.1063/1.1583862}. At the end of the suspension system there was a a control mass which interacted with the test mass via niobium flexures~\cite{AIGO_2009,doi:10.1063/1.3250841}. This demonstration from AIGO was not quite at the lower frequencies needed by ELGAR, but the methods presented there could possibly be further extended toward lower frequencies. For example, an Euler-LaCoste linkage has been demonstrated for vertical isolation with a resonance frequency of 0.15 Hz~\cite{274fe9face5f4a5fbcd42fb43eb7db16}; the same group has also constructed a coil spring based LaCoste stage for vertical isolation with a resonant frequency of 0.05 Hz, and indications that it could be further improved~\cite{doi:10.1063/1.3250841}. The same group also reports a resonant frequency of 0.0006 Hz for a Scott-Russell horizontal linkage~\cite{doi:10.1063/1.3250841,WINTERFLOOD1996141}. The pursuit of such low frequency isolation will always face difficulties due to losses and anelasticity~\cite{doi:10.1063/1.3250841,doi:10.1063/1.1144774}.

%It should be noted that a gradiometer rejection of the retroreflection mirror motion will have a limited efficiency. In addition, the interrogation laser frequency noise rejection by the Michelson configuration includes the beam splitter seismic noise into the detector output. 
The lowering of the resonance frequency of the seismic isolation system cannot just be an upgrade of the super attenuators used in optical GW detectors~\cite{accadia:in2p3-00603336}; a new design is required. This is why this vibration isolation project for ELGAR should be dedicated to creating an innovative suspension design for integration into a GW detector. The Roberts linkage (converting a rotational motion to approximate straight-line motion) for horizontal isolation and the Euler-Lacoste linkage (using pre-compressed springs) for vertical isolation were previously proposed and their performance (as noted above) has been subsequently demonstrated~\cite{doi:10.1063/1.1583862,274fe9face5f4a5fbcd42fb43eb7db16}. A possible strategy could be the optimization of these designs to make them robust by reducing the number of degrees of freedom. For ELGAR the goal is to produce an upgraded innovative design to push the vibration isolation system to an even lower resonance frequency. Whereas the Roberts linkage can have, by its design, as low frequency a mechanical adjustment as can be accurate, the Euler-Lacoste linkage is based on a low resonance frequency mechanical spring. A new design can then be proposed and its feasibility fully studied. It is through this effort that the vibration isolation around 1 Hz can be such that
the sensitivity of the GW detector can be achieved. A vertical antenna configuration, where the laser propagates between vertically displaced atomic probes, would require a different geometry for the retro-reflecting mirror suspension. 

In addition, a trade-off between active / passive systems, post-correction techniques, and their combinations must be studied.
%and if feasible, an implementation for the horizontal or vertical design.
Lessons can be learned form the Advanced LIGO seismic isolation system~\cite{Matichard:2015eva}. The optics for Advanced LIGO hang from suspension systems which are mounted on a pair of tables that are actively isolated. Attached to the ground is a hydraulic platform, or a Hydraulic External Pre-Isolators (HEPI). On top of the HEPI is an Internal Seismic Isolation within what is called the Basic Symmetric Chamber (BSC-ISC). Finally, a quadruple pendulum suspension supporting the test mass is attached to the BSC-ISC. The HEPI platform is used to align and position the entire system, while the BSC-ISC uses active isolation down to 0.05 Hz~\cite{Matichard:2015eva}. This combination system of active isolation and passive suspension provides an example of what ELGAR might be able to use, although research and development work will be necessary to decrease the operating frequency band.

\subsection{Control Strategy}
The control strategy for maintaining the position and operation of the optical components, and providing seismic isolation, will be necessary and inherently complicated. For the mirrors and beamsplitter it will ultimately be necessary to use sensing and feedback systems to control all degrees of freedom. It is likely that actuation forces will be necessary at all stages of the isolation system, from pre-isolation tables to the final suspension stages holding the mirror or beamsplitter. It will be necessary to maintain a locked laser and optical system while maintaining the alignment of the various components, accounting for drifts, and damping excitations. Pushing these requirements to the low frequency band for ELGAR will require research and development work.

The control system must slow-down and align the optical components, putting their operational dynamics within the gain range of the servo systems and the error signals generated by wavefront sensing for angles,  and locking for longitudinal position. This should be done in such a way that only internal forces are used.

There are different options available for applying actuation forces within the suspension system. For gravitational wave detectors there is typically the use of magnetic or electrostatic actuation. Magnetic actuation is typically done with coil and permanent magnetic combinations; for example, note how they are used in Advanced Virgo~\cite{Acernese2014}. A small adjustment, done quickly, of the position of the optical component can be made; one must account for the mechanical transfer function of the system. When forces are applied to the suspension system the energy associated with the actuation will be primarily absorbed at mechanical resonant frequencies; this creates potential difficulties. To apply actuation on a mirror will require the use of a reaction mass. Similar issues were considered for the proposed seismic isolation system of the Einstein Telescope~\cite{Punturo2010}. There is a down-side to magnetic actuation, namely the susceptibility to noise coupling from magnetic fields~\cite{Cirone_2019}. This can be especially problematic with respect to the globally coherent Schumann resonances, which can be a source of coherent noise between gravitational wave detectors at different locations around the world~\cite{PhysRevD.87.123009,PhysRevD.97.102007}.

Electrostatic actuation is another possibility for use in the control of the mirrors and beamsplitter for ELGAR. This is what is used for actuation of the Advanced LIGO test masses~\cite{TheLIGOScientific:2014jea,Carbone_2012}. There are benefits to using electrostatic actuation. No magnets need to be attached to the mirrors, and in that way, this reduces thermal noise by not deleteriously affecting the quality factor of the mirror. There is also no susceptibility to magnetic fields with the electrostatic actuators. A down side would be when the mirror acquires stray charge, which is known to happen~\cite{Martynov:2016fzi}, then additional noise can be induced, although modulating the sign of driving actuation field has been show potentially alleviate this problem~\cite{Acernese_2008}.

%Get rid of resonance. Have to bring energy somewhere. Control like Virgo. Blair to use magnetic induction and dissipation (resistivity).  Problem of control at low frequency without adding noise. Goal of paper is to define the issues. LIGO and Virgo subtract the effect of the control.  Measure what you are doing and subtract it from the data. Event harder at low frequency since electrical measurement is harder. 

A control scheme for the suspension system must mitigate mechanical resonances if the system is to be brought into lock. The attenuation of resonances by the control system will be necessary, but these techniques have already been demonstrated in other systems~\cite{doi:10.1063/1.3250841}. However, when this is done the energy must be extracted from the system. Techniques to do this have used magnetic induction and dissipation (resistivity)~\cite{BRAGINSKY1996164,AIGO_2009,doi:10.1063/1.3250841}. It is known that these suspension control systems can add noise to GW detectors. Advanced LIGO and Advanced Virgo measure this control noise, and subtract it when producing their calibrated strain data~\cite{Meadors_2014,Davis_2019,Acernese:2018bfl}.

In summary, ELGAR will require a sophisticated control systems to align and lock its retro-reflecting mirrors and the beamsplitter. Only in this way can the detector be properly operated and achieve its desired sensitivity. One can imagine a system of pre-isolation tables, similar to what is used by Advanced LIGO~\cite{Matichard:2015eva}. The suspension system will reside on a pre-isolation table, and the optical component (mirror or beamsplitter) will be suspended and have seismic noise filtered to an appropriate level (as discussed above). Finally, the optical component will be controlled via magnetic or electrostatic actuation. Such actuation and control is necessary to keep the detector locked. %and the GW detector operational. 
Clearly the ultimate seismic isolation system for ELGAR will be complex, and to arrive at its construction and installation will take much research. Such a low-frequency isolation system would likely have additional applications outside of GW detection~\cite{Harms2019}.

%... subtraction of control noise in Advaned LIGO~\cite{Meadors_2014,Davis_2019}.
%Presentation of low frequency control scheme of suspended cavities for matter wave interferometers 
%Elaboration of standards for cavity control (local control, linear alignment, mirror accuation)

%A GW detector using atom interferometery will occupy a critical frequency band between LISA (0.1 Hz upper limit)~\cite{2017arXiv170200786A} and LIGO-Virgo (10 Hz lower limit)~\cite{TheLIGOScientific:2014jea,TheVirgo:2014hva}. Although atoms are free falling, reducing the detector sensitivity in the 0.1-10 Hz frequency band will require significant frequency seismic isolation. Innovative techniques will be required to achieve the desired noise level of $x(f) = 10^{-15} \textrm{m}/\sqrt{\textrm{Hz}}$.

\externaldocument{seismic_isolation}

\section{Newtonian-noise reduction}\label{sec_NN_reduction}

Local gravity perturbations can create a spurious atom phase variation on the signal of an atom gradiometer, also known as Newtonian noise~(NN), which can limit the sensitivity of a GW detector~\cite{Saulson}. Indeed, in the gradiometric configuration considered in Sec.~\ref{sec_Detector_configuration}, the freely falling atoms play the role of reference masses corresponding to the suspended cavity mirrors in an optical GW detector: the equivalence principle implies that atom gradiometers will be equally sensitive to NN as an optical GW detector.

Newtonian noise has long been identified as important issue for optical GW detectors and has been thoroughly studied since their first generation \cite{Beccaria1998, Hughes1998, Creighton2008, Harms2019}. Up to now, Newtonian noise has not been revealed as a limiting effect, given the sensitivity and the bandwidth of state-of-the-art detectors, which currently operate above 10~Hz~\cite{PhysRevD.86.102001}. However, for instruments planned to operate below this frequency, studies have shown that NN will start to be limiting \cite{Beker2012,Fiorucci2018}. The operating window of the ELGAR project (0.1~Hz -- 10~Hz) makes NN a very important topic to address. We present here a preliminary study based on existing work applied to the ELGAR configuration.

\subsection{Sources of Newtonian noise}
Two main categories of sources can been defined: moving masses close to the test masses and density variations of the surrounding medium of the detector \cite{Harms2019}. In the first category fall all perturbations linked to human activity such as movement of vehicles or vibration from close-by apparatus, but also different geophysical processes such as for example hydrological effects accounting for water transfers. 
% in the the surrounding medium.  
Human activity can be constrained, and defining a no disturbance perimeter for the antenna will be a necessity~\cite{Thorne1999} to limit its impact on the antenna. For what concerns geophysical effects, their impact for a given potential candidate site must be carefully evaluated: in particular, hydrological effects can be characterized using superconductive gravimeters, which enables the development of precise models of the induced gravity variations~\cite{Rosat2018}. In the second category stands the variation of density of the medium surrounding the antenna which could arise from local seismic activity and perturbations of the atmospheric pressure, contributions that were previously identified as the dominant effects for GW detectors. In the following, we briefly introduce the models used to calculate the gravity perturbation and the strain limitation that both effects will induce on the antenna.

The ground is in continuous motion, driven by the deformation of the Earth criss-crossed by seismic waves propagating within the upper mantle. Beside inducing a direct displacement noise felt by objects mechanically linked to the ground, which may directly impact the sensitivity of an antenna (see discussion in Sec.~\ref{sec_seis}), seismic activity also creates a density variation of the surrounding medium that modifies the local gravitational field leading to a spurious displacement of the detector test masses. In the frequency window of interest of ELGAR (0.1~Hz -- 10~Hz) lies the oceanic microseismic peak, which can show strong seasonal and regional variations. Supposing that within this window, the vertical motion of the ground can be modelled as an incoherent sum of Rayleigh waves~\cite{Beccaria1998, Hughes1998}, one can derive the expression of the seismic NN introduced into a single gradiometer of the detector. The PSD of the difference of the local gravity field induced between the points $X_i$ and $X_j$ near the surface where the gradiometer test masses are located can be written~\cite{Junca19}:
%The seismic contribution within the term $\Delta a_x(d_{ij},\omega)$ in Eq.~(\ref{eq:Sh}) can then be written~\cite{Junca19}:
\begin{equation}\label{eq:SDaSeismic}
S^R_{\Delta a_x}(L,\omega)=\kappa_R^2(k_R)S_{\xi_z}(\omega) (1-J_0(k_R L)+J_2(k_R L))\, ,
\end{equation}
where $\kappa_R(k_R)$ is a function of the mechanical properties of the ground and the frequency of the seismic wave,
$J_n$ the $n^{\text{th}}$ Bessel function of the first kind, $S_{\xi_z}(\omega)$ the power spectral density of the vertical displacement of the ground.

From the same mechanism, fluctuations of atmospheric pressure can create density variations within the atmosphere that may also impact ELGAR's measurements.
Considering that pressure variation inside the atmosphere are adiabatic and described by an isotropic superposition of acoustic plane waves, the PSD of the difference of the local gravity field induced on the test masses of a single gradiometer can be written~\cite{Junca19}:
\begin{equation}\label{eq:SDaAtmo}
S^I_{\Delta a_x}(L,\omega)=\kappa_I^2(k_{I})S_{\delta p}(\omega) \big\langle 2(1-\cos(k_I L\sin\theta\cos\phi)) e^{-2hk_I\sin\theta}\sin^2\theta\cos^2\phi \big\rangle_{\theta,\phi}\, ,
\end{equation}
where: $\kappa_I(k_{I})=\frac{4\pi G\rho_0}{\gamma p_0 k_{I}}$, $S_{\delta p}(\omega)$ is the power spectral density of the pressure noise considered constant around the whole detector and $\langle.\rangle_{\theta,\phi}$ defines an averaging over all the propagating directions $(\theta,\phi)$ of the sound waves.
 \begin{figure}[t!]
  \centering
    \includegraphics[width=0.8\textwidth]{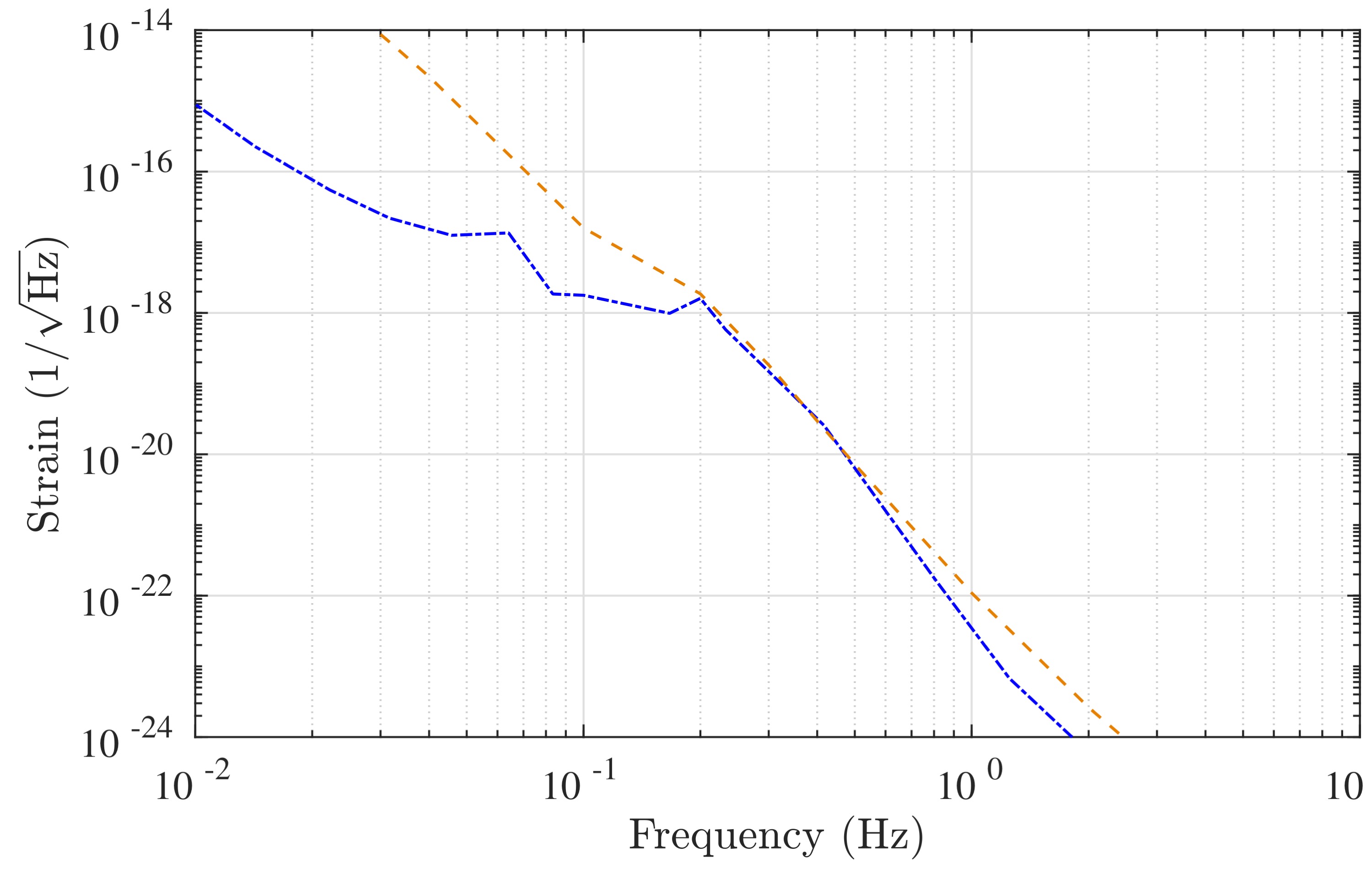}
    \caption{Comparison of NN from seismic and atmospheric sources. % with the atom shot noise of a gradiometer. 
    Dot-dash line (blue) represents seismic NN using Peterson's low-noise model (NLNM) \cite{Peterson1993} for the vertical displacement noise, dash line (orange) represents atmospheric NN using mid-Bowman model \cite{Bowman2005} for the pressure noise.}%, plain line (green) represents the atom shot noise for a gradiometer with characteristics similar to Fig.~\ref{fig:atom-optics-strain-sensitivities} with double-loop interferometers.}
    \label{fig:NN1gradio}
\end{figure}

%\subsubsection{Newtonian Noise projections}
%\medskip
Fig.~\ref{fig:NN1gradio} shows the strain limitation induced by seismic and acoustic NN on a single gradiometer of baseline $L=16.3$~km. The strain projections are derived from Eq.~(\ref{eq:Sh}), using the NN acceleration PSDs of Eq.~(\ref{eq:SDaSeismic}) and Eq.~(\ref{eq:SDaAtmo}), and the relation: 
\begin{equation}\label{eq:Sdeltax}
S_{NN_1}(\omega)=\frac{S_{\Delta a_x}^{R,I}(\omega)}{\omega^4} \, ,
\end{equation}
We observe that both contributions induce a strong limitation for the detection of GWs in the band $0.1$ -- $1$\,Hz where it stands well above the level of $10^{-21}/\sqrt{Hz}$. In the following section, we show how the geometry of the ELGAR detector can help reducing the NN impact by averaging over an array of single gradiometers.

As a final note, it should be expected that other ambient fields produce significant NN including seismic body waves \cite{CoEA2018b} and advected, atmospheric temperature fields \cite{Fiorucci2018}. While seismic body waves are typically weaker than Rayleigh waves and therefore produce a minor contribution to NN, they can interfere significantly with any of the mitigation methods discussed in the following. Atmospheric temperature fields can be considered stationary perturbations, but they convert into fluctuations at a fixed point when this field is transported by wind, which leads to NN. 

\subsection{Mitigation of Newtonian noise using an atom-interferometer array}
\begin{figure}[htbp]
  \centering
	\includegraphics[width=0.8\linewidth]{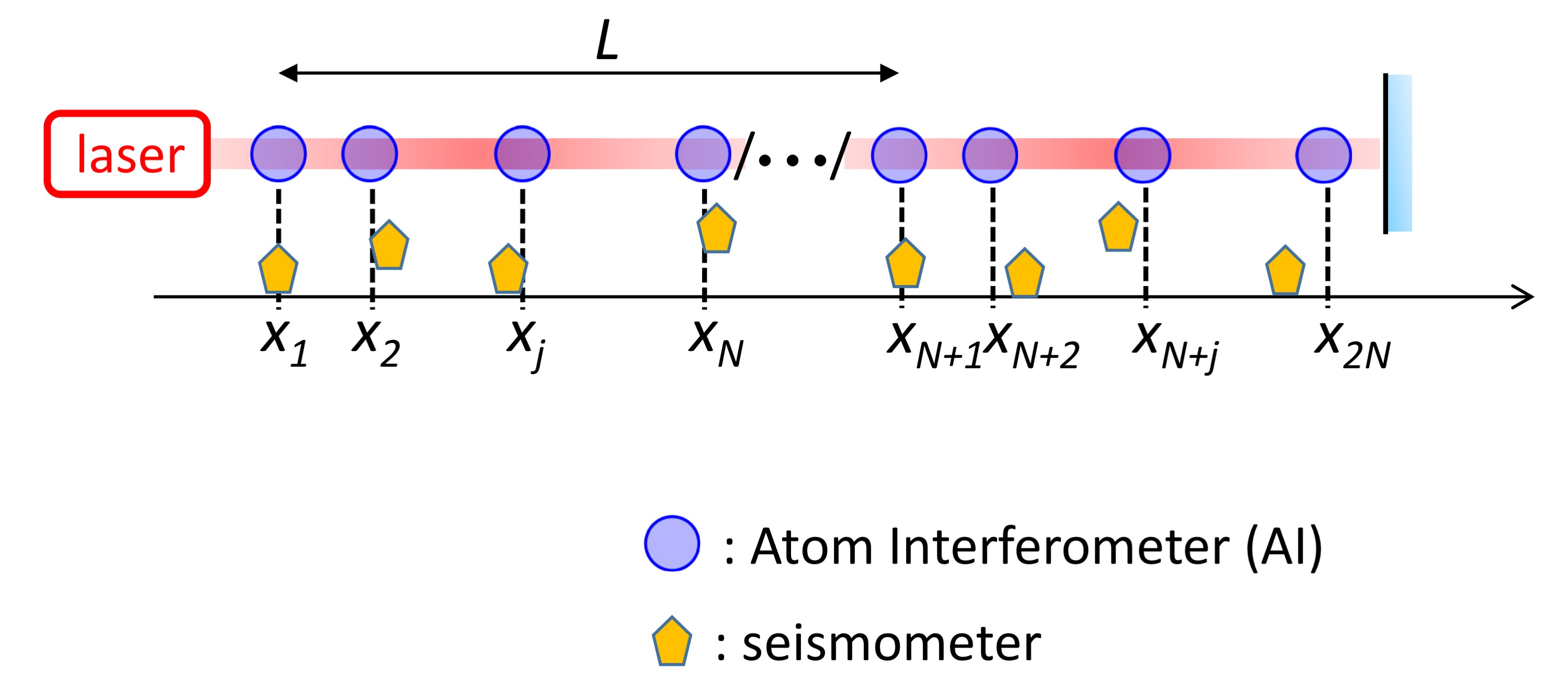}
  \caption{Arm geometry of the ELGAR detector. Each arm composed by N gradiometers regularly spaced over the total detector baseline.}
  \label{AI_network}
\end{figure}

As shown in Sec.~\ref{sec_Detector_configuration}, we propose for ELGAR a geometry using a 2D array of atom interferometers where each arm of the detector is composed by $N$ gradiometers regularly spaced over the total baseline. We now explain how the average signal of the different gradiometers enables to reduce the influence of Newtonian noise. For sake of clarity, we will limit this discussion to the averaging of a single arm of the detector, shown in see Fig.~\ref{AI_network}; the two-arm case being mostly similar.
In the following, we assume that the fields are stationary, homogeneous, and leading to Gaussian fluctuations. This means that we can describe the relevant properties of the field and its impact on the GW detector in terms of two-point correlations. Stationarity and homogeneity of the field $\zeta(\vec x,t)$ generating NN imply
\begin{equation}
\label{Correlation}
\mathcal{C}\left(\tau\right) = \left\langle\zeta\left(t\right)\zeta\left(t+\tau\right)\right\rangle\hspace{0.5cm}\textrm{and}\hspace{0.5cm}\Gamma\left(\vec{\xi}\,\right) = \left\langle\zeta\left(\vec{X}\right)\zeta\left(\vec{X}+\vec{\xi}\,\right)\right\rangle
\end{equation}
Averaging the signals from different gradiometers suppresses the contribution of NN at the same time maintaining the GW signal, because GWs and NN have different spatial signatures: the GW is a plane wave and its effect on a gradiometer is uniform on the earth scale whereas NN has characteristic scale of the order of the wavelength $\lambda$ of the considered source (seismic waves of different types, infrasound). 

The network configuration of $N$ aligned AIs, shown in Fig.~\ref{AI_network}, offers the possibility to repeat the GW measurement with different NN contributions without adding any significant noise in the considered frequency range. We consider the average signal:
\begin{equation}
H_N^X\left(t\right)=\frac{1}{N}\sum^{N}_{i=1} \psi(X_i,X_{N+i},t)\, ,
\end{equation}
where $\psi(X_i,X_{N+i},t)$ is the differential signal of the AIs placed at positions $X_i$ and $X_{N+i}$. We have:
\begin{equation}
H_N^X\left(t\right)=2n k_l\int^{\infty}_{-\infty}\left(\frac{Lh(\tau)}{2}+NN^X(\tau)\right)g'(\tau-t) d\tau\, ,
\end{equation}
where $g'$ is the time derivative of the AI sensitivity function ~\cite{Cheinet08}, and: 
\begin{equation}
NN^{X}(\tau)=\frac{1}{N}\sum^{N}_{i=1}
\Big[\delta x_{at}(X_j,\tau)-\delta x_{at}(X_i,\tau)\Big]\, .
\end{equation}
In this last equation, $\delta x_{at}$ is the displacement of the atom test mass with respect to the interrogation laser.

The average signal $H_N^X$ represents an approximation of the GW signal $Lh/2$. The residual NN on this measurement procedure is given by:
\begin{equation}
\sqrt{S_{H_N^X}\left(\omega\right)}=2n k_l|\omega G(\omega)|\sqrt{S_{NN^X}\left(\omega\right)}
\end{equation}
According to Monte-Carlo theory~\cite{Caflisch1998}, we have:
\begin{equation}
\sqrt{S_{NN^X}\left(\omega\right)}=\frac{K\left(\omega\right)}{N^{1/2}}\sqrt{S_{NN_1}\left(\omega\right)}
\end{equation}where $K\left(\omega\right)$ is the Monte-Carlo variance reduction factor that drives the NN reduction gain $K\left(\omega\right)/N^{1/2}$ of the network with respect to the one of a single gradiometer. This factor, depending on the Fourier frequency through the residual correlation of NN between the considered AIs locations, ranges from 0 to $N^{1/2}$ :         
\begin{itemize}
\item when successive AIs are uncorrelated: $K = 1$, the NN is reduced by a gain $1/N^{1/2}$
\item when successive AIs are correlated: $K>1$, the gain diminishes till the minimal value of 1, reached when correlations are maximum: $K = N^{1/2}$. 
\item when successive AIs are anti-correlated, $K<1$, the NN is reduced by more than $1/N^{1/2}$, which is the best configuration.
\end{itemize}
The Newtonian force is a short range interaction. Hence, the spatial correlation of gravitational acceleration in a given direction simply reproduces spatial correlation of its source. NN is related to the fluctuations of mass density surrounding the AIs and if the field of density fluctuations is homogeneous, and isotropic, correlation between two points only depends on the separating distance $d$. In homogeneous media, this correlation roughly evolves as follows~\cite{Mykkeltveit1983} : (i) $\Gamma\left(d<\lambda/2\right)>0$, (ii) $\Gamma\left(\lambda/2<d<\lambda\right)<0$ and (iii) $\Gamma\left(d>\lambda\right)\simeq 0$ and so does NN. Since $\lambda$ depends on $\omega$, one needs to optimize the distribution of AI along the laser optical axis in order to reduce the $K$ factor on the frequency band of interest. While highest anti-correlations observed in natural seismic fields impose a limit $K(\omega)\gtrsim 0.5$, interesting mitigation factors of NN can be reached using adequate configuration by noise averaging as shown in Ref.~\cite{Chaibi2016}. 

The case of an inhomogeneous medium is much more difficult, because the spatial correlation function depends on the considered points~\cite{Braun2008}. Hence, NN correlation between AIs is \textit{a priori} unknown and can only be accurately determined once the detector is settled and running. As for Wiener filters applied for NN mitigation~\cite{Harms2019}, estimation of correlation functions requires a wide distribution of sensors which is difficult to conduct for an underground detector at sub-Hz frequency. Therefore, the distribution of AIs does not offer any degree of freedom to compensate the inhomogeneity of the spatial correlation function. For example, positive correlation between AIs might dominate and the factor $K$ remains above 1 for the whole frequency band of interest. Intuitively, one can think of applying weights when combining gradiometer signals,
\begin{equation}
\label{average_generalized}
H_N^X\left(t\right) = \sum^{N}_{i=1}\alpha_i\psi(X_i,X_{N+i},t)\hspace{0.5cm}\textrm{with}\hspace{0.5cm}\sum^{N}_{i=1}\alpha_i=1,
\end{equation}
in order to promote the effect of anti-correlations. The factors $\alpha_i$ are determined using Wiener filters except that the corresponding optimization procedure requires Lagrange multiplier in order to fulfill the condition $\sum \alpha_i = 1$. Further simulations are required to prove the efficiency of this procedure and determine its limitation.

\subsection{Coherent cancellation of Newtonian noise}
It was proposed that NN can be suppressed by coherent cancellation using environmental sensors like seismometers and microphones~\cite{HuTh1998}. One possibility is to calculate a Wiener filter based on correlations between environmental sensors and ELGAR data~\cite{Cel2000}. The Wiener filter is a linear filter that takes data from all environmental sensors as input, and it outputs an estimate of the ELGAR instrument noise that is correlated with the environmental data. This output is subtracted from ELGAR data ideally providing a time series without significant NN. However, there are practical limitations of the cancellation performance including numerical and statistical limits, incomplete information content in environmental data, and limited sensitivity of the environmental sensors~\cite{BaHa2019}. It might be possible to overcome some of these limitations with more sophisticated filtering techniques, e.g., based on Kalman filters or machine-learning algorithms, which can deal in a straight-forward way with non-stationary data or reducing statistical errors. 

Newtonian-noise cancellation may already help with geophysical observations~\cite{BaCr1999,JuEA2018a}, and a few orders of magnitude NN reduction is required to sufficiently reduce NN for GW observations~\cite{HaEA2013,FiEA2018}. This challenge can only be partially ameliorated when combined with other mitigation approaches. Newtonian-noise cancellation is therefore an essential technology for the realization of ELGAR as GW detector. 

The cancellation of NN below 10\,Hz from seismic fields was recently investigated for the Einstein Telescope (ET)~\cite{BaHa2019}, which is the European concept of a large-scale underground laser-interferometric GW detector, and for a sub-Hertz GW detector concept~\cite{HaPa2015}. One important aspect of these studies is that they focus on different types of seismic waves. The most relevant types of waves are the body waves (shear and compressional), which can propagate along all directions inside Earth, and Rayleigh waves, which are evanescent surface waves~\cite{AkRi2009}. While Rayleigh waves typically dominate surface displacement since they are produced efficiently by surface sources including human activities and weather phenomena, body waves can still make important contributions to the seismic displacement at underground sites~\cite{OlEA2015} and even to surface displacement~\cite{CoEA2018b}. A NN cancellation system needs to address contributions from all relevant wave types.

One of the major challenges of NN cancellation in ELGAR is the configuration of the environmental monitoring system. This comprises the choice of sensors and the array configuration. Seismic sensors can be simple broadband seismometers common in earthquake monitoring, but it can also be advantageous to consider other types of sensors such as tiltmeters~\cite{HaVe2016}. The optimal choice of sensors is still an open problem for ELGAR, but it seems possible in principle to provide sufficient reduction of seismic NN albeit at substantial cost for the installation of (likely) hundreds of sensors at the surface and inside deep boreholes (compare with~\cite{BaHa2019}). 

Far more challenging is the monitoring of atmospheric temperature, humidity, and pressure fields, which all give rise to NN~\cite{Cre2008,Har2015}. Microphones can in principle be used to monitor the pressure fluctuations, but data quality is severely degraded by wind noise, i.e., local pressure fluctuations from wind-driven turbulence around the microphones~\cite{WaHe2009}. Microphones have the additional disadvantage that they cannot be easily deployed at greater heights to form 3D arrays. It seems that the only reliable future option of atmospheric monitoring is through LIDAR; a laser-based observation of atmospheric disturbances. Current systems have enough sensitivity to monitor in a scanning fashion temperature fields~\cite{HaEA2015b}, humidity fields~\cite{SpEA2016}, and velocity fields~\cite{CLN2004}. However, the smallest perturbations of atmospheric density associated with pressure fluctuations, e.g., sound waves, cannot be resolved with LIDAR yet. It is expected though that acoustic NN needs to be substantially reduced by NN cancellation for ELGAR~\cite{FiEA2018}. This means that technological developments of LIDAR are still required, or other options for atmospheric sound monitoring are being developed that do not suffer from significant wind noise.

\subsection{Compatibility of mitigation methods}

Mitigation of NN on a single reference mass using an array of sensors can be very effective: each sensor measures its surrounding density fluctuation whose contribution to the mass gravitational motion is estimated using a Wiener filter~\cite{Cel2000}. The efficiency grows rapidly with the number of sensors $N_s$ and is above all limited by the sensor noise, ability of an array configuration to extract information about the field, and statistical estimation errors of the Wiener filter~\cite{Harms2019}. The diameter and density of the array determine the frequency band where noise suppression is effective.

Having an underground detector offers the advantages to significantly reduce atmospheric NN. However, NN from seismic body waves remains important at any depth, and also Rayleigh-wave NN is only suppressed to some extent underground. In this case, a three dimensional distribution of the sensors is required~\cite{BaHa2019}, which is technically difficult to achieve at low frequency ($<1\,\textrm{Hz}$) with a necessary array diameter of about $0.3\lambda$, i.e., more than a kilometer.

Nevertheless, a Wiener filter based on sensor arrays can be combined with the AIs averaging procedure. This process is shown in Fig.~\ref{AI_network} and requires the following steps:
\begin{itemize}
\item distribution of the sensors along the optical axis.
\item estimation of NN gravitational effect for each AI using a Wiener filter based on surrounding sensors.
\item use of an averaging procedure on the Wiener filter residuals.               
\end{itemize}
In the general case, the detector output is inferred from Eq.~\eqref{average_generalized}:
\begin{equation}
\label{Combination_Wiener_average}
H_N^X= \sum^{N}_{i=1}\alpha_i\left(\psi(X_i,X_{N+i})-\sum^{N_s}_{j=1}\sum^{3}_{k=1}\beta_{i,j,k}\,\left[\eta_{j+N_s,k}-\eta_{j,k}\right]\right)\hspace{0.5cm}\textrm{with}\hspace{0.5cm}\sum^{N}_{i=1}\alpha_i=1
\end{equation}where $\eta_{j,k}$ designates the $j$th sensor output in the direction $k$. $\beta_{i,j,k}$ are Wiener filter coefficients optimized to reduce the NN residual. 

The last equation can be approached in a simple case : (i) $N=N_s$ and to each AI corresponds a single sensor; (ii) all AIs are uncorrelated and so are the sensors: then $\alpha_i=1/N$; (iii) the media is homogeneous and all sensors are placed at the same position with respect to the AIs; (iv) only one sensor output is relevant, for example $k=1$. Hence, all AIs are equivalent and so are the sensors, i.e.  and $\beta_{i,j,k} = \delta_{i,j}\delta_{k,1} \beta$. Eq.~\eqref{Combination_Wiener_average} becomes:
\begin{equation}
\label{Combination_Wiener_average_uncorrelated}
H_N^X= \frac{1}{N}\sum^{N}_{i=1}\left(\psi(X_i,X_{N+i})-\beta\left[\eta_{i+N,1}-\eta_{i,1}\right]\right)
\end{equation}
In this case, the optimization procedure to reduce the NN contribution, i.e. $S_{NN}^X\left(\omega\right)$ at a given frequency is straightforward and gives :
\begin{equation}
\beta\left(\omega\right) = 2n k_l|\omega G(\omega)|\frac{\left\langle\delta x_{at}(X_i)\times\eta_{i,1} \right\rangle_\omega}{S_\eta\left(\omega\right)}\hspace{0.5cm}
\end{equation}for which
\begin{equation}
S_{NN}^X\left(\omega\right)=\frac{S_{NN1}\left(\omega\right)}{N}\left(1-\frac{\left\langle \delta x_{at}(X_i)\times\eta_{i,1} \right\rangle_\omega^2}{S_{\eta}\left(\omega\right)S_{\delta x_{at}(X_i)}\left(\omega\right)}\right)
\end{equation}
Hence, Eq.~\eqref{Combination_Wiener_average_uncorrelated} represents an observable, which combines the average procedure and the Wiener filter at the same time with a combined mitigation efficiency. For a correlation coefficient $\left\langle \delta x_{at}(X_i)\times\eta_{i,1} \right\rangle_\omega/\sqrt{S_{\eta}\left(\omega\right)S_{\delta x_{at}(X_i)}\left(\omega\right)}\simeq 0.5$, the NN is only reduced by $13\%$. This factor should be improved when several sensors are correlated with a single AI, also their distribution geometry is not optimal. The general case described by Eq.~\eqref{Combination_Wiener_average} gives a mitigation of the form:
\begin{equation}
\sqrt{S_{NN}^X\left(\omega\right)}=K\left(\omega\right)\sqrt{R\left(\omega\right)}\frac{\sqrt{S_{NN1}\left(\omega\right)}}{N^{1/2}}
\end{equation}
where $R\left(\omega\right)$ represents the residual of the Wiener filter and $K\left(\omega\right)$ represents the Monte-Carlo variance reduction factor. Sensor positions, their number $N_s$, $\alpha$ and $\beta$ factors are to be optimized in order to reduce the product $K\left(\omega\right)\sqrt{R\left(\omega\right)}$. A complete set of simulations are required to confirm this combination of NN mitigation methods and gives a realistic order of magnitude for its efficiency.

\subsection{Site evaluation criteria with respect to NN}

Based on the analysis described above, its appears that the following properties of the site are crucial to ensure the efficiency of NN mitigation :
\begin{itemize}
\item Reducing the correlation length $\lambda$ offers the possibility to increase the number $N$ of AIs for a given antenna length. For an 
underground detector for which air contribution to the NN is naturally reduced, this concerns a low sound propagation velocity. In practice, this is the case for media of low density which in this same time has high seismic noise. Seismic NN being given by the product of the density by the seismic noise, it clearly appears that site choice must balance between the sound velocity and the seismic noise.
\item Inhomogeneities within the site medium increases the probability to have pure positive correlations between AIs and $K<1$ might not be reached for any frequency $f$. Therefore, homogeneity must also be considered as a crucial factor for the detector site choice.
\end{itemize} 

\section{Noise couplings}
\label{sec_other_noise_couplings}

 Eq.~\eqref{eq:SHnetwork} in Sec.~\ref{sec_Detector_configuration} gives the link between the strain sensitivity of the whole detector and the  phase noise level of a single atom interferometer forming the array as:
\begin{equation}\label{sensitivity_link}
\sqrt{S_h(\omega)}=\frac{1}{\sqrt{N}}\times\frac{1}{2nk_{l}L|\omega G(\omega)|}\times2\sqrt{S_{\epsilon}(\omega)},
%\sqrt{S_h(\omega)}=\frac{2\sqrt{S_{\epsilon}(\omega)} }{\sqrt{N}2nk_lL|\omega G(\omega)|}\,
\end{equation}
The target floor sensitivity level of $3.3\times 10^{-22}/\sqrt{\rm Hz}$ for the array of $2N$ uncorrelated gradiometers corresponds to the strain sensitivity for a single gradiometer of $4.1\times 10^{-21}/\sqrt{\rm Hz}$ which is reached  at the peak frequency of 1.7~Hz. This, in turn, sets the differential atom phase sensitivity limit of $\sqrt{2S_{\epsilon}^{\rm{lim}}}= \sqrt{2}\times 10^{-6}~\rm{rad}/\sqrt{\rm{Hz}}$, with $\sqrt{S_{\epsilon}^{\rm{lim}}}$ being the atom shot noise limited phase sensitivity of a single interferometer. In the following, we will restrict the analysis to the case of a single  interferometer and assume uncorrelated phase noise between the interferometers forming the array. We set a conservative requirement for each of the spurious  phase noise contributions to not exceed the level of $0.1~\mu\rm{rad}/\sqrt{\rm{Hz}}$ at 1.7~Hz for a single atom interferometer.

\subsection{Couplings of rotations, gravity gradients, beam misalignment and beam pointing jitter to the initial position and velocity and to the gravitational acceleration.}

The phase-shift contributions due to rotations and gravity gradients in light-pulse atom interferometry have been calculated in Refs.~\cite{Audretsch1994,Borde2001,Antoine2003,Bongs2006,Hogan2009,Roura2014,Kleinert2015,Bertoldi2019}.
Here we will make use of the convenient theoretical tools developed in Refs.~\cite{Roura2014,Roura2020} to derive the relevant terms for the interferometry scheme introduced in Sec.~\ref{sec_double_loop}.
We will first focus on those contributions that depend on initial central position and velocity of the atomic cloud and then consider the couplings to the gravitational acceleration $g$.
Cross-couplings between the different contributions considered here are typically further suppressed and will not be explicitly discussed.

\subsubsection{Rotations.}\label{sec_coupling_rotation}

The proposed atom-interferometer geometry can successfully suppress phase-shift contributions of order $\big( \Omega\, T \big)$ that depend on the initial velocity of the atomic cloud. However, some dependence on the initial conditions remains at higher orders in $\big( \Omega\, T \big)$. The corresponding contributions to atomic phase shift are given by
\begin{align}
\Delta \phi_\Omega &= -\,6\, \mathbf{k}_\text{eff}^\perp \cdot \mathbf{v}_0\, T \, \big( \Omega\, T \big)^2
+ 4 \Big( \big( \bold{\Omega}\, T \big) \times \mathbf{k}_\text{eff} \Big) \cdot \mathbf{v}_0\, T \, \big( \Omega\, T \big)^2 \nonumber \\
&\quad + 2 \Big( \big( \bold{\Omega}\, T \big) \times \,\mathbf{k}_\text{eff} \Big) \cdot \mathbf{x}_0 \, \big( \Omega\, T \big)^2
\,+\, O \Big( \big( \Omega\, T \big)^4 \Big) ,
%\\
%&\quad + 2 \Big( \big( \bold{\Omega}\, T \big) \times \,\mathbf{k}_\text{eff} \Big) \cdot \mathbf{g} \, T^2
%\,+\, O \Big( \big( \Omega\, T \big)^2 \Big)
\end{align}
where $\mathbf{k}_\text{eff}^\perp \,=\, - \, \hat{\bold{\Omega}} \times \big( \hat{\bold{\Omega}} \times \mathbf{k}_\text{eff} \big)$ is the component of $\mathbf{k}_\text{eff}$ perpendicular to the direction of Earth's angular velocity, characterized by the unit vector $\hat{\bold{\Omega}}$. The dominant contribution comes from the first term, which gives the maximum allowed amount of initial-velocity jitter. For the interferometer configuration under  consideration, %in particular $L = 10\, \text{km}$
and given Earth's angular velocity $\Omega \approx 73 \, \mu\text{rad/s}$, an initial-velocity stability of $\sqrt{S_{v_0}} \lesssim 0.05 \, \mu \text{m/s} / \sqrt{\text{Hz}}$ is required for each velocity component in order to reach the targeted sensitivity.

\subsubsection{Relative beam misalignment.}\label{sec_coupling_beam_misalignment}

A small misalignment between the two interferometer beams also leads to the following velocity-dependent phase-shift contribution \cite{Savoie2018,Altorio2020}:
\begin{align}
\Delta \phi_\theta \, &= \, 4\, k_\text{eff}\,\Delta\hat{\mathbf{n}} \cdot \mathbf{v}_0\, T
= \, 4\, k_\text{eff}\,\big( \sin \delta\theta \ v_0^{y'} T + (\cos \delta\theta -1)  \ v_0^{x}\, T \big)
%+ \ldots
\nonumber \\
&\approx 4\, k_\text{eff}\,\big( \delta\theta_y\, v_0^{y} + \delta\theta_z\, v_0^{z} \big) \, T ,
\label{eq:misalignment}
\end{align}
where $v_0^{y'}$ in the second equality is the initial velocity component along the $\Delta\hat{\mathbf{n}}$ direction projected onto the $y-z$ plane
and in the last equality we have kept just the lowest-order terms corresponding to $\Delta\hat{\mathbf{n}} \approx  \big( 0, \delta\theta_y, \delta\theta_z \big)$.
Assuming the stability requirement for the initial velocity obtained in the previous paragraph,
the maximum allowed relative beam misalignment is  $\delta\theta \lesssim 0.32 \, \text{nrad}$.

\subsubsection{Beam pointing jitter.}\label{sec_coupling_beam_jitter}
Small fluctuations in the direction $\hat{\mathbf{n}}_j$ of each laser pulse (with $j=1,\ldots,4$) can be characterized by $\delta\hat{\mathbf{n}}_j \approx \big( 0,\boldsymbol{\xi}^\perp_j \big) = \big( 0,\xi^y_j,\xi^z_j \big)$, where $\boldsymbol{\xi}^\perp_j$ are stochastic processes. For \emph{low-frequency} pointing jitter with $\omega\, T \ll 1$ the directions of the two pulses associated with each beam are correlated, so that $\boldsymbol{\xi}^\perp_1 = \boldsymbol{\xi}^\perp_4$ and  $\boldsymbol{\xi}^\perp_2 = \boldsymbol{\xi}^\perp_3$. This reduces then to shot-to-shot fluctuations of the relative beam misalignment considered in Sec.~\ref{sec_coupling_beam_misalignment}, but with $\Delta\hat{\mathbf{n}} \approx \boldsymbol{\xi}^\perp_2 - \boldsymbol{\xi}^\perp_1$, and leads to the following coupling between $\boldsymbol{\xi}^\perp_j$ and the mean initial velocity $\langle \mathbf{v}_0 \rangle$ that would result from averaging over many shots:
\begin{equation}
\Delta \phi_\xi \,\approx\, 4\, k_\text{eff}\, \big( \boldsymbol{\xi}^\perp_2 - \boldsymbol{\xi}^\perp_1 \big) \cdot \langle \mathbf{v}_0 \rangle\, T
\label{eq:pointing_jitter1} .
\end{equation}

On the other hand, for \emph{higher-frequency} pointing jitter with $\omega\, T \gtrsim 1$ the above correlations no longer hold and one gets the following coupling with the mean initial position and velocity:
\begin{equation}
\Delta \phi_\xi \,\approx\, \sum_{j=1}^{4} \varepsilon_j \, k_\text{eff}\, \boldsymbol{\xi}^\perp_j \cdot
\Big(  \langle \mathbf{x}_0 \rangle + \langle \mathbf{v}_0 \rangle\, t_j \Big)
\label{eq:pointing_jitter2} ,
\end{equation}
where $t_j$ is the time at which the $j$-th pulse is applied and we have introduced $\varepsilon_1 = -\varepsilon_4 = 1$ and $-\varepsilon_2 = \varepsilon_3 = 2$. From Eq.~(\ref{eq:pointing_jitter2}) one can derive bounds on the amplitude of the power spectral density for beam pointing jitter depending on the values of the mean initial position and velocity. However, since it is important to take into account that beam pointing jitter also couples to the local gravitational acceleration and to consider their combined effect, we defer this discussion to Sec.~\ref{sec_coupling_g}, where the stability requirements for a differential measurement compatible with the targeted sensitivity will be obtained.

\subsubsection{Gravity gradients.}\label{sec_coupling_grav_grad}
The phase-shift for the double-loop interferometer geometry proposed in Sec.~\ref{sec_double_loop} is free from any coupling of the gravity gradient to the initial position, but not to the initial velocity, which contributes as follows:
\begin{align}
\Delta \phi_\Gamma = - 2\, \mathbf{k}_\text{eff}^\text{T}\, \big( \Gamma\, T^2 \big) \, \mathbf{v}_0 T
\,+\, O \Big( \big( \Gamma\, T^2 \big)^2 \Big), 
\end{align}
where the transposed vector $\mathbf{k}_\text{eff}^\text{T}$ is introduced to underline the vector-matrix notation conventionally employed in the literature. Given Earth's gravity gradient in the horizontal direction, $\Gamma_{xx} = \Gamma_{yy} \approx 1.5 \times 10^{-6}\, \text{s}^{-2}$, this would lead to more stringent (by several orders of magnitude) requirements on the initial velocity, than those obtained above as a consequence of Earth's rotation.

Fortunately, in order to substantially relax these requirements, one can make use of the compensation technique proposed in Ref.~\cite{Roura2017PRL} and involving in this case a single-photon frequency change $\Delta\nu \sim 12\, \text{MHz}$ for the two intermediate pulses. Indeed, taking the stability requirement for the initial-velocity jitter that follows from the effect of rotation, one finds that the gravity gradient needs to be compensated at  $1 \%$ level. Compensation at this level has already been demonstrated experimentally in both gradiometry measurements \cite{d_Amico2017} and tests of the universality of free fall \cite{Overstreet2018}.

It is, however, important that the gravity gradient is the same (within that level of accuracy) for every atom interferometer on the baseline so that the sensitivity to their independent initial velocities can be simultaneously minimized \cite{Roura2018a}. If necessary, this can be achieved by making small adjustments in the local mass distribution. Furthermore, in the whole process (including the determination of the appropriate value of $\Delta\nu$) the compensation technique itself can be employed for self-calibration \cite{Overstreet2018}.

\subsubsection{Couplings to the local gravitational acceleration $g$.}\label{sec_coupling_g}

Time variations of the local gravitational acceleration $\mathbf{g}$ at the positions of each interferometer also couple to rotation, gravity gradient, relative beam misalignment and beam pointing jitter.
The corresponding phase-shift contributions are given by
\begin{align}
\Delta \phi_g \,&=\, 2 \Big( \big( \bold{\Omega}\, T \big) \times \,\mathbf{k}_\text{eff} \Big) \cdot \mathbf{g} \, T^2
\,+\, 8\, k_\text{eff}\,\Delta\hat{\mathbf{n}} \cdot \mathbf{g}\, T^2 
+ \frac{1}{2} \sum_{j=1}^{4} \varepsilon_j \, k_\text{eff}\, \boldsymbol{\xi}^\perp_j \cdot \mathbf{g}\, t_j^2
\nonumber \\
& \quad\quad 
- 4\, \mathbf{k}_\text{eff}^\text{T}\, \big( \Gamma\, T^2 \big) \, \mathbf{g}\, T^2
+\, O \Big( \big( \Omega\, T \big)^2 \Big) \,+\, O \Big( \big( \Gamma\, T^2 \big)^2 \Big)
\label{eq:grav_coupling}.
\end{align}
Here we have neglected any time dependence of $\mathbf{g}$ within a single shot for simplicity. This can be taken into account, but merely gives rise to lengthier expressions without substantially %qualitatively
altering the main conclusions. Below we discuss the stability requirements implied by each term.

(\emph{i}) Shot-to-shot variations of $\mathbf{g}$ with characteristic frequencies comparable to the GW frequencies of interest can give rise to spurious signals. However, such variations are directly related to the Newtonian noise discussed in Sec.~\ref{sec_NN_reduction}. Thus, any methods employed to mitigate the effects of Newtonian noise would also help in this case.
Moreover, assuming that Earth's \emph{angular velocity} is sufficiently stable, the first term is suppressed by a factor $\big( \Omega\, T \big) \sim 10^{-5}$ compared to standard Newtonian noise.

(\emph{ii}) On the other hand, shot-to-shot variations of the \emph{relative beam misalignment} also couple to the static part of $\mathbf{g}$.
Taking into account that a similar coupling to the mean initial velocity $\langle \mathbf{v}_0 \rangle$ appears in Eq.~(\ref{eq:pointing_jitter1}), both contributions can be combined to give $4\, k_\text{eff}\,\Delta\hat{\mathbf{n}} \cdot \!\big( 2\, \mathbf{g}\, T + \langle \mathbf{v}_0 \rangle \big)\, T$
with $\Delta\hat{\mathbf{n}} \approx \boldsymbol{\xi}^\perp_2 - \boldsymbol{\xi}^\perp_1$.
%$4\, k_\text{eff}\,\Delta\hat{\mathbf{n}} \cdot \!\big( \langle \mathbf{v}_0 \rangle\, T + 2\, \mathbf{g}\, T^2 \big)$.
Therefore, since it is dominated by the directions orthogonal to the interferometer beam, the net contribution can be minimized by choosing $\langle \mathbf{v}_0^\perp \rangle = - 2\, \mathbf{g}^\perp T$ for those directions.
In practice this optimal value of the initial velocity can be determined by varying $\Delta\hat{\mathbf{n}}$ on purpose by a substantial amount and finding the value of $\langle \mathbf{v}_0^\perp \rangle$ that minimizes the dependence on $\Delta\hat{\mathbf{n}}$ \cite{Savoie2018,Altorio2020}.
If one could achieve this at the p.p.m.\ level, so that $\big| 2\, \mathbf{g}^\perp T + \langle \mathbf{v}_0^\perp \rangle \big| \lesssim 10^{-6} g\, T$,
a beam-alignment stability of $\sqrt{S_{\delta\theta}} \lesssim 4\, \text{prad} / \sqrt{\text{Hz}}$ would be required in order to reach the target sensitivity.

(\emph{iii}) In contrast, for higher-frequency \emph{beam pointing jitter} (with $\omega\, T \gtrsim 1$) the cancellation between the contributions from Eqs.~(\ref{eq:pointing_jitter1}) and (\ref{eq:grav_coupling}) no longer holds because the coefficients will be different in general and one needs to consider both contributions separately.
As far as the contribution involving the static part of $\mathbf{g}$ is concerned, the relevant quantity for the differential measurement of two interferometers $A$ and $B$ is proportional to $k_\text{eff}\,\,\boldsymbol{\xi}^\perp_j  \!\cdot \!\big( \mathbf{g}_A - \mathbf{g}_B \big) \, T^2$. Hence, assuming $\big| \mathbf{g}^\perp_A - \mathbf{g}^\perp_B \big| \lesssim 10^{-5} g$, which is compatible with a 30-km total baseline%
%between the two interferometers%
\footnote{The difference due to Earth's curvature and projected onto the transverse plane will typically be of this order for baselines of $30\,\text{km}$.}, one concludes that a beam-alignment stability of $\sqrt{S_{\boldsymbol{\xi}^\perp_j}} \lesssim 0.3\, \text{prad} / \sqrt{\text{Hz}}$ is required at those frequencies.
This conservative estimate is obtained by considering the dominant contribution to the third term on the right-hand side of Eq.~(\ref{eq:grav_coupling}), which corresponds to that from the third pulse ($j=3$).
Similar conclusions are reached for the contribution involving the mean initial velocity if one assumes $\big| \langle \mathbf{v}_0^\perp \rangle_A - \langle \mathbf{v}_0^\perp \rangle_B \big| \lesssim 10^{-5} g\, T$. This could be accomplished by linking $\langle \mathbf{v}_0^\perp \rangle$ with $\mathbf{g}^\perp T$ independently for each interferometer as described in the previous paragraph.

In addition, there is a third contribution involving the mean initial position transverse to the interferometry beam. If one wants to have comparable requirements on the beam pointing stability, the condition $\big| \langle \mathbf{x}_0^\perp \rangle_A - \langle \mathbf{x}_0^\perp \rangle_B \big| \lesssim 10^{-5} g\, T^2$ needs to be fulfilled. In this case, however, one will need to vary the relative alignment of the two laser baselines for different laser pulses within a single shot %(i.e.\ a single atom interferometry realization)
in order to calibrate that.

(\emph{iv}) Finally, for the static part of the \emph{gravity gradient} similar considerations to those made above for Earth's angular velocity apply. In this case there is an additional suppression factor $\big( \Gamma\, T^2 \big) \sim 6 \times 10^{-8}$ with respect to standard Newtonian noise. Furthermore, although time variations of $\Gamma$ can couple to the static gravitational acceleration $\mathbf{g}$, preliminary estimates suggest that this contribution is smaller than the regular Newtonian noise and one expects that it can be tackled through the same means as the latter.

\subsection{Magnetic fields, electric fields, blackbody radiation}\label{sec_mag_elect_temp}
In this section we estimate the limiting noise levels on various background fields (magnetic field, black body radiation and electric field) that produce a local acceleration noise $a_x$ along the direction of the Bragg beams at the position of the atoms in each AI forming the array of the antenna. An acceleration noise of PSD $S_{a_{x}}(\omega)$ translates into an interferometer phase noise given by:
\begin{equation}\label{eq:anoise}
    \sqrt{S_{\phi}(\omega)}=2nk_{l}\frac{|\omega G(\omega)|}{\omega^2}\sqrt{S_{a_{x}}(\omega)}\,
\end{equation}
We assume uncorrelated acceleration noise between the atom interferometers forming the antenna.
In this section, we estimate a requirement on the acceleration noise level  at the peak frequency of 1.7~Hz in order to ensure an  associated phase noise contribution  below $0.1~\mu\rm{rad}/\sqrt{Hz}$. When going to lower frequencies, the contribution to phase noise of a white acceleration noise reduces due to the factor $H_a(\omega)=\frac{|\omega G(\omega)|}{\omega^2}$ that diminishes when going to lower frequencies. 
%In order to ensure that the associated phase noise contribution is below $0.1~\mu\rm{rad}/\sqrt{Hz}$ at 0.1~Hz, the level of acceleration noise at 0.1~Hz has to be divided by a factor of $\frac{H_a(2\pi\times 0.1)}{H_a(2\pi\times 1.7)}\simeq 4.6$ compared to that at 1.7~Hz.  
%Assuming uncorrelated acceleration noise of the two AIs forming a gradiometer we recover the connection between strain sensitivity for a single gradiometer and the PSD of the acceleration noise as $\sqrt{S_{h}(\omega)} = \frac{2}{L\omega^2} \sqrt{2S_{a_{x}}(\omega)}$.

\subsubsection{Second order Zeeman force.}
We consider rubidium atoms prepared in the first order Zeeman-insensitive ground state ($F=1$,  $m_{F} = 0$). In the presence of a fluctuating, non-homogeneous linearized magnetic field $B(x,t)=B_{0}(t) + \beta x$, the second order Zeeman shift induces an acceleration $ a_{x}(t)=\frac{2\alpha B_{0}(t) \beta }{m_{\rm{at}}} $ which results in a sensitivity to magnetic field fluctuations :
\begin{equation}
    S_{a_{x}}(\omega) \simeq\left(\frac{2\alpha \beta }{m_{\rm{at}}}\right)^{2}S_{B_{0}}(\omega)\, ,
\end{equation}
where $\alpha \simeq \frac{h}{2} \cdot 575$ Hz$/$G$^{2}$ is half the second order Zeeman shift on the rubidium clock transition and $m_{\rm{at}}$ is the atomic mass. For a magnetic field gradient of $\beta = 1$ $\mu$G/cm, the magnetic field stability requirement is then $\sqrt{S_{B}(\omega)}\le 1$ $ \mu$G/$\sqrt{\rm{Hz}}$ at $1.7$ Hz.

\subsubsection{Blackbody radiation force.}

Blackbody radiation emitted from the walls of the vacuum chamber induces an AC Stark shift on the atomic levels. If we consider a thin horizontal cylindrical chamber surrounding the interferometers, a temperature gradient along the horizontal axis will induce a gradient of potential energy and thus an acceleration for the atoms of \cite{haslinger_BBR_2018}: 
\begin{equation}
    a_{x}=\frac{\partial}{\partial x} \frac{2 \alpha_{\rm{at}} \sigma T^{4}(x,t)}{m_{\rm{at}} c \epsilon _{0}}\, ,
\end{equation}
where $\alpha_{\rm{at}}=h \times 0.0794$ Hz$/$(V$/$cm)$^2$ is the atom's static polarizability in the ground state, $\sigma$ is the Stefan-Boltzmann constant, T is the temperature of the walls and $\epsilon_{0}$ is the vacuum permittivity. If we consider a small linear temperature gradient $\tau$ and small temperature fluctuations $\delta T$ so that $T(x,t) = T_{0} + \delta T(t)+\tau x$, we can write at first order:
\begin{equation}
    S_{a_{x}}(\omega) \simeq \left(\frac{24\alpha_{\rm{at}} \sigma T^{2}_{0} \tau}{m_{\rm{at}} c \epsilon _{0}}\right)^{2}S_{\delta T}(\omega)\, .
\end{equation}
Using Eq.~\ref{eq:anoise} we can then deduce that for rubidium atoms in a vacuum chamber at $ T_{0}=300$ K with a gradient of $\tau = 0.1$ K/m the temperature stability requirement at $1.7$ Hz is $\sqrt{S_{\delta T}}\le 2 \times 10^{-3} $ K/$\sqrt{\rm{Hz}}$.

\subsubsection{DC Stark force.}
A non homogeneous linearized electric field $E(t)=E_{0}(t)+\epsilon x$ induces an acceleration on atoms in the ground state $a_{x}(t)=\frac{\alpha_{at} E_{0}(t) \epsilon }{2m_{\rm{at}}}$ which results in a sensitivity to electric field fluctuations : 
\begin{equation}
    S_{a_{x}}(\omega)=\left(\frac{\alpha_{at} \epsilon }{2m_{\rm{at}}}\right)^{2}S_{ E_{0}}(\omega)\, .
\end{equation}
For an electric field gradient of $\epsilon = 0.1$ V/m$^{2}$, the electric field stability requirement is then $\sqrt{S_{E}(\omega)}\le 0.2 $ (V/m)/$\sqrt{\rm{Hz}}$ at $1.7$ Hz.

\subsection{Differential wavefront distortions}\label{sec_coupling_wavefront}

Wavefront distortions in light-pulse atom interferometry have been theoretically and experimentally studied in the context of inertial sensors \cite{fils_influence_2005, louchet-chauvet_influence_2011, schkolnik_effect_2015, trimeche_active_2017, karcher_improving_2018}. The sensitivity to wavefront distortions  stems from the coupling of the atomic sample's time-varying spatial extend due to its finite initial size and temperature with the laser beam's non-uniform phase front. This effect has usually been considered as static and is the dominant systematic effect for state-of-the-art atomic gravimeters. In these cases, the baseline of the atom interferometer is  on the order or below \SI{1}{\meter} so changes of the wavefront distortion along the direction of propagation of the laser beams is usually not considered. The effect of wavefront propagation has been studied theoretically in the context of dual-species atom interferometer of longer baselines for testing the weak equivalence principle \cite{hu_analysis_2017, schubert_differential_2013}. Wavefront aberrations have furthermore been discussed in the context of satellite-based atomic gravitational wave detectors \cite{dimopoulos_atomic_2008, bender_comment_2011, dimopoulos_reply_2011,Hogan2011}.
We give here an estimate of the impact of wavefront distortions in a simple yet representative case, in order to derive requirements on the initial position and velocity jitter of the atom source. 

\subsubsection{Effect of curvature.}
\label{subsec:wf_curvature}
We consider two atom interferometers separated in x-direction by a baseline $L$, one near the beamsplitter and one near the retro-reflecting mirror. Assuming a  waist of the incident beam of $w_0$=50 mm at the position of the beamsplitter, corresponding Rayleigh length 1s $x_R \simeq \,$10 km, the radius of curvature at position $x$ along the baseline is given as $R(x) = x \sqrt{1+ (x_R/x)^2}$. The wavefront of the beam in the direction transverse to the propagation axis ($z$) is given by $\phi(r,x)=k_{l}r^2/[2R(x)]$. The  atom interferometer at the position $z$ is driven  by two counter-propagating beams of different curvatures, corresponding to a relative wavefront $\Delta\varphi_{las}(r,x)=\varphi_{las}(r,x)-\varphi_{las}(r,2L_{T}-x)=k_{l}r^2/(2R_{\text{eff}})$ with  $R_{\text{eff}}^{-1} = R(x)^{-1} - R(2L_{T}-x)^{-1}$. We will consider the largest possible contribution of this effect to the gradiometer phase shift, which corresponds to the relative phase between the most distant interferometers located at $x\simeq 0$ (close to the input telescope where $R_{\rm{eff}}\simeq R(2L_{T})$) and $x\simeq L_{T}$ (close to the retro-mirror where $\Delta\varphi_{las}(r,L_{T})\simeq 0$). To simplify notations, we drop out the $x$ dependence below.
The phase shift of the 4-pulse interferometer in the limit of infinitely short interrogation pulses (see Eq.~1) is given by 
\begin{equation}
   \Delta\phi=\Delta\varphi_{las}(r(0))-2\Delta\varphi_{las}(r(T))+2\Delta\varphi_{las}(r(3T))-\Delta\varphi_{las}(r(4T))\, ,
\end{equation}
with $4T$=800 ms the total interrogation time and $r(t)=r_0+v_0 t$ the transverse position of the atom in the laser beam at the time $t$ of the light pulse. In the case where the curvature of the two (lower and upper) beams is the same and where the light pulses occur well at the center ($r=0$), the contribution to the interferometer phase shift vanishes. 
A phase shift can appear in two cases: \textit{(i)} if the center of the beams are not well aligned with the atom trajectory, i.e. if the atom interacts with the beams away from the center of curvature (we denote these offsets as $\epsilon_{1,2}$) ; \textit{(ii)} in the case of a different curvature between the two beams ($R_{\text{eff},1}=R_{\text{eff},2}+\Delta R$), due, for example, to different waists $w_{0,1}=w_{0,2}+\Delta w$. Here, the subscript $1$ ($2$) refers to the bottom (top) beam where the $\pi/2$ ($\pi$) transitions are driven at $t=0$ and $t=4T$ ($t=T$ and $t=3T$). Finally, the atomic phase shift associated to the wavefront curvature becomes:
\begin{equation}
%   \Delta\Phi=N_{\text{LMT}} k \times \frac{\Delta R}{R_{\text{eff},1}^2} \Big[ 8(v_0T)^2 +4r_0 v_0T \Big].
\Delta\phi_{\rm{wf}} = 2nk_{l} \times \Big(-\frac{1}{R_{\text{eff},1}}[4 v_0 T(\epsilon_1+r_0)+8v_0^2 T^2] + \frac{1}{R_{\text{eff},2}}[4 v_0 T(\epsilon_2+r_0)+8v_0^2 T^2] \Big).
\end{equation}
As expected from the quadratic dependence of the wavefront, all the terms represent second-order corrections scaling as the product of either of two small parameters: $\epsilon_{1,2}$ (offsets from the center), $r_0$ (jitter in initial position) and $v_0$ (jitter in initial velocity).

For a numerical estimate with $2n=1000$, we calculate the effect of the coupling between centering error and velocity jitter (term  $\sim \epsilon_{1,2} v_0 T$).
In that case, constraining the differential phase noise at a level below $0.1\ \mu$rad/$\sqrt{\rm{Hz}}$ in the target frequency band then  requires  a control of atomic velocity jitter at the level of $v_0<0.8$ (nm/s)/$\sqrt{\rm{Hz}}$ for a centering error $\Delta\epsilon=|\epsilon_1 - \epsilon_2|=0.5$~mm and $w_{0,1}=w_{0,2}=50$~mm. The fine-tuning of a difference of waists $\Delta w$ might further reduce the total value of the phase shift $\Delta\phi_{\rm{wf}}$ and thus relax the requirement on atomic velocity jitter.

\subsubsection{Higher order effects.}
Higher order (than curvature) effects will occur as the laser beam reflects off a mirror of imperfect flatness, which causes a differential phase shift between the two distant atom interferometers as the  phase field evolves upon propagation.
Estimation of this effect highly depends on the phase map of the field after its reflection on the mirror \cite{karcher_improving_2018}, which is outside of the scope of this manuscript. We can nevertheless point towards a possible procedure to study this effect: \textit{(i)} characterization of the surface of the retro-mirror; \textit{(ii)} simulation of the propagation of the field with the phase map from the mirror imperfect planarity (e.g. with the angular spectrum method), to obtain the relative phase map at various locations along the detector baseline; \textit{(iii)} Monte-Carlo simulation of the sampling of this phase map by an atomic cloud, taking into account position and velocity initial jitters. The result of such a simulation would yield requirements on the initial position and velocity jitters as well as on the planarity of the retro-mirror.

%Here, we only considered spherical wavefront curvature in paraxial approximation, i.e. defocus aberrations. In this simplified case the dependence on the atom's center-of-mass position vanishes. This will not be true if higher order wavefront aberrations (typically modeled by Zernike polynomials) are considered and will put restraints on the initial position of the atomic sample. Moreover, previous studies \cite{louchet-chauvet_influence_2011, schkolnik_effect_2015,karcher_improving_2018} have shown that in order to fully model the effect of wavefront aberrations the exact shape of the wavefront needs to be known. While the influence of speckle on atom interferometers have been qualitatively studied before \cite{peters_high-precision_2001}, the effect of scattered light needs quantitative modeling  given the targeted precision and the vastly different experimental apparatus. One potential benefit of cavity assisted atom interferometry is the suppression of higher order spatial modes. However, the goal to drive large-momentum beamsplitters with $N=1000$ over a baseline of $L=\SI{10}{\kilo\meter}$ prohibits the use of high-finesse cavities  due to the bandwidth limit \cite{dovale-alvarez_fundamental_2017}.

\subsection{Scattered light and diffraction phase shifts}
\label{subsec:diff_phase_shifts}
Large momentum transfer atom interferometers are impacted by diffraction phases which originate from the non-resonant couplings between momentum states during Bragg diffraction \cite{Buechner2003,Estey2015}. These diffraction phases depend on the local effective Rabi frequency and therefore on the local laser  intensity. Inhomogeneities in intensity arise from the Gaussian profile of the laser beam, which can be mitigated by the use of a top-hat laser beam \cite{Mielec2018}. Still, intensity inhomogeneities remain because of scattered light in the process of beam shaping or upon propagation along the baseline \cite{Vinet1996coherent,Vinet1996stat}. The effect of light scattered off the walls of the vacuum tube or off the optics depends on the vibrations of these elements, which renders the evaluation of its impact on a particular detector a challenging task. Dedicated working groups tackle this challenge in the case of ground-based laser interferometers \cite{Ottaway2012,Canuel:13} or for the LISA mission \cite{Spector2012}.
Estimating the contribution of scattered light in the ELGAR detector scheme is therefore outside the scope of this paper.

Still, to provide an estimate of diffraction phase shifts in a simple configuration and hence derive a requirement for the intensity homogeneity of the laser beam, we numerically investigated diffraction phases for a Mach-Zehnder geometry and compared two different realizations of large momentum transfer beam splitter, implemented with either sequential Bragg transitions or combined Bragg transitions and Bloch oscillations. The numerical simulations were carried out in a position space approach solving Schr\"odinger's equation with the help of the split operator method \cite{javanainen2006symbolic}. The atom-light interaction describing Bragg diffraction as well as Bloch oscillations was modeled as described in Ref.~\cite{muller2008atom}.

%The sequential Bragg geometry consists of a sequence of seven Bragg pulses given by the standard $\frac{\pi}{2}, \pi, \frac{\pi}{2}$ pulses ($\Omega_{\frac{\pi}{2}}=1.05 \, \omega_r$, $\Omega_{\pi}=2.1 \, \omega_r $, $\tau$=25 $\mu$s) and four additional $\pi$ pulses ($\Omega_{\pi}=2.1 \, \omega_r $, $\tau$=25 $\mu$s), where $\omega_r$ is single-photon recoil frequency. These pulses are used to give the atoms an additional momentum kick of $2\hbar k$, such that the overall transferred momentum is $4 \, \hbar k$. For the second studied interferometer the four $\pi$ pulses are replaced by four Bloch sequences, each consisting of an adiabatic loading of the atoms into the Bloch lattice ($\Omega = 4 \, \omega_r$, $T_{\rm{loading}}$=500 $\mu$s), a frequency chirp to imprint a momentum of $2 \, \hbar k$ ($\Omega = 4 \, \omega_r$, $T_{\rm{chirp}}$=500 $\mu$s) and an adiabatic unloading to reach the $4 \, \hbar k$ momentum transfer ($\Omega = 4 \, \omega_r$, $T_{\rm{unloading}}$=500 $\mu$s). Both considered interferometers have by construction the same space-time area, in order to ensure a meaningful quantitative comparison. The numerical simulations are carried out in a position space approach solving Schr\"odinger's equation with the help of the split operator method \cite{javanainen2006symbolic}. The atom-light interaction describing Bragg diffraction as well as Bloch oscillations is modeled as described in Ref.~\cite{muller2008atom}.

We analyzed the influence of a change of the Rabi frequencies of the first $\pi$ pulse and the first Bloch sequence by $2\%$ and calculated the resulting phase shift by numerically performing a phase scan. For the sequential Bragg interferometer we found a phase shift of $\Delta \phi$=9.3 mrad and for the Bragg+Bloch geometry a shift of $\Delta \phi$=1.3 rad.
The larger Bragg+Bloch phase shift can be partly explained by the fact that in this geometry the atoms are trapped in the optical lattice for $1.5$ ms. This directly leads to an energy shift for the upper arm of the atoms, but not the lower one and therefore to an extra relative phase shift. 
This shows that when considering the effects of the diffraction phase the usage of sequential Bragg pulses is the preferable way to imprint momentum on the atoms. 

Scaling up the momentum splitting to the desired $1000 \, \hbar k$ (i.e. 500 sequential $2 \, \hbar k$ pulses), a coarse extrapolation leads to a phase bias $\Delta \phi$=4.6 rad (for a $2\%$ change of laser intensity). We emphasize that this number is very specific to the geometry studied. This estimate is anticipated to be  reduced (at least factor 10) by the integration over the atomic ensemble which averages the local laser intensity profile. When considering the differential phase between two distant AIs, the relative phase  bias will result from the difference in intensities of the common interrogation laser beam at the two positions along the baseline. Assuming a difference of $2\%$ in local laser intensities then yields a relative phase bias of the order of 460~mrad. The phase noise can then be estimated by studying the  mechanism  underlying relative laser intensity fluctuations (e.g. vibration noise on the walls of the vacuum chamber that affect the locally scattered light). Although we gave here a  coarse estimate, our simple calculation highlights the challenge in controlling diffraction phases at the $\mu$rad$\sqrt{\rm{Hz}}$ level  and conceiving mitigation schemes.
%These results are preliminary and further investigation is needed to develop a thorough understanding of the concrete nature of these phase shifts as well as possible mitigation strategies.

\subsection{Effect of inter-atomic interactions ($^{87}$Rb atoms)}
\label{subsec:atomic_interactions}

Following \cite{debs2011cold}, we estimate the phase noise resulting from interactions between the atoms and the fact that the beamsplitters in the proposed scheme are only characterized up to the shot noise level. This phase noise contribution is directly given by the mean-field energy term of the Gross-Pitaevskii equation averaged over atomic spatial distribution
\begin{equation}
    \Delta \phi_{\rm{MF}}=\frac{1}{\hbar} \int_0^{4T} dt \; \langle \Delta E_{\rm{MF}} \rangle = \frac{4T}{\hbar}\frac{\sqrt{N}U}{V_{\rm{atoms}}}\, ,
\end{equation}
where $U=\frac{4 \pi \hbar^2 a_s}{m_{at}}$, $a_s$ is the s-wave scattering length, and $V_{\rm{atoms}}=\frac{4}{3}\pi((\frac{7}{10})^{1/3}R_{\rm{TF}})^3$ is effective atomic volume with $R_{\rm{TF}}$ being the Thomas-Fermi radius of the BEC. Considering $4T=0.8$ s, $N=10^{12}$, $a_s = 95~a_0$ for a case of $^{87}$Rb ($a_0$ is Bohr radius) and $R_{\rm{TF}}=1 \ $mm we obtain an estimation of $\Delta \phi_{\rm{MF}} = 0.013$~rad.
A reduced phase noise can be reached by using bigger atomic clouds with smaller atom number, which would compromise the shot noise limited sensitivity. Alternatively, with 10 interleaved interferometers \cite{Savoie2018} and a squeezed atomic source by $20$ dB \cite{Hosten2016}, an atomic source with $N=10^9$ atoms would allow for equivalent sensitivity but significantly reduced phase noise.  An interaction-induced phase noise of $\Delta \phi_{\rm{MF}} =0.1 \ \mu\rm{rad}/\sqrt{Hz}$ would then correspond to a Thomas-Fermi radius of about $R_{\rm{TF}}=16$~mm. This requirement on the cloud size, while being challenging for BEC, seems feasible for a sufficiently cold thermal atomic ensemble with an expansion frozen by delta-kicked cooling technique.   
%These numbers are relaxed by the level of fluctuations of the atomic cloud size between the two different sensors of the gradiometer.

\externaldocument{noise_couplings}

\section*{Conclusions}
We have discussed the main technological bricks that constitute a preliminary design for the large scale atom interferometer ELGAR, that envisions to observe of gravitational waves in the deciHertz band. 
Based on the latest results in atom optics, we detailed the technologies of the main systems of ELGAR and pointed out the main developments needed in cold atom technology, suspension systems and Gravity Gradient Noise reduction techniques. We also realized new metrological studies to evaluate the impact of the main noise couplings to this detector.

The work presented here could be the base of a design study that will refine the requirements of the different systems of ELGAR, thanks to the development of a complete model of the antenna, from its metrology to its observable astrophysical sources. 
This work should also include a trade-off in terms of cost and performances for the realization of ELGAR on different candidate sites, such as the LSBB in France or the Sos Enattos and Seruci/Nuraxi sites in Italy. 
Together with a design study that would form a road map of ELGAR and identify the key experimental developments required, the realization of the antenna could be sustained from a growing number of national initiatives in Europe for the study of large scale atom interferometry, such as MIGA~\cite{Canuel2018} in France, VLBAI~\cite{Schlippert_2020} in Germany, or AION~\cite{Badurina_2020} in United Kingdom. 

\section*{Acknowledgments}
This work was realized with the financial support of the French State through the ``Agence Nationale de la Recherche'' (ANR) in the frame of the ``MRSEI'' program (Pre-ELGAR ANR-17-MRS5-0004-01) and the ``Investissement d'Avenir'' program (Equipex MIGA: ANR-11-EQPX-0028, IdEx Bordeaux - LAPHIA: ANR-10-IDEX-03-02). A.B. acknowledges support from the ANR (project EOSBECMR), IdEx Bordeaux - LAPHIA (project OE-TWR), the QuantERA ERA-NET (project TAIOL) and the Aquitaine Region (projets IASIG3D and USOFF). The work was also supported by the German Space Agency (DLR) with funds provided by the Federal Ministry for Economic Affairs and Energy (BMWi) due to an enactment of the German Bundestag under Grant Nos.~50WM1556, 50WM1956 and 50WP1706 as well as through the DLR Institutes DLR-SI and DLR-QT.
X.Z. thanks the China Scholarships Council (N$^\mathrm{o}$ 201806010364) program for financial support. J.J. thanks ``Association Nationale de la Recherche et de la Technologie'' for financial support (N$^\mathrm{o}$ 2018/1565).
P.A.-S., M.N., and C.F.S. acknowledge support from contracts ESP2015-67234-P and ESP2017-90084-P from the Ministry of Economy and Business of Spain (MINECO), and from contract 2017-SGR-1469 from AGAUR (Catalan government). L.A.S. thanks Sorbonne Universit\'es (Emergence project LORINVACC) and Conseil Scientifique de l'Observatoire de Paris for funding. R.G. acknowledges Ville de Paris (Emergence programme HSENS-MWGRAV), ANR (project PIMAI) and the Fundamental Physics and Gravitational Waves (PhyFOG) programme of Observatoire de Paris for support. We also acknowledge networking support by the COST actions GWverse CA16104 and AtomQT CA16221 (Horizon 2020 Framework Programme of the European Union).
D.S. gratefully acknowledges funding by the Federal Ministry of Education and Research (BMBF) through the funding program Photonics Research Germany under contract number 13N14875.
Sv.Ab., N.G., S.L., E.M.R., D.S., and C.S. gratefully acknowledge support by ``Nieders\"achsisches Vorab" through the ``Quantum- and Nano- Metrology (QUANOMET)" initiative within the project QT3, and through ``F\"orderung von Wissenschaft und Technik in Forschung und Lehre" for the initial funding of research in the new DLR-SI Institute, the CRC 1227 DQ-mat within the projects A05 and B07, the Deutsche Forschungsgemeinschaft (DFG, German Research Foundation) under Germany’s Excellence Strategy – EXC-2123 QuantumFrontiers – 390837967 (B2), and the German Space Agency (DLR) with funds provided by the Federal Ministry for Economic Affairs and Energy (BMWi) due to an enactment of the German Bundestag under Grants No. DLR~50WM1641 (PRIMUS-III), 50WM1952 (QUANTUS-V-Fallturm), and 50WP1700 (BECCAL), 50WM1861 (CAL), 50WM2060 (CARIOQA) as well as 50RK1957 (QGYRO).
\section*{References}
\bibliographystyle{iopart-num}
\bibliography{ELGARTechno}
\end{document}